\documentclass[a4paper,12pt]{article}

\usepackage{amssymb,amsmath,mathbbol,mathrsfs}
\usepackage[usenames,dvipsnames]{color}
\usepackage{hyperref}
\usepackage{xcolor}
\usepackage{stmaryrd}
\usepackage{authblk}
\usepackage{framed}
\usepackage{empheq}
\usepackage{slashed}
\usepackage{hyphenat}
\usepackage{cite}
\usepackage{apacite}
\usepackage{dsfont}
\usepackage{natbib}
\usepackage[utf8]{inputenc} 
\usepackage{pifont}% http://ctan.org/pkg/pifont
\newcommand{\cmark}{\ding{51}}%
\newcommand{\xmark}{\ding{55}}

\usepackage{marginnote}    

\usepackage[left=.8in,right=.8in,top=.8in,bottom=.8in]{geometry}                  % For good copy.

\usepackage{sectsty} 
\allsectionsfont{\sffamily\mdseries\upshape} % (See the fntguide.pdf for font help)
\usepackage{tocloft}

\makeatletter
\renewenvironment{abstract}{%
    \if@twocolumn
      \section*{\abstractname}%
    \else %% <- here I've removed \small
      \begin{center}%
        {\bfseries\sffamily\abstractname\vspace{\z@}}%  %% <- here I've added \Large
      \end{center}%
      \quotation
    \fi}
    {\if@twocolumn\else\endquotation\fi}
\makeatother

\numberwithin{equation}{section}
5

\hypersetup{
	colorlinks=true,         
	linkcolor=MidnightBlue,          
	citecolor=BrickRed,        
	urlcolor=MidnightBlue            
}

\newcommand{\be}{\begin{equation}}
\newcommand{\ee}{\end{equation}}

\newcommand{\F}{{\Phi}}
\renewcommand{\d}{{\mathrm{d}}}
\newcommand{\D}{{\mathrm{D}}}

\newcommand{\U}{{\mathrm{U}}}

\newcommand{\G}{{\mathcal{G}}}

\newcommand{\pp}{{\partial}}

\renewcommand{\bar}{\overline}
\newcommand{\dd}{{\mathbb{d}}}

\newcommand{\RR}{\mathds{R}} %Isham-style

\newcommand{\cint}{{\int\kern-.87em{<}}}
\newcommand{\sint}{{\int\kern-.75em{\sim}}}
\newcommand{\fint}{{\int\kern-1.00em{\int}}}

\newcommand{\bb}{\mathbb}

\renewcommand{\#}{\sharp}
\let\oldmarginpar\marginpar
\renewcommand\marginpar[1]{\oldmarginpar{\color{red}\raggedright\footnotesize #1}}

\newcommand{\old}{\color{red}}

\title{The role of representational conventions in assessing the empirical significance of symmetries}
\author{Henrique Gomes \footnote{\href{mailto:gomes.ha@gmail.com}{gomes.ha@gmail.com}} \\\it University of Cambridge\\ \it Trinity College, CB2 1TQ, United Kingdom}
\begin{document}
\maketitle
\abstract{
This paper explicates the direct empirical significance  (DES)  of symmetries in gauge theory, with comparisons to classical mechanics. Given a physical system composed of subsystems, such significance is to be awarded to physical differences of the composite system that arise from symmetries acting solely on its subsystems. So my overarching main question is: can DES be associated to the local gauge symmetries, acting solely on subsystems? 

In local gauge theories, any quantity with physical significance  must be a gauge-invariant quantity.
 To attack the question of DES from this gauge-invariant angle, we require a split of the state into its physical and its representational content: a split that is relative to a representational convention, or a gauge-fixing. 
 Using this method, we propose a rigorous definition of DES, valid for any state. 
 This definition fills the gaps in influential previous construals of DES, \citep{GreavesWallace, Wallace2019, Wallace2019a, Wallace2019b}. In particular, Wallace's need to specialize to `generic' states is explained and dispensed with.%, which requires states to be \emph{generic}. %Taken at face value, such a definition would eliminate the possibility of DES for internal boundaries, but for external boundaries, we recover the same results. 
% We argue that: (i) DES should not be associated to local gauge symmetries, only to the rigid (or global) components of these symmetries; (ii) in the field theoretic case, DES can be either relational or intrinsic to the boundary; (iii) while  there is no  relational DES in vacuum  in a simply-connected manifold; (iv) in point-particle mechanics, we \textit{do} recover the recent literature's standard notions of DES, as exemplified by the thought-experiment known as `Galileo's ship'; finally (v)   we recover previous construals of DES \citep{GreavesWallace, Wallace2019b}. 
}
\tableofcontents
\section{Introduction}
\subsection{The debate}\label{sec:debate}
Symmetries of the whole Universe are widely regarded as not being directly observable: that is,  as  having no direct  empirical significance. At the same time, it is widely accepted that some of these symmetries, such as velocity boosts in classical or relativistic mechanics (Galilean or Lorentz  boosts),  are observable when applied solely to subsystems.  Thus Galileo's famous thought-experiment about the ship---that a process involving some set of  relevant physical quantities in the cabin below decks proceeds in exactly the same way,  whether or not the ship is moving uniformly relative to the shore---is used to show that subsystem boosts have a direct, albeit strictly relational, empirical significance. For while the inertial state of motion of the ship is undetectable by experimenters confined to the cabin, the entire system, composed of ship and sea,  registers the difference between two such motions, namely in the relative velocity between ship and sea.\footnote{There is always some amount of approximation in these notions: we ignore the ripples produced by the ship,  assume the sailors don't  carry a GPS or look outside for that matter, etc. }

Thus the broad notion of `direct empirical significance'  of a symmetry amounts to the existence  of transformations of the universe possessing the  following two properties (articulated in this way by \citep{BradingBrown}, following \citep{Kosso}): \\
\indent \,(i): \emph{Global Variance}--- the transformation applied to the Universe in one state should lead to an empirically different  state; and  yet\\
\indent  (ii): \emph{Subsystem Invariance}---the transformation should be a symmetry of the subsystem in question (e.g. Galileo's ship), i.e. involve no change in quantities solely about the subsystem. \\
I will take the concept of `directly empirically significant subsystem symmetries' (DES henceforth) to imply observability of those symmetries; but I will prefer the use of the label DES as opposed to `observable' since I  take it to connote an action of a symmetry on a subsystem.

Whether the concept of DES extends to local gauge theories is less settled.\footnote{DES has been discussed in \citep{Kosso, BradingBrown, GreavesWallace, Healey2009, Teh_emp, Friederich2014, Ladyman_DES,  Friederich2017,  GomesStudies, Gomes_new, Teh_abandon, Wallace2019a, Wallace2019b, Seb_ramirez, Chasova}. None of these completely encapsulate my own views, but there are large, though of course varying,  overlaps of agreement with each. % Besides my own previous work, which I still endorse, here I broadly side with \citep{Kosso, BradingBrown, Healey2009, Teh_emp, Friederich2014, Ladyman_DES,  Friederich2017,   Teh_abandon, Wallace2019b} even if I disagree with their reasons.
} 
  Local gauge symmetries are normally taken to encode descriptive redundancy, which suggests  local gauge theories cannot illustrate the concept of DES. For surely, a ``freedom to redescribe''  could not be observable. 
This argument was  developed in detail by Brading and Brown \citep{BradingBrown}. They take themselves---I think rightly, in this respect---to be articulating the traditional or orthodox answer. In the case of gauge theory, this answer differs from  \citet[p. 110]{thooft}'s claim, that applying distinct rigid phase shifts in the two arms of a beam-splitter experiment would alter the interference pattern. Brading and Brown point out that distinct phase shifts would produce a non-continuous gauge transformation: assuming the two subsytems are contiguous, there would be a mismatch of the phase rotation at the interface,  which should not be an allowed operation, on mathematical and physical grounds. 

 Building on \citep{Healey2009} and \citep{BradingBrown}, Greaves and Wallace resist the orthodoxy by articulating DES for a local gauge theory differently  \citep{GreavesWallace}. They point out that, since gauge transformations are not physical transformations, we should not demand that they remain continuous at the interface between contiguous subsystems. Rather, we should demand only a continuous transition between the \emph{states} representing such  subsystems. \cite{Gomes_new} argues that this amendment is correct, but it does not go far enough: what matters is a continuous transition \emph{between the physical states of the two subsystems}. Here I will argue that this last demand is more accurate, and it recovers Greaves and Wallace's focus on states for a large number of cases---\emph{just as long as we fix and keep track of representational conventions for the states}, as I will explain in due course.
 
 And indeed,  Greaves and Wallace's treatment of DES for local gauge theories bears many similarities to ours here. They focus on subsystems as given by regions and they identify transformations possessing properties  (i) and (ii)  by first formulating the putative effects of such transformations on the gauge fields in these regions. A more refined treatment that takes into consideration extensions of the symmetries to the measuring apparatus (or subsystems) was developed in \citep{Wallace2019} and applied to particle mechanics in \citep{Wallace2019a} and field theory in \citep{Wallace2019b}. To settle the question of whether local gauge symmetry
can be said to have relational `empirical significance' in the sense of Galileo's ship---viz. in the sense that certain subsystem symmetries can be used to effect a relational difference between a
subsystem (e.g. the ship system) and an environment (e.g. the shore)---I will in this paper follow fairly similar routes as Wallace. The two main differences are that:\\
\indent (1): I will be explicit about the need for, and use of, representational conventions. This first demand  is in line with \cite{Gomes_new}: it is a consequence of the   focus on the physical, as opposed to the representational, content of the states---a focus that is necessary in order to assess physical significance.\\
\indent (2): My  treatment of the boundary of subsystems---in particular the relation between non-asymptotic and asymptotic boundaries---is different. I believe we should first understand how gauge symmetry behaves in the non-asymptotic case and then translate that understanding to the asymptotic case; whereas \cite{Wallace2019, Wallace2019b} goes in the other direction.\\
 So naturally, my conclusions will differ somewhat from the previous literature.\\

That is the debate  this paper aims to resolve. I will show that there is a general, coherent formalization of  DES which yields (a) the aforementioned Galilean  symmetries in the ship-scenario (and its relativistic analogue), but which (b) yields no non-trivial realizer of the concept in the case of \textit{local} gauge symmetries.  DES appears only in certain circumstances, and then only related to global---which we will here call `rigid', as introduced in \citep{Gomes_new}--- gauge transformations.\footnote{ Those transformations depending on a finite number of parameters; i.e. not definable point by point, are the ones called rigid, or global. Their status as regards DES  is (much) less disputed. } Since the case of local gauge theories is the contentious one, it will be my focus in this paper.

\subsection{Roadmap}
 In Section \ref{sec:fund_sym}, I  will lay out some of the conceptual background assumptions, required for the rest of the paper. This Section will consider definitions of subsystems, what notion of isolation is required, and how we construe, mathematically and conceptually, the symmetries that should act intrinsically on the subsystem in comparison with those that act on the universe as a whole.
   
 In  Section \ref{sec:gen_DES}, I  will give all the general ingredients for DES. In particular I will show why the use of representational conventions is necessary for articulating DES. In this Section  I will  provide the general structure of gauge symmetry and representational conventions that I will need; I will justify the unobservability of quantities that are variant under symmetries of the entire universe; I will discuss the composition of subsystems using representational conventions; and then finally I will show how to extract empirical significance for the entire universe from subsystem symmetries. 
 
  In Section \ref{sec:gt},  I summarize  the treatment of gauge theories of fields. %\footnote{The specific form of the action functional will not play a role in what follows, since the statements I am interested in  occur already at the level of kinematics: gauge-invariance of the action is all that matters.}
  I give more detail about the obstructions to defining regional gauge-invariant dynamical structure, and discuss in brief two ways to overcome this obstruction: edge modes and a careful use of representational conventions that are not taken  a priori as anchored to the boundary. I then summarise the details of DES in this context,  but leave details of particular examples to the numerous appendices. In Section  \ref{sec:conclusions} I conclude.  
  
   In Appendix \ref{sec:gf_Lorentz} I  show that no non-trivial realizer of DES exists   for electromagnetism in a simply-connected Universe in the  vacuum sector of the theory (i.e. no matter field). It exists in the sector in which there is matter field but not in the interface between the regions.  In Appendix \ref{sec:holonomy}, I will treat the same sectors using holonomy variables, and find the same  conclusions. In Appendix \ref{sec:ext_sub}, I look at what I will call the `externalist case'.  In Appendix \ref{sec:particle}, as a consistency check,   I  apply the same criteria for DES  to  theory of particles; and obtain both the Gallilean symmetries of Galileo's ship thought experiment, and the uniformly accelearated solutions representing the Einstein's elevator type of scenario.%Section  \ref{sec:comparison}, I reassess the standard derivation of DES in the same context---vacuum, simply-connected Universe, under the internalist's notion of subsystem---and flag its problematic assumptions.

\section{Background assumptions  }\label{sec:fund_sym}
In the received view,  gauge theory accords  physical reality only to  certain quantities: those that are invariant under a  class of transformations labeled `gauge'. These transformations are usually construed as mere redescriptions of the same physical state of affairs. While this construal of gauge theory  is business as usual,  it might seem at first sight inimical to gauge symmetris having DES. 
 
However, according to Section \ref{sec:debate}'s condition (ii): \emph{Subsystem Invariance},  empirical significance must involve subsystems, and subsystems introduce a crucial novelty:  we leave  ``God's vantage-point'' for a more regional one. That is: we  assume our access is restricted   to quantities solely about the subsystem---as illustrated by the sailors being cooped up inside the cabin of Galileo's ship. In this context,  subsystems bring  in what amount to epistemic considerations additional to ontological ones.%\footnote{Nonetheless, subsystems themselves may still be characterized from the ``God's-eye''  point of view. For instance, by introducing a spacetime internal boundary and stipulating the flux of some select range of physical quantities through this boundary. } 
 
 But `subsystem' is a vague concept. In this Section, we will constrain that concept so that it is suitable for our investigations of DES. Section \ref{sec:sub_rec} gives a first gloss on the idea of a subsystem as reflecting important kinematical features of the larger system of which it is a part. Section \ref{sec:bdary_sym} describes the relationship between symmetries and subsystems that are defined by boundaries. Section \ref{sec:rep_conv_DES} describes the necessity and use of representational conventions in assessing DES. 
 
 \subsection{Kinematical subsystem recursivity}\label{sec:sub_rec}
 
 In general, observations are  modelled as being made from outside the subsystem being studied.  Therefore, the importance of  subsystems  to understanding the observability of symmetries is relatively uncontroversial. As \citet[p. 4]{Wallace2019a} points out: 
\begin{quote}Observations are physical processes, but they are not normally modelled
explicitly within the system being studied, but are considered as external
interventions. [... Then  a]  dynamical symmetry has implications for observability of
physical quantities [...] when the symmetry can be extended so as
to apply also to the dynamics of those interventions. [...]
\end{quote}
I think this is right, and it leads to a natural understanding of Section \ref{sec:debate}'s requirements (i) and (ii):  \emph{Global Variance} and \emph{Subsystem Invariance}.  Indeed, when focusing on subsystems of the Universe, Wallace says (p. 4, ibid): ``it becomes relatively simple to understand  modal questions in more
directly empirical terms: is a situation where the symmetry transformation
is applied to this system, but not to other systems, the same as or
different from the original situation?''. The answer to the question is that the situation should be the same when that subsystem symmetry is part of a global symmetry, and different when it is not.

Thus the empirical significance of a symmetry hinges on how a  symmetry, when applied to a subsystem, extends to a larger system of which it is a part; the complementary subsystem is then interpreted as representing the `environment', or a `measuring apparatus'. The main question surrounding empirical significane  then is about how global symmetries relate to subsystem symmetries. And here I will consider only those theories which satisfy \emph{subsystem recursivity}, i.e. theories that  (p. 5-9, ibid) 
\begin{quote} have the remarkable and
underappreciated feature of being able to reinterpret subsystems of their models, when dynamically isolated, as other models of the same theory. [... in these cases]  any model can be interpreted [...as a]
dynamically isolated subsystem under certain idealizations about its environment
and where, if we want to remove those idealizations, we can embed the
model in a model of a larger system within the same theory---and where that
larger system in turn is interpretable in the first instance as a subsystem of a
still-larger system.
\end{quote}
As Wallace argues, `dynamical isolation' is a term of art in physics, but we will not need to be more precise about this, except that we need to assume  that isolation entails a weak form of dynamical autonomy. 

 Unpacking Wallace's definition of dynamical autonomy, I take it to mean that the dynamical equations governing the motion for the subsystem, up to the level of approximation required by the situation at hand,  does not depend on the details of the rest of the system, except insofar as the rest of the system defines \emph{initial} boundary conditions for the subsystem. %The framing of (1) allows a much broader definition of isolated subsystem, in which one also includes a possibly non-trivial evolution of boundary conditions on the subsystem.   %: ``we can write down closed-form equations of motion for the subsystem that, to a good degree of approximation, govern the dynamics of its constituents, with the details of the rest of the system irrelevant except insofar as they do or do not satisfy the isolation condition.'' here are essentially two meanings of dynamical autonomy, one more inclusive than the other. Namely, d either: (1)  do not depend  on  the details of the rest of the system,  except insofar as they do or do not satisfy the isolation condition; or  the isolation condition is defined by a boundary condition whose evolution

But here I am only interested in the behavior of symmetries of the laws, at both subsystem and global levels. And since there is a sense in which symmetries can be seen as `laws on laws', or `metalaws' (see \cite{Lange_meta}), my requirement about dynamical isolation and subsystem recursivity can  be weaked in two senses. 

First, I will only be interested in whether the subsystem enjoys the `same type' of symmetries as the larger system in which it is embedded. This will be labeled \emph{downward consistency}. This weaker requirement allows evolving boundary conditions, if they are symmetry-invariant.

Second,  an isolation condition may only hold   for a certain interval of time, $I\subset \RR$. But I do not want to focus here on the loss of autonomy over time, and so I will only require some small $I\neq 0$.   Thus, differently from Wallace, I will focus on the relation between system and subsystem symmetries \emph{for initial states}; assuming only that some $|I|=\delta> 0$ exists in which downward consistency is satisfied. 

In the case of gauge theories, the arguments of this paper will require only such kinematical considerations. %More specific isolation conditions would become important once we attempted to extend  our conclusions about initial states to some given interval of time.  
Due to the locality of the interactions,  the weakened form of kinematical isolation---that is only required to  obey downward consistency---can always be satisfied by any subsystem defined through a partition of space, as we will see in Section \ref{sec:gt}. %That is, downward consistency can hold even if boundary conditions on the subsystem change over time. 

In the particle theory case, the assumption that the subsystem dynamics inherits the symmetries of the larger universe requires stronger isolation conditions, but these can be encapsulated in our embedding of the subsystem into the larger universe, as done in Section \ref{sec:particle}.

I believe such a kinematical understanding of subsystem recursivity about symmetries can accommodate our intuitions about, and the familiar examples of, direct empirical significance. Consider, for simplicity, a  Galileo's ship scenario with the shore (not the sea) taken as the environment, in which the subsystem at $t=0$ is inertial and at a finite distance $d$ from the shore. Now, for a fixed time interval $I$, the boosts must be pared down to a scale given by $d/I$. But we are not concerned with these `practical matters' when describing the subsystem symmetries; we use certain idealizations, e.g. that the shore is infinitely far away, so that $d\rightarrow \infty$. Here  I prefer a different idealization, in which I take $I$ to be  small. Thus the kinematical understanding of subsystem recursivity   avoids some of the fuzziness of dynamical isolation, and yet has the resources to articulate a fruitful construal of DES.
 \\

%Nonetheless, the question os kinematical subsystem recursivity remains, and it will be central in what follows. 

With Wallace, I will take subsystems to be represented as elements $X$ of  a collection  $\Xi$, so $X\in \Xi$. The collection $\Xi$ is partially ordered by inclusion, and bounded by a minimal and a maximal element---representing the empty set and the entire universe, respectively. And we define a state space for each $X$, $\Phi_X$, such that the state spaces respect the partial ordering. Namely,  for $X\subset Y$, we define  $\iota_{XY}$ as the inclusion map (or embedding), $\iota_{XY}:X\rightarrow Y$, and, schematically,  $r_{YX}$ as the restriction map: $r_{YX}:Y\rightarrow Y_{|X}$, with  $\iota_{XY}\circ r_{YX}=\mathsf{Id}_X$. 
The idea is that the restriction on the subsystems gets `pulled-back'  to a restriction on the state spaces, which we can here schematically denote: 
\be\label{eq:restriction} \iota^*_{XY}:\Phi_Y\rightarrow \Phi_X.\ee  
We denote it thus since, in cases of interest in field theory, this ``restriction map'' is really a type of pull-back.\footnote{ For cases of interest our state space will be the space of sections of some vector bundle (cf. footnote \ref{sec:sub_rec}). If $M$ is the base space of a vector bundle $E$, and $\iota:N\rightarrow M$ is an embedding map, then $\iota^*E$ defines a vector bundle over $N$ by pull-back (i.e. the fiber over $x\in N$ is the fiber over $\iota(x)\in M$).}  And also with Wallace, we  assume ``upwards consistency'': given $X\subset Y$, for a given $\varphi_X\in \Phi_X$, there exists a $y\in \Phi_Y$ such that $r_{XY}(\phi_Y)=\varphi_X$. This means that any subsystem state is compatible with some global state. %An isolation condition would allow us to maintain the same collection of subsystems throughout the time interval $I$. 

Differently from Wallace, I will further demand that  the restriction $r_{XY}$ of \eqref{eq:restriction}  co-varies with the symmetries of $Y$, if there are any. Namely, if $g_Y$ is any symmetry of $\Phi_Y$, i.e. a certain type of automorphism $g_Y:\Phi_Y\rightarrow \Phi_Y$ (cf. footnote \ref{ftnt:syms_def}), then the composition is also a symmetry of the subsystem. In other words, I will demand that subsystems 
 satisfy \emph{downward consistency:} For $g_Y$ a symmetry of $\Phi_Y$ and any $\varphi_Y$ that restricts to a $\varphi_X\in \F_X$ of the subsystem, i.e. $\varphi_X=  \iota^*_{XY}\varphi_Y$, 
\be\label{eq:dwc}\iota^*_{XY}(g_Y(\varphi_Y)) \,\text{ and}\,\,  \varphi_X\,\, \text{are symmetry-related in}\,\,\Phi_X,\ee 
where $\Phi_X$ is understood to have its own dynamics, possibly with time-varying boundary conditions. We can equivalently rewrite \eqref{eq:dwc} as: 
\be \iota^*_{XY}g_Y=g_X\iota^*_{XY}, \quad\text{for some}\,\,g_X\,\,\text{symmetry of $\F_X$.}
\ee 
This is a watered-down version of subsystem-recursivity: all we require from our definitions of subsystem is that the  symmetries are recursive in this way. 

  Downward consistency demands  that the embedding should be symmetry-invariant from the perspective of the entire universe. But gauge theories are local field theories with no action at a distance, and thus already have in-built a weak notion of dynamical isolation of disconnected subsystems.  Thus, in their case, I need only demand that  subsystems that are demarcated by boundaries (as we will define them in  Section \ref{sec:two_bdaries}) satisfy \eqref{eq:dwc}. This is a necessary and suffient condition for my weaker notion of subsystem recursivity. And indeed, if it is satisfied, due to the locality of interactions in gauge theory,  the regional dynamics can be reinterpreted as the dynamics of other models of the same theory (even if we allow general  evolution of boundary conditions). In other words, if \eqref{eq:dwc} is satisfied, the \emph{local} equations  governing a subsystem that is demarcated by a boundary are identical to those governing a larger bounded system of which it is a part; and, to the extent that boundary conditions differ, that difference does not pare down the symmetry group of the subsystem equations of motion.\footnote{ For instance, in these subsystems, one can always find a representational convention in which the evolution equations are hyperbolic.  There are certain complications with elliptic initial value problems, which are to a certain extent non-local. But these complications are under control, as  will be discussed in Section \ref{sec:subs}.} 
 Thus downward consistency provides a consistent---though weaker, in the sense that subsystems do not be idealized as infinitely far-apart---notion of subsystem-recursivity. And just to give an example, it is this weaker notion that would allow us to model the interior and the exterior of black holes as subsystem and environment: a case of great interest. 
  
In the particle theory case, things are subtler, since forces act a distance. %

\subsection{Symmetries and boundaries}\label{sec:bdary_sym}
 In what follows I will mostly concentrate on the case of field theories; the considerations for particle systems will differ and will be left for Appendix \ref{sec:particle}.  In Section \ref{sec:two_syms}, I will classify symmetries into two general sorts, in a way that is useful for the study of subsystems.  In Section \ref{sec:two_bdaries} I similarly assess two notions of boundaries, that are naturally paired with the two notions of symmetry. And in Section \ref{sec:int_bdary}, I discuss the interplay between these notions of boundary, subsystems and symmetry.% In particular, in the gauge theory case,  I will introduce a weaker notion of boundary, that only requires that the subsystem symmetries are the ones it inherits from the symmetries of the larger universe. 

\subsubsection{Two notions of symmetry}\label{sec:two_syms}

First, it is helpful to distinguish two types of  symmetries:\footnote{There are many precursors to this distinction. \cite[Section 5.1]{ButterfieldHaro_sym}'s idea of stipulated vs accidental symmetries for instance. These roughly correspond to fundamental and dynamical, but are seen as mutually exclusive since the label refers to their origin only.  Or \cite{Dasgupta_sym}'s formal and ontic symmetries. Formal  definitions ``define [the notion of symmetry] in purely
formal, set-theoretic terms'', p. 861; while  an ontic ``definition of  symmetry
[...] requires a symmetry to preserve the laws and preserve certain
privileged physical features'' p. 862. I would add that, with regards to scientific practice, a hard and fast distinction would  over-simplify depiction of how  science historically homes in on suitable Lagrangians and associated symmetries. In practice, symmetries that are eventually classified as  `fundamental' can first appear dynamically through the invariance of a Lagrangian but are then elevated to fundamental status and serve as a guiding principle.}  \\

\noindent$\bullet$\textit{`Fundamental symmetries'}: The symmetry is given in purely formal terms. A symmetry group   is \textit{defined} as being gauge. So invariance under transformations of the states  constrains the laws to respect the symmetries. In this case, there is a symmetry principle, simpliciter, in play; dynamics comes only after. \\

\noindent$\bullet$\textit{`Dynamical symmetries'}:  we define the symmetry transformations as those that leave relevant structures---the state space and  the Lagrangian or the Hamiltonian of the theory--- invariant.\footnote{This is a simplification: under this general definition we run the risk of allowing  models which we would intuitively take to depict physically distinct situations as  nonetheless symmetry-related.  \citep{Belot_sym} gives an exposition of the obstacles  to a general definition.  My definition is closest to what \cite[p. 3]{Wallace2019} dubs the `representational strategy', which ``instead builds the representational equivalence of symmetry-related models into the definition [of symmetry], usually by requiring that symmetries
are automorphisms of the appropriate mathematical space of models (hence
preserve all structure, and thus all representation-apt features, of a model)''. As discussed in \cite[Section 3.3]{Samediff_1}, this definition is still not ideal, since it is slightly circular:  structure can be defined implicitly by the symmetry-relation, whatever that is. More generally, I endorse the  account of dynamical symmetry in 
\cite[Section 1.2]{Samediff_1}. For our purposes in this paper, the vaguer definition above suffices. \label{ftnt:syms}} In this case, the symmetries are subservient to, i.e. entirely determined by, the particular features  of the state space (e.g. phase space) and of the action functional. Broadly, the laws determine the symmetries; here there is no `symmetry-principle', simpliciter, in play.\footnote{In \cite[p. 8]{Teh_abandon}, this distinction has a different label: (A) and (B), with (A) corresponding  to `Fundamental' and (B)  (roughly) to  `Dynamical'. They describe the latter as ``a more refined [...] notion according to which an (A)-type gauge symmetry is further required to encapsulate redundancy for a particular subservient system, whose states can only be defined after fixing specific boundary conditions''. They make a slightly different categorization of (B): it is a subset of (A) that is required to obey boundary conditions. %They say (p. 8): ``(A) a generic---possibly non-dynamical---notion of gauge symmetry (in which any map from spacetime to a structure group acting on fields in a certain way is a gauge symmetry); and (B) a more refined dynamical notion according to which an (A)-type gauge symmetry is further required to encapsulate redundancy for a particular dynamical system, whose states can only be defined after fixing specific boundary conditions''. 
\label{ftnt:syms_def}} 
\\

% Although such a simplistic depiction is not always faithful to how  science historically homes in on suitable Lagrangians and associated symmetries, the broad characterization is useful. 

So,  given a fundamental symmetry Lie group $G$, acting on some fields over a space or spacetime manifold, $M$ with value space $F$---here taken to be a vector space---the fundamental symmetry will be deemed to act uniformly over $M$. Thus I want to highlight an asymmetry: fundamental symmetries are  judged to be dynamical, for the appropriate dynamical structures. But dynamical symmetries are not necessarily fundamental: for example, dynamical symmetries of a field theory may be  different on the bulk and boundary of a manifold, and in this case they should not count as fundamental. Indeed, one of the central points I want to argue for here is that for field theories there are two notions of boundary; and for one of them only the dynamical type of symmetries is a natural notion. 

%And although only dynamical symmetries should have any use in modelling physical situations. 
Another central point that I want to argue for is that theoretical parsimony and consistency between system and subsystem push us to formulations of subsystems in which the two types of symmetry match. Thus, clearly, downward consistency will have consequences for how we understand DES. But first,  we need to develop our ideas about symmetries and boundaries.

%\subsection{Two ways of introducing boundaries}\label{sec:bdary_disc}{sec:gt}

\subsubsection{Two notions of boundary}\label{sec:two_bdaries}\label{sec:ext_bdary}
 
Both the Lagrangian and the Hamiltonian formulations of field-theory refer to the fields over the entire universe; we are at first not given any   subsystems that DES can latch on to. Subsystems must be somehow ``conjured into being'', and there are two general ways of doing this in field theory: one  by introducing a boundary internal to the entire universe; and the other by introducing a boundary  `external' to it. In other words, we can introduce subsystems either by boundaries that define an inside and an outside---the boundaries have two sides---or by ones that define only an inside; boundaries that are one-sided, so to speak. 
 
Thus an\\
\indent \emph{External boundary}: imposes a  boundary on the whole universe; and an \\
\indent \emph{Internal boundary}: imposes  boundaries within the bulk of the universe.\\
  In the former case we have a bounded manifold representing the entire universe, and in the latter case we get complementary  subsystems demarcated by internal divisions of the entire universe, which, as a whole, is assumed  boundary-less.  
  
  In the first, external way,   the entire universe is taken as a type of subsystem: the `environment' label can be loosely attached to the boundary itself. %Moreover, in the external case, the boundary is not just a lower-dimensional dynamical system: it will be taken to have no dynamics of its own. The boundary states are just postulated as frozen. 
  In the second, internal way, if one has only two such subsystems, we can label them `subsystem and environment'. % Here I will assume the boundary has no dynamics of its own. If the boundary has its own dynamics, and since there is no boundary of the boundary,  then by including the boundary dynamics we are again dealing with a closed, boundary-less  universe. So my notion of external boundary here should really be seen as yielding fixed boundary conditions, and is in contrast with the internal notion of boundaries, whose dynamics are whatever it inherits from the ambient dynamics, and is thus not fixed or frozen. 
  
 %A boundary with its own dynamics is taken to be a very special and strange case of the internal view. 
 
\paragraph*{The external boundary}

In the first, external way, downward consistency, \eqref{eq:dwc}, is satisfied vacuously. That is,  since the entire universe is considered as a subsystem, no consistency conditions with the symmetries of a larger system can arise. In other words, downward consistency is a condition about the boundary as seen from both the inside and the  outside; therefore it does not substantively apply to a one-sided boundary. 

Thus, in specifiying the state space and dynamics to which the dynamical symmetries are subservient, there is no obstacle to imposing a fixed representation of the states at the boundary. For instance, one could say: ``the configuration space with which I am dealing possesses only one representative of the gauge potential at the boundary''. As we will discuss in the next section, one could not have the same type of restriction for an internal boundary without flouting downward consistency. 
  
 Of course, if boundary states are pared down, or restricted, they offer an anchor to the representational conventions of the rest of the system. Namely, if  the state itself is fixed at the boundary, gauge transformations there are also constrained to preserve that state. The boundary state itself would be gauge-invariant and thus, in the familiar interpretation of gauge theories,  accorded physical status.
 
%  either the Lagrangian or the Hamiltonian would then have to specify how the configurations (or the phase space) would behave at such a boundary. The full set of local gauge transformations that is available elsewhere may no longer represent symmetries at the boundary. This set of transformations can most readily be filed under `dynamical', where one takes the boundary conditions.  

On the fundamental view, the main issue with pared down states at the boundary is that they will impose boundary conditions that would not be gauge-invariant (or even covariant). Thus we would have to allow  a subsystem-quantity that is in this view gauge-\textit{variant}---a quantity such as the boundary value of the gauge potential---to acquire physical significance in the dynamical view. That is, these realizations of the externalist notion of subsystems  ascribe   gauge-\textit{invariance} under  the dynamical view  to a quantity that would be viewed, under  the fundamental view, as  gauge-\textit{variant}.\footnote{In \citep{Teh_abandon}, the externalist account of DES is labeled `Type II'. As far as I am aware, they give the only other consistent description  of DES in these circumstances. } 

 On the other hand, according to the dynamical view of symmetries, there is no conflict:  while external boundaries may curtail or pare down the full set of gauge transformations and gauge representatives, they do not \textit{break} gauge-invariance (nor do they flout downward consistency). For there were no gauge transformations acting on the boundary to begin with.% For  we would  necessarily have all the same symmetries at the boundary, only  under the fundamental view of symmetries.
 
The external boundary is familiar in the treatment of spatial infinity for field theories (cf. \cite[Ch. 7]{3+1_book}). Spatially asymptotic boundaries are usually construed as  boundaries of the entire universe, and  the representations of the states can be asymptotically pared down, so as to have a different behavior at those  boundaries (\cite{Belot50} gives a philosophical treatment of this idea).

  At this point, I should make a disclaimer. Although I will analyze the externalist notion of subsystem within the dynamical view of symmetries (in Section  \ref{sec:comparison}), I do not believe this description is as physically relevant as the internalist notion of subsystem. Of course, no one is forbidden from specifying a system where gauge symmetries act differently at the boundary by fiat---as they can in the externalist's notion of subsystem---but the status of such boundaries is not very clear. It is hard to see how such boundaries have ontological significance: even asymptotic boundaries are but a convenient idealisation, and are normally interpreted
as describing the way in which a system embeds into a larger system. (And if the notion is epistemic, it should still allow for the possibility that the universe extends beyond the boundary.)

\paragraph*{The internal boundary}
 
  Let us now suppose we would like to introduce subsystems in field theory in the internal way, by embedding a given system into a larger system. Suppose moreover that the entire universe has no boundary, and that the dynamical symmetries of the whole universe are  also fundamental: they are given by a symmetry group that acts on some value space, pointwise on spacetime.  Following \eqref{eq:dwc}, if a subsystem is to be demarcated by a boundary of space or spacetime, I will require the boundary conditions to have physical significance. That is,  according to the theory as applied to the entire universe, the boundary conditions must be gauge-invariant, or leave the representative of the state \emph{unfixed} there (i.e. subject to gauge transformations). 
   In contrast to the externalist view, in the internalist view, one is not given any boundary-anchor for the representational conventions.

For local field theories such as  Yang-Mills gauge theories and general relativity, a subsystem whose boundaries do not break the symmetries of the larger system respects downward consistency. And thus the subsystem inherits the local (fundamental) symmetries of the global system of which it is to be a part.\footnote{In the case of general relativity, downward consistency would require us to demarcate subsystems using diffeomorphism-invariant conditions; such as Komar-Bergmann scalars \citep{Bergmann_Komar}. This is easy to do asymptotically, and indeed this is one of the great advantages of the treatment of asymptotic infinity through Penrose compactification, see e.g. \citep{AshtekarStreubel} and \cite[p. 52]{AshtekarNapoli}. There are also many characterizations of black holes that are diffeomorphism-invariant in this way (see e.g. \cite[Chs. 5, 8 and 9]{Hayward_book}). I will have more to say about this in Section  \ref{sec:conclusions}. \label{ftnt:Ashtekar}} As mentioned in Section \ref{sec:sub_rec}, this is in fact the only feature of subsystem-recursivity that we will require in this paper: that the subsystem enjoys the same type of symmetries as the larger system in which it is embedded. 

In the internal boundary case, assuming downward consistency, if the universe as a whole is boundary-less, there is not much to say about the choice of state space: it depends only on the value space of the fields and on the underlying topological properties of spacetime. A gauge-invariance condition of isolation may furthermore be implemented through appropriate sectors of the theory: for example, by saying that the  boundary is free of matter. %All we are given are the physical states of the subsystems, and we must choose how we are going to represent states by a choice of convention that is not pre-fixed at the boundary.
  %But as stated in Section  \ref{sec:sub_rec}, my argument for gauge theories will be independent of such conditions; they will enter once we attempt to extend my conclusions to some interval of time in the  evolution of the system. 

\subsubsection{Symmetries and internal boundaries}\label{sec:int_bdary}
 
 But there are subtleties in reconciling the fundamental symmetries of  a subsystem defined by an internal boundary with its dynamical symmetries. 
 In particular, there are subtleties about the symmetry-invariance of  a bounded subsystem's own dynamical structures, such as its intrinsic Hamiltonian, symplectic structure, and variational principles in general. Until recently, subsystems that were so defined were \emph{not} supplied with gauge-invariant boundary conditions. The reason for this was the existence of an  obstacle towards a gauge-invariant formulation of subsystems: gauge theories manifest a type of non-locality. Thus the global, physical phase space (or the corresponding global physical Hilbert space) is not factorizable into the physical phase spaces over regions  (see footnote \ref{ftnt:edge_nonlocal} for references and more remarks on this issue). 
 
 That means that the standard manner of specifying the field dynamics of a subsystem would  not be  fully gauge invariant if we viewed symmetries as fundamental. The usual response is to pare down gauge symmetries at the boundary. In this way,  the boundary conditions and the  boundary contributions to the dynamics remain symmetry-invariant, but only  in the pared down dynamical view (see e.g. \cite{ReggeTeitelboim1974} for the first paper to enforce this approach explicitly, and, e.g.  \cite[Section 2]{Harlow_cov} and \cite[Section 2]{Geiller_edge2020} for  more modern  treatments). I will return to this issue in greater detail in Section \ref{sec:subs}. 
 
That  standard approach treats the lack of fundamental invariance of the subsystem similarly to the one of external boundaries, usually idealized to be infinitely far away, or asymptotic. We saw in Section \ref{sec:ext_bdary} that  if the whole universe is bounded---there are external boundaries---there is exceptional behavior of symmetries at the asymptotic boundary. So the standard approach takes this to be reflected  on  subsystems demarcated by internal boundaries.\footnote{\citet[p. 11]{Wallace2019b} endorses this view of subsystems,  because he takes subsystems  as sufficiently isolated so as to warrant an asymptotic-like treatment. So, for instance, the treatment would find no extension to spatially closed manifolds. \label{ftnt:Wallace} }
%Thus there is a sense in which we can enforce subsystem-recursivity: if the asymptotic boundaries of the entire universe enforce a notion of dynamical isolation, we can try to reproduce that for subsystems. %That such isolated sectors require the fundamental symmetries is then just par for the course. In this case internal boundaries within a bounded universe could still allow downward consistency: the internal boundary is, by the accounts of the theory, symmetry-invariant, and also subsystem-recursive, since the same considerations apply to a boundary seen as asymptotic---of the entire universe---or of an intrinsic subsystem.  Nonetheless,  this notion of subsystem---which we will label `\textit{externalist}'---allows gauge theories that can be realized within the dynamical, but not  the fundamental, view of symmetries. 

But if the subsystem symmetries are not fundamental, there is a clear conflict with downward consistency,  reflecting the incompatibility between an inside and an outside perspective of the internal boundary. For local field theories, how does  the environment, i.e. the entire universe, `see' the symmetries of the subsystem? Even for a bounded universe and on a dynamical view, the symmetries of the theory that act far away from the asymptotic boundary are unconstrained: they are not pared down. So how should observers from the environment construe a definition of subsystem---a sector of the theory, in Wallace's nomenclature---that does not support the full action of the  dynamical symmetries? % In other words,  symmetries on the internal boundary that would differ  from the inside and the outside perspective breaks downward consistency, given in Equation \eqref{eq:dwc}.
 The standard treatment of bounded subsystems in gauge theory breaks downward consistency, given in Equation \eqref{eq:dwc}: the action of the universal symmetry on the subsystem is \emph{not} a subsystem symmetry. %Ignoring this failure would imply that, somehow,  we should construe gauge symmetries as pared down from the inside even if they are not pared down from the outside.
 
 Recently, this pared down treatment of internal boundaries of subsystems has been called into  question  (cf. \cite{Teh_abandon, Geiller:2017xad, DonnellyFreidel, GomesRiello_new, Riello_symp, GomesStudies, Geiller_edge2020, carrozza2021}). New geometrical structures, for instance, `edge-modes', have been devised to maintain the gauge invariance of the internal boundary under the symmetries of the entire universe.\footnote{ In the symplectic case, a resolution requires extensions of the original phase space, to include facts about the representational convention and relational facts about the embedding of subsystem into system \cite{RovelliGauge2013,GomesStudies, carrozza2021}. In worked out examples, cf. \cite[Section 6]{GomesRiello_new}, one can however show that the composition of subsystems (which we will tackle on Section \ref{sec:gluing_field}) depends only on the symmetry invariant content of each region, and does not depend on any extra, symmetry-variant quantities on the interface of the subsystems.}  Though far from trivial, I will assume that the subsystems in gauge theory that are defined by regional restrictions have fully gauge-invariant  dynamical structures. I will briefly return to this topic, in Section \ref{sec:gt} for gauge theories and in Appendix \ref{sec:particle} for particle theories. \\

   % given the unobservability thesis, discussed in Section  \ref{sec:unobs}, symmetry-invariant. We will go back to this point in Section \ref{sec:particles}. 

 There are two main upshots of this Section. The first is to suggest a different treatment of asymptotic boundaries, that maintains invariance under symmetries of boundary states. Though this was long ago achieved for null asymptotic infinity through Penrose compactification \citep{AshtekarStreubel} and \cite[p. 52]{AshtekarNapoli} (see footnote \ref{ftnt:Ashtekar}), it has also been  developed in the case of Yang-Mills theory for spatial slices in \citep{RielloSoft}, where the spatial subsystem is extended asymptotically. %The suggestion is that we should reconcile fundamental and dynamical symmetries for both internal and external boundaries. 
 
This resolution is at the crux of my disagreement with Wallace, who  \cite[p. 11]{Wallace2019b} endorses a pared-down version of symmetries on internal subsystems. That is because he takes subsystems  as sufficiently isolated   to warrant an asymptotic-like treatment, and for external boundaries, there is really no conflict with downward consistency. But recent developments in gauge theory---which will be further discussed in Section \ref{sec:subs}---have shown that we can have finitely bounded subsystems in which e.g. the field-strength $F_{\mu\nu}$ is non-vanishing everywhere, and which still enjoy the same set of fundamental symmetries for their intrinsic dynamics.\footnote{And this is true even if these subsystems require time-varying boundary conditions! See e.g. \cite{Geiller_edge2020, carrozza2021, Teh_abandon}.} And although Wallace does not   encompass this possibility under his notion of subsystem recursivity, these recent developments  in gauge theory show that there is a good notion of subsystem recursivity for  subsystems---namely, downward consistency---that does not mimic the asymptotic ideal of  perfect isolation. Conversely, there are asymptotic treatments that do not require an anchor state at the boundary, paring down symmetries. I thus conclude that   a treatment of internal  subsystems  in gauge theories that respects the downward consistency of symmetries is conceptually and technically justified.

The second, more practical upshot of this Section, is that, from here on, in the internal boundary case, I will  assume a fundamental notion of symmetries, acting intrinsically on the subsystems as well as on the entire universe. In particular, this implies that   conventions about the representation of the state are not anchored at the boundary.

\subsection{Representational conventions and DES}\label{sec:rep_conv_DES}
The great obstacle in assessing the observability of subsystem gauge symmetries (DES) is that physical facts and representational facts come to us highly entangled. This is of course, a common theme. It occurs in  the logical positivists’ aim of presenting physical theories with a once-and-for-all division of fact and convention; and it was the center of a dispute between Carnap and Quine. I reject this once-and-for-all distinction, both in gauge theory and in the broader philosophical context (for familiar reasons, that I take to be  best articulated by \cite{Putnam_analytic}). But I judge that we can nonetheless assess matters of physical fact. The trick is to anchor these facts to an analogue of a Carnapian framework, that we call a \emph{representational convention}. 
 Each representational convention will have a unique representation of the physical facts. And as long as we stick with a single convention---whatever that is---we can compare and count different physical possibilities unambiguously. Like any good anchor, it will only serve its function if it doesn't move about.

 Of course, in the highly regimented  domain of mathematical physics we have much better control of the interchangeability of frameworks than we do in the purely philosophical debate. Here we can explicitly articulate which  quantities will be independent of the representational convention---the  \emph{gauge-invariant}  quantities. 
The existence of these invariant quantities may suggest formalisms that explicitly eliminate the need for conventions---e.g. the holonomy formalism, discussed in Appendix \ref{sec:holonomy}.\footnote{This discussion is familiar from the debate between the `eliminativist' and the `sophistication' approach to symmetries; see e.g. \cite{Dewar2017, ReadMartens, Samediff_1}.} 

But such formalisms inevitably carry several explanatory and pragmatic deficits (see e.g. \cite[Section 4.2]{Samediff_2}). More importantly, these formalisms are inadequate to deal with subsystems of the Universe, in the following sense: the set of invariant quantities of the whole universe does not equal the union of the sets of invariant quantities of  partition of the universe into a set of mutually exclusive, jointly exhaustive subsystems. Gauge theories involve a type of holism, or non-separability (cf. \cite{Gomes_new, GomesRiello_new,  GomesStudies}, and references therein).

 This is often noted even in the classical domain,  where it is expressed by the Gauss constraint.  For this constraint implies that by simultaneously measuring the electric field flux on all of a large  surface surrounding a charge distribution, and integrating, we can ascertain the total amount of charge inside the sphere {\em at the given instant}. In its quantum version, the non-locality implies the total Hilbert space of  possible states is not factorizable.%, even without taking into account the fully quantum sources of entanglement such as appear in EPR.
 \footnote{This type of holism, or non-locality is a well-known issue for theories with elliptic initial value problems: e.g. Yang-Mills theory and general relativity. For a reference that explores this in the context of the holonomy formalism, in the spirit of Appendix \ref{sec:holonomy}, see \cite{Buividovich_2008}. For an attempt to find  symmetry-invariant variables in gauge theory, see \cite{Elements_gauge}, and for general relativity see e.g. \cite{Donnelly_Giddings}; for more recent use of this non-factorizability in the black hole information paradox, see \cite{Jacobson_2019}.  For a discussion of the relation between the factorizability of Hilbert spaces and the augmentation of the phase space with `edge-modes', see \cite{Teh_abandon, Geiller_edge2020}.\label{ftnt:edge_nonlocal}} If we seek to employ in our theories only  invariant quantities of the subsystems, we may miss important physical  facts about the whole universe. In other words, there is a possible gap  between regional and universal gauge-invariant quantities.

So we can limit the domain in which the use of representational conventions is necessary as follows. Suppose first that there is \emph{no} concrete, unambiguous, choice-free representation of the physical state-of-affairs. Even, then, in the study of a single physical possibility---describing features of a given solution of the equations of motion for example---a representational convention may be left as implicit. Nothing physically important turns on which representational convention was used, though some conventions may be more convenient than others. On the other hand,  if we are to compare different physical possibilities, we must ensure the comparison is made under a fixed representational convention. We will return to this point in Section \ref{sec:rep_conv}, once we have introduced some notation.

  In sum, even if it is not always inevitable, the use of representational conventions in gauge theory is extremely useful. Moreover, it is not only useful but \emph{necessary} when dealing with subsystems and counting possibilities; as we must to assess DES. In this assessment, we need to keep careful track of which convention we use to anchor our representations; and we must keep track at both the level of  the subsystems and of the entire universe. For on both levels we will have to compare alternative  possibilities, and this comparison is only meaningful if made under a fixed   convention.

%This move does not yet alleviate the tension between gauge and DES: if we assume that the regional subsystems in question are intrinsically described by local gauge theory, their physically  significant quantities must also be suitably gauge-invariant. But now
  Thus, by carefully employing representational conventions at both the subsystem and universal level, we will completely characterize the gap between the subsystem and universal symmetry-invariant quantities. And it is this gap that enables  a well-defined type of  DES for local gauge theories, which we will label as `relational'. In line with the topic of holism (i.e. the question ``does the state of the parts determine the state of the whole?''), the question of relational DES turns on whether or not  the (union of)  gauge-invariant quantities of the subsystems determine those of  the entire system.
  
  This was the basis of the broad argument advanced and explored in \citep{Gomes_new}. Here we develop in more detail the use of representational conventions in different settings and with a greater focus on properties of  boundaries.\footnote{In \cite{Gomes_new} it was found that in the presence of matter, a non-trivial relational DES exists for local gauge theories, but only under conditions allowing conserved regional charges, and for  \textit{rigid} symmetries. In the Abelian theory, this would   require that the charged matter fields are present solely within each region, and would thereby include scenarios like  `t Hooft's beam splitter experiment; see \cite[p.110]{thooft}  and  \cite[p. 651]{BradingBrown}. In the non-Abelian case, relational DES also requires the regional or boundary field to be highly homogeneous (see \citep{GomesRiello_new}). For a more complete discussion of  circumstances which \textit{would} allow a non-trivial realization of DES in local gauge theories, see \citep{Gomes_new}.  I will return to these issues in Section \ref{sec:DES_matter}. \label{ftnt:GWDES_claim}}\\

 % the present paper adds  a seemingly innocuous remark: local gauge-invariance is transparent within a gauge-fixed treatment, and  we can describe any  quantity that is invariant under local gauge symmetry as a \textit{gauge-fixed} quantity.\footnote{This is not to say that gauge-fixing is the only way of describing the gauge-invariant content of the theory. In certain cases one may have direct access to the gauge-invariant degrees of freedom.}

%The remark is, in fact, consequential for our topic, for it implies that an approach that stresses gauge-fixed quantities can lay bare the concept of DES, relational or not. That is, once we assume empirical significance should involve only gauge-invariant quantities, we conclude from the remark that if any empirical significance is to be found, it can be found within a gauge-fixed treatment, and this will allow us to move beyond the domain of relational DES.\footnote{This is true for both indirect and direct significance. E.g. conservation laws all hold within a gauge-fixed formalism \citep{Strocchi_phil}, as does the Aharonov-Bohm effect, etc. }

Thus the question of DES in gauge theory will require us to first investigate what constitutes a representational convention in the presence of boundaries. For it is often assumed that, in a division of the universe into subsystem and environment, the latter already comes equipped with its own representational convention. Thus it is assumed that the very existence of an environment serves to anchor representation at the `edges' of our subsystem. This position is  defended in detail in  \cite{Belot50} for asymptotic boundaries, and also in \cite{Wallace2019b}. But although endowing environment with a ready-made convention  is very often  useful, it is in  tension with downward consistency, of  Equation \eqref{eq:dwc}, as I discussed in Section \ref{sec:int_bdary}.

 \section{General structure of DES}\label{sec:gen_DES}
I will here present the main ingredients for the analysis of DES (and I will I apply these ingredients to field theory and particle mechanics in detail, in the appendices).  In Section \ref{sec:prems}, I introduce basic notation about the action of symmetry groups on state spaces globally, or for the entire Universe. There I also discuss representational conventions and the unobservability thesis: that symmetries of the entire universe are not observable.  In Section \ref{sec:gluing}  I analyse the internalist notion of subsystem and derive DES in this general   scenario.  In Section \ref{sec:comparison} I provide a criticism of a previous derivation of DES, which takes the environment to come equipped with its own representational convention. 

\subsection{Preliminaries about symmetry}\label{sec:prems}
 I start  in Section \ref{sec:group} by describing a group action on a general  state space and defining the space of physical states. In Section \ref{sec:intro_gf}, I discuss representational conventions, in the generality. Along with  \cite[p. 18]{Wallace2019}, I argue  for the importance of a fixed representational convention when assessing differences in states due to the action of a symmetry. In Section \ref{sec:unobs}, I use the representaional convention to, following \cite{Wallace2019}, demonstrate the unobservability thesis: that the symmetry-invariant degrees of freedom are completely autonomous from the symmetry-variant features of the states. 

\subsubsection{Group action}\label{sec:group}
 Take a system $X$, with an associated state space $\F_X$, on which  a group of symmetries, $\G_X$, acts. Omitting the subscript $X$, we have, for $g\in \G$ and $\varphi\in \F$,  a map
 \begin{eqnarray}\mu:\G\times \F&\rightarrow&\F\nonumber\\
(g,\varphi)&\mapsto&\mu(g, \varphi)=: \varphi^g. \label{eq:group_action}
\end{eqnarray}
The symmetry group partitions the state space into equivalence classes, $\sim$, where $\varphi\sim \varphi'$ iff for some $g$, $\varphi'=\varphi^g$. We denote the equivalence classes under this relation by square brackets $[\varphi]$ and  the orbit of $\varphi$ under $\mathcal{G}$ by $\mathcal{O}_\varphi:=\{\varphi^g, g\in \mathcal{G}\}$.

 For the purposes of this paper, we could assume that the state space is phase space; but I will further assume that the  symmetries acting on phase space are inherited by symmetries acting on the configuration space of the system under consideration. So, I will take $\Phi$ to be  configuration space, with the cotangent bundle $T^*\Phi$ its associated phase space, and the symmetry $\mu$ from \eqref{eq:group_action} then induces an action  (for which I use the same label) $\mu:\mathcal{G}\times T^*\Phi\rightarrow T^*\Phi$ that preserves the symplectic structure and leaves invariant the Hamiltonian of the system, which is a function $H:T^*\Phi\rightarrow \RR$ that determines the dynamical laws (cf. footnote \ref{ftnt:syms}).  For this paper, these assumptions suffice: I will not need to provide details of the dynamics (through a specification of the Hamiltonian of the system or otherwise).

  Presaging the conclusions of this Section, we call  $[\varphi]$ the  \textit{physical state}, and $\varphi'\in \mathcal{O}_\varphi$ is its \textit{representative} (when there is no need to  emphasise that $\varphi$ involves a choice of representative, we call it just `the state'  for short). We call the collection of equivalence classes, $[\Phi]:=\{[\phi], \phi\in\Phi\}$,  \emph{the physical state space}. As written, this is an abstract space, i.e. defined implicitly by an equivalence relation. 
  
  Eliminativism about symmetries is a position that seeks an  intrinsic parametrization of $[\F]$. But such parametrizations are hard to come by, or have serious deficits. In their absence, we opt for a representational convention, that uniquely picks out members of each orbit.

\subsubsection{Representational convention, aka gauge-fixing}\label{sec:intro_gf}\label{sec:rep_conv}
Suppose we choose one representative  per gauge-orbit  for each $[\varphi]$. That is, an injective map $\sigma:[\Phi]\rightarrow \Phi$ that takes each equivalence class to a member of the respective orbit. Then, \emph{armed with such a choice of representative for each orbit}, a generic state $\varphi$ could be written uniquely as the doublet $([\varphi], g)_\sigma$, i.e. $\varphi=\sigma([\varphi])^g$.  That is, we identify $\Phi$ with $[\Phi]\times \G$ via:
\be\label{eq:doublet} \text{For all}\,\, \varphi, \,\,\exists!\,([\varphi], g)\in [\Phi]\times \G\,\,\text{such that}\,\, \varphi=\sigma([\varphi])^g=:([\varphi], g)_\sigma.
\ee
This representation is guaranteed to satisfy:
\be\label{eq:equivar} \varphi^{g'}=\sigma([\varphi])^{g'g}=([\varphi], g'g)_\sigma.\ee
To be able to use such doublets in assessing dynamical statements about the action of symmetries, we must moreover assume that  the map $\sigma:[\Phi]\rightarrow \Phi$ respects the required mathematical structures of $\Phi$ (cf. footnote \ref{ftnt:syms_def}), e.g. smoothness or differentiability. 
In more formal language, \eqref{eq:doublet} provides a  structure-preserving map (e.g. a diffeomorphism) from $[\Phi]\times \G$ to $\Phi$.\footnote{In the case of non-Abelian field theories, such a global representation of the state space does not exist due to the Gribov obstruction (see \cite{Gribov:1977wm, Singer:1978dk}). We will have more to say about this when we introduce the notion of a gauge-fixing in the case of Yang-Mills theory, in Section \ref{sec:gf}. \label{ftnt:Gribov}}

 It is convenient to have a separate label for the state  that is in the image  of $[\Phi]\times \mathrm{Id}$, where $\mathrm{Id}$ is the identity of $\G$:
\be\label{eq:conv} \varphi_\sigma:=\sigma([\varphi]), 
\ee
so $\sigma:[\Phi]\rightarrow \Phi;$ acting as $[\varphi]\mapsto \varphi_\sigma$, is a diffeomorphism onto its image. 
Then any state $\varphi'\in \mathcal{O}_\varphi$, including those not in the section, can be written as in \eqref{eq:doublet}: $\varphi'=\varphi_\sigma^g=([\phi], g)_\sigma$, for some $g\in \G$.

Now, as I mentioned, the space $[\F]$ is abstract, or only defined implicitly. Therefore it is convenient, if not necessary, to have a definition of the image of $\sigma$ that only traffics in $\F$. That is achieved by \emph{a projection operator} from $\Phi$ to the image of $\sigma([\Phi])$: 
\begin{align}
h_\sigma:\F&\rightarrow\F\nonumber\\
\varphi&\mapsto  h_\sigma (\varphi)=\sigma([\varphi])\label{eq:proj_h}
,\end{align}
which must of course be invariant, i.e. such that $h_\sigma(\varphi^g)=h_\sigma(\varphi)$. In practice, we only have a concrete or direct implementation of the projection operators,  not of $\sigma$. \\

Here the projection is, essentially, an interpretation of the idea of a \emph{gauge-fixing}, which we will develop in the case of field theories in Section \ref{sec:gf}.\footnote{Our notation is slightly different than \citet[p. 9]{Wallace2019}'s, who denotes these doublets as $(O, g)$ (in our notation $([\varphi], g)$), and  labels the choice of representative (or gauge-fixing) as $\varphi_O$ (our $\varphi_\sigma$). We prefer the latter notation, since it makes it clear that there is a choice to be made. As with coordinate systems, the interesting quantities will be invariant under these choices; nonetheless, we need to keep them fixed. This requirement will become nuanced when we are comparing different subsystems, with each other and with the joint system.\label{ftnt:Wallace_equi}}

And it is important to note that the decomposition of a given state $\varphi$ into a doublet, consisting of an equivalence class and a group element is not unique, which is why we have keep the subscript $\sigma$ on the doublet, indicating this choice (cf. footnote \ref{ftnt:Wallace_equi} for the analogous construction and notation in \citep{Wallace2019}).
That should be clear from the fact that,  if one is to change the choice of representative $\sigma$, the same state can be represented by different doublets, or, conversely, different states can be represented by the same doublet. 
That is, we can have 
\be([\varphi], g)_\sigma\neq ([\varphi], g)_{\sigma'}, \quad \text{or}\quad ([\varphi], g')_\sigma= ([\varphi], g)_{\sigma'},\ee for $g\neq g', \sigma\neq \sigma'$. 
It is important to remember this when comparing states at a common boundary, where group elements can match without a matching of the doublet, or vice-versa. 
In other words, given just the state, $\varphi$, we cannot discern any symmetry transformation that has been  applied to it. But armed with a choice of representative as in \eqref{eq:doublet}, we can do exactly that.
Thus, as a general principle, any physical significance that we attribute to group elements, or functions of group elements, must make reference to such a choice.% which should remain invariant across use. %This fact will be of paramount importance when we formally discuss DES, below. 

Equally important is the fact emphasized in Section \ref{sec:rep_conv_DES}: that we require a representational convention when combining subsystems. 
\citet[p. 18]{Wallace2019} highlights this same point:\begin{quote}
given configurations $(q; q')$ of the systems separately, we have
not been given enough information to describe their joint configuration: that
requires, in addition, a representational convention as to how points in the two
configuration spaces are to be compared. Such a convention is inevitably required
whenever we combine subsystems into a joint system. (In practice, the
convention is often given by a choice of coordinate systems, and/or of reference
frames, in the two subsystems.)
Prior to stipulating any such convention, there is no sense in which $(q, q')$ specifies a different configuration from $(R(g)q, q')$, since $q$ and $R(g)q$ are 
representationally equivalent.\footnote{In our notation, introduced below in Section \ref{sec:subs_loc}: $q\equiv\varphi_+, q'\equiv \varphi_-, R(g)q\equiv\varphi_+^{g_+}$} Given a choice of representational convention [i.e. $\sigma$],
though, it is clear that applying the symmetry transformation to one system
gives rise to a different total configuration (and that this is true independent of
what the actual representational convention is). So: symmetry-related configurations
can be understood as representing different possible configurations \emph{if we
hold fixed the choice of representational convention.} [my italics] \end{quote}

The requirement of a fixed representational convention is paramount for DES, since it discloses whether a symmetry transformation has been applied to a given state. %Without it, as described above, one cannot tell whether two states represent the same physical situation or not, and thus cannot adjudicate whether a transformed world is observationally/physically/empirically distinct.
But it is easy to see that one cannot just leave all these choices implicit when composing subsystems. For instance, the representational convention of the universe may not, when restricted, respect the representational convention of its subsystems. To give a simple example, in the non-relativistic particle case:  if the convention employs the center of mass, there will be a conflict between the center of mass of the subsystems and of the whole. A similar issue appears in gauge theory.  

\subsubsection{Unobservability and other theses about symmetry}\label{sec:unobs}
 The central idea of dynamical symmetries (cf. footnote \ref{ftnt:syms_def}) can now  be put as follows: given some notion of dynamical evolution, $U$, then $\varphi(t)$ satisfies the evolution equation, $U(t')\varphi(s)=\varphi(t'+s)$, if and only if $g(t)\varphi(t)$ also satisfies it.  Once we assume a well-defined gauge-fixing exists, we can translate the central idea of dynamical symmetries from a statement about the dynamics of $\varphi$ to one about the dynamics of $([\varphi], g)_\sigma$. Then, from \eqref{eq:equivar} it is easy to show (see, for example, \cite[p. 10]{Wallace2019}) that for a dynamical symmetry the future evolution of $\varphi_\sigma$ depends only on the present value of $\varphi_\sigma$, with no additional dependence on $g$. Since the map $[\Phi]\times \mathrm{Id}\rightarrow \mathrm{Im}(\sigma)$ is a diffeomorphism, we get to translate these statements into ones about the equivalence classes: the future evolution of $[\varphi]$ depends only on the present value of $[\varphi]$---which is how it is is stated by \cite[p. 10]{Wallace2019} (where this last step of translation from Im$(\sigma)$ to $[\F]$ is omitted). 
 
The natural interpretation is that there is ``a self-contained dynamics
for the invariant degrees of freedom of the system that is quite independent of
the $\G$-variant features'' \citep[p. 10]{Wallace2019}. If one moreover assumes that ``the system under investigation is rich enough to model
its own dynamics, and that the system is measuring itself rather than being
observed from outside,'' this demonstrates 
\noindent \textit{the unobservability thesis}: given a family of models of a global system which
are related by a symmetry transformation, it is impossible to determine
empirically which model in fact represents the system.

In a similar spirit, \citet[p. 7-8]{Wallace2019} provides four theses about symmetries in general, and I completely endorse his demonstrations regarding these. In particular, I will further jointly assume that: 
``given a family of models of a
theory which are related by a [dynamical] symmetry transformation, insofar as one
model successfully represents a system, so do all the others''; and that ``two states of affairs related by the action
of a symmetry transformation are really the same state of affairs, differently
described.''

\subsection{DES and gluing}\label{sec:gluing}
 
 In section  \ref{sec:debate},  we defined DES as  transformations of the Universe possessing the  following two properties: \\
\indent (i): \,(\emph{Global Variance}) the transformation should lead to an empirically different  scenario, and\\
\indent (ii):\,  (\emph{Subsystem Invariance}) the transformation should be a symmetry of the subsystem in question.

We also saw that physical quantities in gauge theories are characterized as gauge-invariant quantities, and that this obtains for both the subsystems and for the entire system.  

Therefore, to earn the labels `direct' and `empirical', DES must be  construed as referring solely to universal and subsystem gauge-invariant concepts. 

Here, properties (i) and (ii) will be  taken to apply to a Universe composed of a subsystem and an environment (as two subsystems). Following the internalist's symmetric treatment of subsystem and environment, (ii) will be taken to apply to all subsystems, i.e. to subsystem  and environment.\footnote{Couldn't we allow for a transformation that also changes the physical state of the environment?  Invoking a physically significant change in the environment leads to the
concern (voiced in, for example, \cite[p. 544]{Friederich2014}) that the empirical significance intended for the subsystem gauge symmetry is in fact completely due to the
change in the environment state, thus leaving no room for the gauge symmetry to do any
work. For example, in the Galileo's ship thought experiment,  a transformation that leaves the ship and its relation to the shore as they are but changes a grain of sand on the other side of the Universe satisfies (i) and (ii). This is an observable change perhaps, but it has little to do with symmetries. \citet[p. 68, 86 and 87]{GreavesWallace} react to this concern by requiring a further condition: there should be a `principled connection' between the putative change in the environment 
 and the gauge symmetry. But they  say nothing further
about what such a connection might be.  
  On the other hand, if (ii) applies symmetrically to all subsystems,  empirical significance is encoded solely in the relations between them.  In \citep{Teh_abandon}, a `principled connection' is taken to exist when the changes in the subsystem are taken to be generated by a charge on the boundary. Here, we will be able to make some sense of such `principled connections' in the externalist case, where the environment is just the boundary (see section \ref{sec:ext_sub}). In later work, \cite{Wallace2019} also restricts attention to extendible symmetries, and thereby to a relational understanding of the observability thesis (which is tantamount to the question of DES). Cf. the following footnote \ref{ftnt:Wallace_DES}.\label{ftnt:Teh}  } \\
  
  In Section \ref{sec:subs_loc} I will develop the definition of the internalist subsystem begun in Section \ref{sec:int_bdary}. In Section \ref{sec:DES_phys} I describe how DES emerges from the gluing of physical states, using representational conventions. 
  
 \subsubsection{Internalist subsystems}\label{sec:DES_gi}\label{sec:subs_loc}
In our discussion of subsystems it is important to note that, in the internalist case, none of the symmetries here are, in the words of \cite[p.12]{Wallace2019}, `subsystem-specific'. That means that the symmetries of a subsystem are extendible to the symmetries of other subsystems of the same universe. So given two subsystems, $\Phi_1, \Phi_2$, and $\G_1$, there is an action $\G$ defined on $\Phi_1\times \Phi_2$ that extends elements of $\G_1$. This is in line with what I labeled downward consistency in Section  \ref{sec:two_bdaries}.

 The extension may be unique or not. Wallace calls symmetries with unique extensions  `subsystem-global';  I call them \emph{rigid}. In the rigid case, for each $g_1$, we get a unique $g_1\times g_2=g$. But instead of taking the (extendible) alternative to these---what he calls \textit{subsystem-local}---to be ones in which the extension $g_2$ is  given by an independent action of the same symmetry group, he defines them as one in which, for any action of the symmetry in one subsystem, a composition of that symmetry with the identity on the second subsystem is still a symmetry of the universe. Namely, for him, a subsystem-local symmetry is one for which, for every $g_1\in \G_1$, $g_1\times\mathrm{Id}=g$ is a symmetry of $\Phi_1\times\Phi_2$. 

In the case of field theory, the malleable symmetries  on two subsystems that lie on \emph{completely disjoint subsets} of spacetime (i.e. ones whose closure are nowhere intersecting) are independent, and thus conform to this definition. But  when two subsystems are contiguous, this definition of subsystem-local symmetries is in clear tension with my assumption of downward consistency, as defined in Section  \ref{sec:two_bdaries}. Indeed,  I understand Wallace's construal of subsystem-local symmetries to be imposing an unnecessary further restriction on the behavior of symmetries at the common boundary of the subsystems; and the tension with downward consistency will be carried forward to a tension between the representational conventions for each subsystem. This will become clear, below, in Section \ref{sec:gluing}, when we learn how to compose physical states belonging to the subsystems.

Adopting the internalist perspective, we do not require such a restriction. We must carve up the system into two (mutually exclusive, jointly exhaustive)  subsystems whose state spaces we label $\F_+$ and $\F_-$, or $\F_\pm$, for short (a mnemonic notation to think of the subsystems as complements of one another, and intersecting only at a common boundary, e.g. 0). When these subsystems are made to correspond to regions, we will name the regions $R_\pm$. These are taken as subsets of the spatial manifold $M$, i.e. such that $R_+\cup R_-=M$, and I will moreover assume that the intersection of the closure of the regions is a \emph{boundary manifold}, $S$, i.e. 
\be \bar R_+\cap \bar R_-=S.
\ee

 As discussed in Sections \ref{sec:sub_rec} and  \ref{sec:int_bdary}, under the assumption of downward consistency  \eqref{eq:dwc}, the universal symmetries  bequeath  symmetries, through the split, to the subsystems, by mere restriction. Thus we  write $\G_\pm$; and similarly, we extend the use of the equivalence class notation and of the square brackets: 
 $\varphi_\pm\sim_\pm\varphi'_\pm$ iff $\varphi'_\pm=\varphi_\pm^{g_\pm}$ for some $g_\pm$, in which case $\varphi'_\pm\in[\varphi_\pm]$. Note that \emph{no} extra conditions on the gauge transformations at the boundary are imposed, thus in particular these symmetries are not required to be subsystem-local in the sense of \cite{Wallace2019}. Subsystem symmetries are just the symmetries obtained in each subsystem through the restriction of the symmetries of the larger system; this is the assumption of downward consistency of Section  \ref{sec:two_bdaries}.

\subsubsection{DES in terms of the physical states}\label{sec:DES_phys}\label{sec:gluing_field}
 We can now translate: \\
 \noindent $\bullet$\,\,\textit{Global Variance:}~~~ $[\varphi]\neq [\varphi'] $: the two physical states of the Universe are distinct according to the $\sim$ relation.   \smallskip
 
\noindent $\bullet$\,\,\textit{Subsystem Invariance:}~ $[\varphi_\pm]= [\varphi'_\pm] $: regionally the states are physically indistinguishable according to the $\sim_\pm$ relation; that, is for each ($\pm$)  subsystem, the primed and unprimed states are symmetry-related according to their internal models. 
Two subsystem physical states $[\varphi_\pm]\in [\Phi_\pm]:=\Phi_\pm/\sim_\pm$ are composable iff they jointly descend from a global state, $[\varphi]$.

Note that only the `physically significant', i.e. gauge-invariant, content of the subsystem and Universe states is relevant in the characterization of DES. 
The physical difference between $[\varphi]$ and $[\varphi']$ clearly must lie in the different possibilities for composing the two regional states, $[\varphi_\pm]$. The transformation must leave the subsystems physical content alone, but change their relation. 
 This is possible because   there are different domains for the equivalence relations---subsystem or universe---and therefore a Universal empirically significant difference may arise from a transformation  that doesn't change the subsystem states, but \textit{does} change their relation.\footnote{Here I disagree with \citep{Friederich2017}, who seems to exclude the possibility of DES almost by assumption. He demands that no difference in relations should be present. E.g. on page 155: ``The present article explores the idea that two subsystem ontic variables designate one and the same physical subsystem state only if the states designated by them are empirically indistinguishable both from within the subsystem and from the point of view of arbitrary external observers." and then again (p. 157): ``In other words, s and s' designate the same physical state if they are empirically equivalent both from within the subsystem and from the perspectives of arbitrary external observers."}  This idea is further explored in \cite{Gomes_new}, under the lable of \emph{holism}. Let us see how it plays out in more detail. \\

 By introducing some, yet-to-be-defined, composition of physical states, $\boxplus$, and writing 
\be\label{eq:phys_var} [\varphi_+]\boxplus[\varphi_-]=:[\varphi]\neq [\varphi']:=[\varphi'_+]\boxplus'[\varphi'_-]=[\varphi_+]\boxplus'[\varphi_-]
\ee
we indicate more clearly that the very concept of DES needs to be gauge-invariant, i.e. physical. Note also that the \textit{subsystem states} are intrinsically identical between the $ [\varphi]$ and the $ [\varphi']$ Universes, i.e. between the left and right hand sides of \eqref{eq:phys_var}. Therefore the difference between the two sides of the equality must lie in the relation between the subsystems; this is signalled by \eqref{eq:phys_var}'s use of  $\boxplus$ as well as $\boxplus'$. Thus here DES appears when there is a type of \emph{holism}: when the subsystem physical states $[\varphi_\pm]$ do not suffice to determine the physical state of the joint system.\footnote{In later work, \citet[p.13-14]{Wallace2019} has a similar characterization:
focusing on the extent to which the orbits of the subsystems determine the orbit of the joint system, and attributes failures of this determination to relational information.\label{ftnt:Wallace_DES} }

 Of course, as mentioned Section \ref{sec:prems}, equivalence classes are abstract and implicit, and  notoriously resistant to explicit mathematical manipulation. In particular, we cannot articulate a notion of composition using only equivalence classes (see e.g. \citep{Dougherty2017, Teh_surplus, GomesStudies}). To analyze \eqref{eq:phys_var} explicitly, we must  refer back to local representatives, i.e. to representational conventions, as argued in Sections \ref{sec:rep_conv_DES} and  \ref{sec:rep_conv}.
 
 For  local, smooth representatives in field theory, there is a straightforward definition of composition, as smooth composition, or gluing. 
More especifically, in the field-theories we will study here, we are given a Lie group $G$ and a gauge transformation is a map from the spatial  manifold $M$ to $G$, i.e. $\G:=C^\infty(M, G)$. It is a group on its own right, whose structure is inherited pointwise from the composition properties  of $G$.    $M$ is also the manifold on which the global states of $\Phi$ are represented, usually as maps $\varphi: M\rightarrow V$, where $V$ is some value space of the field.\footnote{More accurately, we would have a fiber bundle over $M$, given by a manifold $E$ with a well-defined, surjective operator $\pi:E\rightarrow M$, such that $\pi^{-1}(x)=V_x$ are the isomorphic fibers of $E$, i.e. $V_x\simeq V_y\simeq V$, for all $x, y\in M$. A field as we are defining it above would then be a section: $\varphi:M\rightarrow E$ such that $\pi\circ \varphi=\mathrm{Id}_M$. This is useful to construe gauge transformations as certain automorphisms of the bundle (e.g. spacetime dependent changes of bases for $V_x$ that are representations of $G$). On the other hand, writing the fields just as maps, as I have done above, requires many other assumptions, e.g. about the topology of $M$ and $E$. Nonetheless, I judge these issues to be unimportant to this paper, and will thus proceed with the simplified presentation above. \label{ftnt:vector} }

Suppose the regional state spaces are given by $\F_\pm$ and the regional gauge transformations are given by $\G_\pm$.\footnote{Once downward consistency is respected, given regions $R_\pm$ and the restriction maps $r_\pm: M\rightarrow M|_{R_\pm}$, or alternatively  the embedding maps $\iota_\pm: {R_\pm}\rightarrow M$, we would write $\G_\pm:=\iota_\pm^*\G=\G\circ \iota_\pm$ and $\Phi_\pm:=\iota_\pm^*\Phi$. }
Then we can write the conditions on the composition operation, $\boxplus$, for physical, i.e. symmetry-invariant states as follows: 

In field theory the two physical states are composable iff there exist states in each orbit, $\varphi_\pm\in \mathcal{O}_{\varphi_\pm}$, such that the value of $\varphi_+$ and all its derivatives at the boundary $S$ match those of  $\varphi_-$. We call such a notion of composition \textit{gluing}. (see Appendix \ref{sec:particle} for subsystem composition in the case of particles).
 
  Given  $[\varphi_\pm]$, and representational conventions $\sigma_\pm$, the condition of composition can thus be translated into the following gluing condition: there exist  gauge transformations, $g_\pm\in \G_\pm$, such that:
\be\label{eq:bdary_cont}\sigma_+([\varphi_+])^{g_+}=_{|S}\sigma_-([\varphi_-])^{g_-},
\ee
where the subscript ${}_{|S}$, restricting the equality to $S$, is understood as also matching derivatives.\footnote{For the standard notion of continuity, i.e. when all we require is the value of $f$ at the boundary, and not also of its derivatives, we employ no bar, i.e: $f=_{S}f'$ iff $f(x)=f'(x)~\forall x\in S$. \label{ftnt:C0}} 

I will use the notation  $\oplus$ with the  meaning of `composition of \textit{representatives}'; I do not restrict $\oplus$ to mean `direct sum'. So, if \eqref{eq:bdary_cont} is satisfied, we translate the physical compositions of \eqref{eq:phys_var}, i.e. $ [\varphi_+]\boxplus[\varphi_-]$,  into: 
\be\label{eq:comp_simp} \sigma_+([\varphi_+])^{g_+}\oplus \sigma_-([\varphi_-])^{g_-}
\ee 
In the field theory case, we can usually understand $\oplus$ just as addition in some vector space of smooth functions. 

Two important things to note: though we are not specifying the choice of convention, we label each choice and do not leave implicit the fact that one is being made. Note also that we cannot eliminate either of $g_\pm$ in \eqref{eq:bdary_cont} and \eqref{eq:comp_simp}, since they act on different spaces and are therefore not subject to the same representational convention. 
 
 For point-particle systems, as we will see in section \ref{sec:particle}, $\oplus$ requires an embedding of the subsystems into a common Euclidean space, and then it signifies vector addition. %And, like for $\boxplus, \boxplus'$, I will signal different ways of combining---different relations between---representative states by $\oplus, \oplus'$.
%Even if $\oplus$ elicits the classical idea of the direct sum of the vector spaces in which fields take their values, e.g. for $L^2$ functions on intervals, $ L^2([0,1])\oplus L^2([1,2])=L^2([0,2]).$   
\\

Again, the way to make the two conditions for DES precise and clear is by using fixed representatives, for $\F_\pm$ and also $\F$. Namely, \emph{Subsystem Invariance} just means that we have just two Subsystem Invariance classes, $[\varphi_\pm]$,  that are composable. Global Variance means then that there exist $g_\pm$ and $g_\pm'$ such that, given the same representational convention for the global state, $\sigma$, the glued states will differ. Simplifying the notation by writing $\sigma([\varphi])=:\varphi_{\sigma}$ as in \eqref{eq:conv}, the condition for DES is simply:
\be\label{eq:final_gluing}\varphi_{\sigma_+}^{g_+}+\varphi_{\sigma_-}^{g_-}=:\varphi_\sigma\neq\varphi'_\sigma:=\varphi_{\sigma_+}^{g'_+}+\varphi_{\sigma_-}^{g'_-}
\ee

This is the most important equation for the matter of DES. This rendering of  the physical significance  of symmetries employs  fixed representational conventions; it is this convention that allows us to unambiguously compare different possibilities, as is required in our construal of DES (cf. Sections \ref{sec:rep_conv_DES} and \ref{sec:rep_conv}).

 \subsection{An incomplete derivation of DES}\label{sec:comparison}
This paper started with the question: it is widely ackowledged that rigid symmetries in particle mechanics can have a (relational) DES when applied to subsystems; do local gauge theories also realize the concept of DES? 

%Gauge symmetries are normally taken to encode descriptive redundancy: a view I endorse. This descriptive redundancy means that the natural answer to our question is `No'. This `No' answer was   developed in detail by Brading and Brown \citep{BradingBrown}. The `Yes' answer has been argued for by Greaves and Wallace \citep{GreavesWallace}. Building on \citep{Healey2009}, they articulate DES for gauge theory differently,  fostering their `Yes' answer to the above.
 Although our answers differ, the  treatment of this question by  \cite{GreavesWallace, Wallace2019, Wallace2019b} bears many similarities to ours here: they focus on subsystems as given by regions; they think of subsystems as given by a splitting of the universe;  they identify transformations possessing properties  (i) and (ii) in Section s \ref{sec:debate} and \ref{sec:DES_gi} by first formulating the putative effects of such transformations on the gauge fields in these regions; and they construe DES essentially as a relational property.\footnote{Although Greaves and Wallace allow for the larger, non-strictly relational quotient group,  of all subsystem symmetries quotiented by the interior ones, at the theoretical level they do not investigate this larger (infinite dimensional) group, whose physical meaning---if any---is unclear \cite[p.86,87]{GreavesWallace}. They only mention Einstein's elevator and the Faraday cage as examples of this extension, but have little to say about what are the principled connections that render these, and only these (?) as bona-fide examples of DES, while disallowing examples where only an irrelevant change in the environment accompanies the subsystem symmetry; e.g. a change of a grain of sand on the beach should not be associated to a DES of Galileo's ship. See their footnote 17, p.74.  The later papers \cite{Wallace2019, Wallace2019a, Wallace2019b} are less ambiguous in this respect, and are more aligned with our relational view here, cf.  e.g. \citep[p. 13-14]{Wallace2019}. \label{ftnt:principled}} 
 
 But \textit{unlike} our results,  they claim that there is relational DES transformations  in 1-1 correspondence with the following quotient:
 \be\label{eq:GW_DES}\G^{\text{\tiny{GW}}}_{\text{\tiny DES}}(\varphi)\simeq \mathcal{G}_+^{\varphi_-}/\mathcal{G}_+^{\text{Id}},\ee 
 where  $\mathcal{G}_+^{\varphi_-}$ are the elements of $\G_+$ which are `in the $\varphi_-$--sector', that is, that can be composed with $\varphi_-$ (cf. \cite[p.10-11]{Wallace2019b}). These transformations need   preserve (only) the state $\varphi_+$ \textit{at the boundary} of the region---which we  call the \textit{boundary}-stabilizer group for $\varphi_+$, as in \eqref{eq:stab}---and $\mathcal{G}_+^{\text{Id}}$ are the gauge transformations of the region which are the identity at the boundary (and thus preserve all states at the boundary); the latter symmetries make up what he calls `subsystem-local' symmetries (see Section \ref{sec:subs_loc}). 
 
  %Below, I will describe these claims in detail, comparing the results of this paper with those of \citep{GreavesWallace, Wallace2019b}.\footnote{See \cite[Sec 4.3.2]{GomesRiello_new}, and \citep{Gomes_new} for a more thorough treatment with matter and non-trivial topology.} 
I start in Section \ref{sec:rep_states} by presenting a sketch of the standard derivation, and then I criticize this derivation in Section \ref{sec:gap}. 

\subsubsection{The derivation}\label{sec:rep_states}

 Assuming the subsystem physical states are composable, given two global states
 $$\varphi:=\varphi_+\oplus\varphi_-\quad\text{and}\quad \varphi':=\varphi'_+\oplus\varphi_-',$$
the condition \textit{Subsystem Invariance} translates into:
\be\label{eq:states}\varphi':=\varphi'_+\oplus\varphi_-'=\varphi^{g_+}_+\oplus\varphi^{g_-}_-\quad \text{for some pair of elements}\quad g_\pm\in \G_\pm
\ee
Now,  \textit{Global Variance} demands  that, for DES to be realized, there can be \textit{no} $g$  such that $\varphi'=\varphi^g$. That is, \textit{Global Variance} implies:
\be\label{eq:nog}  \text{there is no universal}\,\, g\,\, \text{such that}\quad g{}_{|R_+}=g_+, \quad g{}_{|R_-}=g_-,\ee
 for otherwise $\varphi'=\varphi^g\sim\varphi$ and the two states are entirely physically equivalent.

The result \eqref{eq:GW_DES} requires an assumption: when looking for the realizers of the conditions \textit{Global Variance} and \textit{Subsystem Invariance} (cf. Section  \ref{sec:DES_gi}), one may keep one of the regional subsystems---labeled `the environment'---not only physically fixed (both are physically `fixed' according to \textit{Subsystem Invariance}), but also representionally fixed.  
In other words, there is an assumption that we have a fixed representative $\varphi_-$ of the physical environment state $[\varphi_-]$ which we can employ  as a reference to externally assess the capacity of regional gauge transformations $g_+$ to produce empirically distinguishable differences. (But no mention of an explicit use of a representational convention is made.) This restricts the states $\varphi_+\in \Phi_+$ to be, in the language of \citet[p. 10]{Wallace2019b}, in the \emph{$\varphi_-$-sector of the theory} (and in the nomenclature of Section  \ref{sec:sub_rec}, that parallels that of Wallace, called $\Phi_{\varphi_-}$). It is thus usually taken for granted that we can assume the environment is in this implicitly  given representation and restrict attention to $g_-$ being the identity transformation. 

Then, \textit{if} some physical states already satisfy \textit{Global Variance} and \textit{Subsystem Invariance}, instead of \eqref{eq:states}, the assumption is that we have representatives of the states  fulfilling:
\be\label{eq:states2}\tag{\ref{eq:states}'} \varphi=\varphi_+\oplus\varphi_-\quad\text{and}\quad \varphi':=\varphi^{g_+}_+\oplus\varphi_-.
\ee If \eqref{eq:states2} is assumed,  and we moreover assume that $\varphi_-$ has only the trivial stabilizer, meaning there are no $g_-\neq$ Id such that $\varphi_-^{g_-}=\varphi_-$ (see Section \ref{sec:stab}),  we can similarly rewrite \eqref{eq:nog} as follows:
\be\label{eq:nog2}\tag{\ref{eq:nog}'}  \text{there is  no}\,\, g\in \G\,\, \text{such that}\quad g{}_{|R_+}=g_+, \quad g{}_{|R_-}=\mathrm{Id}.\ee
Of course, jointly, the assumptions above  would then mean that \emph{Global Variance} requires $g_+{}_{|S}\neq \mathrm{Id}$. By quotienting all the transformations that do have this boundary behavior, namely those that preserve the $\Phi_{\varphi_-}$-sector of the theory, by those such that $g_+{}_{|S}\neq \mathrm{Id}$, we arrive at \eqref{eq:GW_DES}.

\subsubsection{The gap in the previous derivation}\label{sec:gap}
The assumption that $g_-=\mathrm{Id}$ (or equivalently, that the transformation is \emph{subsystem-local} in the narrow understanding discussed in Section \ref{sec:subs_loc})  is consequential for the issue at hand. 

The assumption is that we do not need to make reference to the representational convention for the environment; that it can be left implicit. We are just `given' a $\varphi_-$. It, like the asymptotic states, will therefore pare down the symmetries at the boundary.
Of course, there is no a priori, or canonically preferred, representational convention for the environment:  $g_-=\mathrm{Id}$ is not a representational convention (or a gauge-fixing); it doesn't fix a map from the equivalence classes to the representative states, as in \eqref{eq:conv}. %Given some arbitrary $\varphi_-$, the condition will not say anything about whether $\varphi_-$ lies inside the class or not. 

 What we are in fact  given is an equivalence class, or a physical state (according to the theory as applied to the environment), and we must choose  a representational convention for the environment just as we must for the subsystem in question and as we will have to for the universe as a whole.   But as one can show, once one makes the representational convention explicit, subtleties arise when comparing the global states, for there is the issue of how these conventions mesh. Let me expand this argument in more detail.

 As we agreed in  Sections \ref{sec:rep_conv_DES} and \ref{sec:rep_conv}, we must keep fixed a representational convention in order to evaluate the observability of subsystem symmetries. It is true that in many, if not most, circumstances we need not make that convention explicit: it suffices that we acknowledge one exists and is kept fixed; we often talk about a representative of the physical state without discussion of how that representative is defined. However, when investigating subsystems and their relation to the entire universe more than one representational convention is at play, and they may be incongruous, in the following sense. Suppose one fixes the representational convention for subsystems and universe, $\sigma_\pm$ and $\sigma$, respectively. 
 Still,  the representational convention of the global state may  have its restrictions to the subsystems fail to satisfy the regional  representional convention.  This is very clear in the point particle case discussed in Section  \ref{sec:particle}: we choose subsystem center of mass coordinates, but, upon composition, a new center of mass will emerge, and we will have to `readjust' both our previous representational conventions. 
 
 In the field theoretic case, something similar happens. Using the nomenclature of Section \ref{sec:gluing_field} and the projection operator \eqref{eq:proj_h}, we may have: 
 \be\label{eq:mesh} \iota^*_\pm h_\sigma\neq h^\pm_\sigma.
 \ee
 Thus, in order to count global possibilities given just the physical state, or, equivalently, the $h^\pm_\sigma$, some adjustment between the two states in their regional representational conventions may be allowed or even required; that is, we should allow a $g_-\neq\mathrm{Id}$ (which we did in \eqref{eq:final_gluing}; cf. footnote  \ref{ftnt:gluing} in Section \ref{sec:setup_gluing}). 
 %That is the reason that, in  \eqref{eq:final_gluing}, composing the fixed representatives allows for arbitrary $g_-$ as well as $g_+$.  
 We  will see this issue emerge explicitly in Sections \ref{sec:particle} and  \ref{sec:setup_gluing} (see footnote \ref{ftnt:gluing} and equation \eqref{eq:gluing}).

  Indeed, as one can show, for internal boundaries respecting downward consistency,  by rejecting the `God-given' representation of the environment,  no relational empirical significance  in the vacuum, simply-connected case, can be identified. In this sector of the theory, \cite[Section 4, Equation 4.2]{Gomes_new} provides an explicit counter-example to the definition \eqref{eq:GW_DES}.\footnote{The counter-example is as follows: for electromagnetism, for a configuration (or sector) that happens to be in vacuum, any element of $\G_+$ that goes to a constant $c\neq 1$  at the boundary will provide a representative of $\mathcal{G}_+^{\varphi_-}/\mathcal{G}_+^{\text{Id}}$ in \eqref{eq:GW_DES}. Moreover, this can occur for any notion `isolation' of the subsystem. But in fact, for two states, $\varphi'$ and $\varphi$, as in \eqref{eq:states2}, related by such a a transformation, one is always able to explicitly find a global $g$ such that $\varphi^g=\varphi'$, thus foiling \emph{Global Variance}.  I should note that Greaves and Wallace do not overtly narrow down their  formal prescription for DES to include matter. In particular, their derivation  does not mention matter or the lack thereof. The failure of that derivation  in sectors in which matter is absent is neither explained nor mathematically expected; there is nothing in their definition that gives any hint to why this should be the case.  \label{ftnt:Wallace_gen}}
  
  In more recent work, the type of  counter-example of footnote \ref{ftnt:Wallace_gen} is excluded by narrowing the focus of the definition to `generic' states  \cite[p.9]{Wallace2019b}.\footnote{  A`generic' subset here is not defined as usual: it is defined as the set of states with only  the trivial stabilizer (cf. \eqref{eq:stab}). In field theory, were one to use an actual definition of generic subspaces of $\F$ as dense and open subsets, then there would be no DES. For in the presence or absence of matter,  generic states would have matter on their boundaries, and thus would not have any non-trivial boundary stabilizer, and thus  $\mathcal{G}_+^{\varphi_-}=\mathcal{G}_+^{\text{Id}}$. }  But this assumption is not used or sufficiently justified in the rest of the paper, and thus its imposition seems to me slightly \emph{ad hoc}. To be more precise, in \cite{Wallace2019, Wallace2019b},  the generic property is mentioned at the same stage that I mentioned it: between \eqref{eq:nog2} and \eqref{eq:states2}. The idea is that, if the environment does not have any stabilizer, a gauge transformation that preserves the boundary state and is not the identity will necessarily be continued into a transformation that doesn't preserve the environment state, and is therefore ``witnessed'' by the environment. But I find this confusing, since part of the initial assumption was precisely that the representation of the environment state \emph{is fixed} as $\varphi_-$. \\
  
 %We will go back to this poin in Section  \ref{sec:conclusions}.
  %On the other hand, if the restriction of the global state in its representational convention  was constrained to be always   in the regional representational convention,  the assumption of $g_-=\mathrm{Id}$ would be relatively unproblematic.\footnote{ In particular, this issue is related to the presence of `edge modes': see Section \ref{sec:subs} and footnote \ref{ftnt:ghosts}.} \\

In sum: if all we have access to, according to the theory, are the physical content of the states, then we require a representational convention to represent the physical state. Without such a convention, one is liable to be led astray in the internalist case.  %Indeed, even such expert authors as \citep{GreavesWallace} \textit{were} led astray. 
%The lesson is that, to be sure that the end result is both fully regionally and universally gauge-invariant,  any approach to these questions must take due care to employ representational conventions,  as  I will do in Section  \ref{sec:DES} by the use of gauge-fixed quantities. This will 
Employing representational conventions, in Section \ref{sec:gt} we will assess DES for any state (even non-generic ones, cf. footnote \ref{ftnt:Wallace_gen}).%Let us now try to fit the assumptions above with the gauge-fixing paradigm. 

\section{The gauge theory of fields}\label{sec:gt}
Here I will describe the basic setting with which I will treat the local gauge theory of fields, taking as my model vacuum Yang-Mills theory on a simply-connected manifold, $M$. The stated results should be taken as  applying  to both Abelian and non-Abelian interactions alike, and the extension to non-simply-connected manifolds and to the inclusion of matter are straightforward but notationally cumbersome; exceptions and differences to these generalizations will be explicitly flagged. Having said this, I will, as a simplification,   only \textit{explicitly} treat Abelian gauge fields (like electromagnetism).\footnote{ For a more complete treatment, see \citep{GomesRiello_new, Gomes_new}.} 

In Section \ref{sec:gf}, I develop further the ideas presented in Section \ref{sec:rep_conv}, about representational convention, and describe what those ideas have to do with gauge-fixing, with an eye towards the application to Yang-Mills theory. In Section \ref{sec:subs} I describe in a bit more detail what I will take subsystems to be in Yang-Mills theory and discuss recent developments for gauge-invariant subsystem dynamics.  In Section \ref{sec:premsYM}, I write down the specific field content and its symmetry transformation properties, specializing to the case of electromagnetism, and to subsystems defined by gauge-invariant boundaries. And in Section \ref{sec:DES}, I finally put these constructions to use in finding DES, by unpacking the main equation defining DES,  Equation \eqref{eq:final_gluing}, in the case of electromagnetism.   This viewpoint expresses DES in terms of uniqueness properties of coupled partial differential equations with particular boundary conditions.

\subsection{Gauge-fixing: the general ideas}\label{sec:gf}
%The physical content of the field, $[\varphi]$, as an equivalence class, is highly abstract, and as such, intractable for studying certain concrete questions.\footnote{It also has other disagreeable properties vis \`a vis the composition of subsystems. See \citep{Dougherty2017, GomesStudies}. We will come back to this point in Section  \ref{sec:gf_unbounded}.} But there are many alternative representatives of $[\varphi]$; and so, to obtain  1-1 representatives of the equivalence classes, we must eliminate redundancy.

 In the type of field theories we will focus on in this paper, the procedure for fixing the representative of the state, or finding a representational convention, as in Section \ref{sec:rep_conv} and Section \ref{sec:intro_gf},  is intimately related to a procedure called `fixing the gauge'. The procedure  is necessary to  extricate physically significant properties of the state from the unphysical ones, that are not invariant under the symmetries. In other words, by fixing the gauge, no physical property is lost. Thus important physical effects, such as the Aharonov-Bohm effect, quantum anomalies, interference,  are all perfectly expressible in a gauge-fixed setting, as I define it here\footnote{ Much as in other representtions of  gauge-invariant quantities---such as in the holonomy interpretation---fixing the gauge is  non-local   in the following sense: just as $\int A$ requires the value of of $A$ at several points simultaneously as an input, the projected state $h_\sigma(A)$ requires the value of $A$ throughout the region as in input. This is just a reflection of the non-local aspects of gauge-invariant functions  (cf.  \cite[p. 460]{Earman_local}, \cite[Ch. 4.5]{Healey_book}, and  \citep{Strocchi_phil, GomesStudies}).} 

% This is why,  in order  to investigate DES,  we cannot focus solely on the near-boundary properties of the states.

%  It is important to note that a gauge-fixing is \textit{not} analogous to choosing an arbitrary coordinate representation of a single configuration. The latter is mostly useful when we do \textit{not} need to compare the totality of physical attributes of different configurations. We use a single  arbitrary representative $A$, (and similalry, in the case of general relativity and other spacetime theories, an arbitrary choice of spacetime coordinates),  if all we are interested in is to  describe certain  properties of a  single physical situation, e.g. if there are no counter-factual considerations in play. We can then calculate some quantity and verify whether it was  independent of the coordinate choice we made at the beginning. 

  %These are the kinds  of modally loaded questions that we  face when  investigating DES. 

A gauge fixing provides, in the language of Section \ref{sec:rep_conv}, a \textit{a fixed representational convention} with which to compare different states.  As we saw in that Section   (see Equation \eqref{eq:proj_h}) gauge-fixing can be seen as a sort of projection on state space, which allows us to judge whether two given representatives, $\varphi, \varphi'$, unrelated in principle, are physically the same, i.e. give the same value for \textit{all} gauge-invariant quantities. %The answer is that they are physically identical if and only if they have the same such projection
In other words, two configurations are physically the same if and only if they are \textit{identical} once gauge-fixed. Thus a gauge-fixing resolves problems of physical identity.

 In the language of fiber bundles, a gauge-fixing is  a choice of  \textit{section} of the  configuration space, seen as a (possibly infinite-dimensional) principal bundle. %For example:  given the space of all possible gauge potentials, call it $\mathcal{A}:=\{A\in \Lambda^1(M, \mathfrak{g})\}$, a section gives a local submanifold of $\mathcal{A}$ which intersects each  gauge-orbit, i.e. $\mathcal{O}_A:=\{A^g, g\in \G\}$ for each $A$, only once. This is how we defined $\sigma$ in Section \ref{sec:intro_gf}.
A choice of section is essentially an embedded submanifold on the state space $\F$ that intersects each orbit once. In practice, the gauge-fixing procedure relies on the given representational convention $\sigma([\varphi])$ satisfying some auxiliary condition. That is, we impose further functional equations that the state in the aimed-for representation must satisfy. This is like defining a submanifold indirectly, through the regular value theorem. E.g. defining a co-dimension one surface $\Sigma\subset N$ for some manifold $N$,  as $F^{-1}(c)$, for $c\in \RR$, and $F$ a smooth and regular function, i.e.  $F:N\rightarrow \RR$ such that $\d F\neq 0$. Once the surface is defined, $\sigma$, as defined in Section \ref{sec:intro_gf},   will be the embedding map for one such surface, e.g.  $\sigma:[\Phi]\rightarrow F^{-1}(0)\subset \Phi$.  Once the surface is defined, we can define a projection map, that projects any configuration  to this surface, and this projection will be gauge-invariant.\footnote{ Take $\RR^2$, and a choice of a graph, $y(x)$, defined by some function $F(x,y)=0$. Now we can project any doublet, $(x,y)$ onto $y(x)$, namely, $(x,y)\mapsto (x, y(x))$. The projection is independent of $y$, and, if we identify translations in the $y$-directions as `gauge', the projection is gauge-invariant. }

Now I will describe the two conditions expected of a complete gauge-fixing.
 
 %\subsubsection{Conditions on a complete gauge-fixing}\label{sec:conds_gf}

In general,  we fix the  gauge freedom by  imposing conditions on the representative gauge potential, i.e.  
  by imposing a local functional equation $F(A)=0$, for some  $F$ which, besides being regular,  ideally must satisfy two further conditions: \smallskip
 
 \noindent$\bullet$\,\,\textit{Universality} (or existence):~ 
For all $\varphi\in \F$, the equation $F(\varphi^g)=0$ must be solvable by a functional $g_\sigma(\varphi)$. Here, $g_\sigma(\varphi)$ is a gauge transformation required to transform $\varphi$ to a configuration $\varphi^{g_\sigma(\varphi)}$ which belongs to the gauge-fixing section $\sigma$. So $g_\sigma$
 must be such that   $F(\varphi^{g_\sigma(\varphi)})=0$. That is: 
\be\label{eq:gauge-fixing} g_\sigma:\F\rightarrow \G ,\,\,\text{is such that}  \,\, F(\varphi^{g_\sigma(\varphi)})=0, \,\, \text{for all}\,\, \varphi\in \F.
\ee
This condition ensures that $F$ doesn't forbid certain states, i.e. that each orbit possesses at least one intersection with the gauge-fixing section.

\smallskip

\noindent$\bullet$\,\,\textit{Uniqueness}:~ If  $g_\sigma$ as above satisfies  $F(\varphi^{g_\sigma(\varphi)})=0$, then     $\varphi^{g_\sigma(\varphi)}=\varphi'{}^{g_\sigma(\varphi')}$ if and only if $\varphi\sim \varphi'$. That is, the representatives  coincide iff they represent the same physical state, $[\varphi]$. That is, a gauge-fixing resolves mattes of physical identity between representative states.\footnote{Jumping ahead, in Section  \ref{sec:stab}, I introduce one  subtlety in the concept of gauge-fixing, due to \emph{stabilizers}, which plays an important role in the definition of DES. Certain states are not ``wrinkly enough'', do not have features that are detailed enough, to completely fix the representation. These states have stabilizers. Stabilizers are degeneracies in the representational convention, that foil uniqueness for physical reasons. } \smallskip

Since $g_\sigma$ should act as a projection operator on $\F$, onto the gauge-fixing surface, it is convenient to explicitly define this projection  as in \eqref{eq:proj_h}, but now explicitly including $g_\sigma$:
\begin{align}
h_\sigma:\F&\rightarrow\F\\
\varphi&\mapsto  h_\sigma (\varphi):=\varphi^{g_\sigma(\varphi)}
\end{align}
  And, as expected,    $h_\sigma (\varphi)$ is a gauge-invariant functional, in the sense that $h_\sigma(\varphi^{g})=h_\sigma(\varphi)$, i.e. it is invariant under the group action on $\F$ as its domain. Of course, we can still act on the surface itself, i.e. act with the group  on the image of $h_\sigma$.\footnote{\label{ftnt:subs_int_ext}It is important to stress that $h:\F\rightarrow\F$ is a projection, as opposed to a reduction, $\mathrm{pr}:\F\rightarrow [\F]$. In \citep{GomesStudies, Gomes_new}, the construal of a gauge-fixing as a projection, and not as a quotienting, was argued to be fundamental for the gluing of regions: for both $h$ and $\mathrm{pr}$ are gauge-invariant with respect to gauge-transformations on the common domain, $\F$, i.e. $\mathrm{pr}(\varphi^g)=\mathrm{pr}(\varphi)$ as well as $h(\varphi^g)=h(\varphi)$,  but only the projection $h$ allows further transformations to be enacted on its range, and therefore allows for a change of representational convention. \citet{GomesStudies, Gomes_new} distinguishes between two sorts of action of $\G$: \\
\textit{Subsystem-intrinsic gauge transformations:}
Given $h:\F\rightarrow\F$, a subsystem-intrinsic gauge-transformation acts solely on the domain of $h$. 
The projection $h$ is invariant under subsystem-intrinsic gauge transformations: $h(\varphi^{g})=h(\varphi)$.
The label `intrinsic' stands in opposition to `extrinsic'. Such gauge transformations are all that is needed for unique description of the entire Universe. But if we have more than one subsystem and we want to satisfy the gluing condition \eqref{eq:final_gluing}, we may need to change the representative of $[\varphi]$---from the outside, as it were. \\
\textit{Subsystem-extrinsic gauge transformations:}
Given  $h:\F\rightarrow\F$, we can define \textit{subsystem-extrinsic} gauge transformations $g^{\text{ext}}$, as those transformations which act on the \textit{range} of $h$ as 
\be\label{eq:external_gt}
h(\varphi)\mapsto h(\varphi)^g.\ee 
Of course such a transformed field would no longer satisfy \eqref{eq:h_div}. In what follows we omit the superscript `ext'. There are the transformations that are required when we need to change representational conventions, as we must when we glue subsystems.

 It is instructive to compare the two possibilities of action of $\G$  to the use of homotopy type theory (HoTT) in gauge theory,  as advocated by \citep{Ladyman_DES}. Ladyman says HoTT ``both (a) distinguishes states conceived of differently even if they
are subsequently identified, and (b) distinguishes the identity map from non-trivial transformations that nonetheless might be regarded as delivering an
identical state''. Here we have two sorts of transformations: the subsystem-intrinsic  one, $\varphi\mapsto \varphi^g$, which does not change $h(\varphi)$---satisfying Ladyman's (b)---, and the subsystem-extrinsic one, that does the work of Ladyman's (a). } 

Assume both the \textit{Universality} and \textit{Uniqueness} conditions hold for some choice of $F$. Then, as stated in Section \ref{sec:intro_gf},  we can describe any element of $\F$ as $h_\sigma^g$, where $h$ belongs to the surface in question, that is, satisifies the condition  $F(h_\sigma)=0$; and $g\in\G$ describes a gauge transformation as applied to the given element of the section.

A gauge-fixing  thus yields a one-to-one relation: $[\varphi]\leftrightarrow h_\sigma(\varphi):=\varphi^{g_\sigma(\varphi)}$, which  is what is meant when we say that the entire gauge-invariant content of a configuration is contained in its gauge-fixed form. In other words,  $h_\sigma$  is the representational convention, articulated as a projection from a member of
an equivalence class to a unique representative of that class.

We will see explicit examples of $h_\sigma(\varphi)$ in the appendix. It is also important to distinguish $h_\sigma(\varphi)$, which is itself a gauge potential, from $g_\sigma(\varphi)$, which is a group transformation taking $\varphi$ to $h_\sigma (\varphi)$. To unclutter notation, we will remove the subscript $\sigma$ from all functionals unless explicit reference to $\sigma$ is needed as a reminder.

In the upcoming Section  \ref{sec:subs}, we define the generic subsystems that we will focus on. This Section explains the obstacles towards satisfying  downward consistency when dealing with subsystems of a gauge theory. % In \ref{sec:gf_Lorentz} we work out the details of the familiar notion of Coulomb gauge within the framework proposed here, for the unbounded case. Although the bounded case follows in a similar fashion, it  is more mathematically complex and so I  leave it to Appendix \ref{sec:gf_bounded}. 
\subsection{The subsystems}\label{sec:subs}

%\footnote{I will take $g_{\mu\nu}$ to be of Euclidean signature. That is done to avoid certain complications that only occur for a Lorentzian signature of $g_{\mu\nu}$, but are only incidental, and  do not pertain to the issues at stake in this paper. In other words, essentially the same results would apply if we performed a $3+1$ decomposition of the system, or were in the Hamiltonian formalism, and implemented a Coulomb gauge-fixing, or were in the Lorentzian formalism and implemented future and past asymptotic boundary conditions. To discuss these issues would require some stage-setting, and distract from the main topic of the paper. See \cite[Sec. 2]{GomesRiello_new} and \cite{RielloSoft} for discussions. }

Now, I can briefly, and at a pedestrian level, address the issues posed by the non-locality of gauge theories on consistent definitions of subsystems, as mentioned in Sections \ref{sec:int_bdary} and   \ref{sec:rep_conv_DES}.

First, I will just schematically introduce the issue, as seen through the Lagrangian formalism. In the Abelian case, we define the field strength $F_{\mu\nu}:=\pp_{[\mu}A_{\nu]}$, where square brackets denote anti-symmetrization. The Yang-Mills action in vacuum  then is: 
\be S(A):=\int_{M\times \RR}   F_{\mu\nu}F^{\mu\nu}.
\ee
On a bounded submanifold, say, $R\times \RR$, where $R\subset M$ is a spatial submanifold of $M$, a variation of the action yields, after integration by parts: 
\be\label{eq:deltaS} \delta S(A)= -\int_{M\times \RR} \delta A^\nu(\pp^\mu F_{\mu\nu})+\oint_{S\times \RR} s^\mu  F_{\mu\nu} \delta A^\nu,
\ee
where $s^\mu$ is the normal to the hypersurface $S\times \RR$ in ${M\times \RR}$. %When restricted to an equal time surface $s^\mu$ becomes the spatial normal $n^i$, and $ s^\mu  F_{\mu\nu}$ becomes $n^i E_i$, the normal component of the electric field, also called \emph{the electric flux}, which we denote with an $f$.
 Now, for the first term of \eqref{eq:deltaS} to vanish for arbitrary variations of the gauge potential it suffices that the gauge potential satisfies the vacuum Maxwell equations. But the second term vanishes only if either the electromagnetic field tensor  vanishes along the boundary or $\delta A^\mu$ vanishes at the boundary. The first condition is severely limiting; the second is not a gauge-invariant condition.\footnote{It is important here that these are time-like boundaries; for the spacelike initial and final surfaces, one can implement whatever initial conditions one likes. And the boundary term gives rise to the symplectic potential: $\theta=\int E^i\delta A_i$, which defines the symplectic structure of the theory, $\Omega:=\delta\theta$.  } 

In the symplectic formalism, we witness a similar obstruction: in brief,  denoting the symplectic 2-form by $\Omega$ (i.e. a closed, non-degenerate 2-form on phase space), infinitesimal generators of gauge transformations, $\xi\in C^\infty(M, \mathfrak g)$ are usually  characterized by generating phase space vector fields $\xi^\#$ in the kernel of the symplectic-form, that is, gauge transformations satisfy:\footnote{\label{ftnt:magic}Indeed, the null directions of  $\mathfrak{i}^*(\omega)$, where $\mathfrak{i}$ represents the embedding of the constraint surface into phase space,  are necessary and sufficient to characterise the generators of gauge symmetry. For suppose that what we know is that  a certain class of vector fields $X_I$ is such that $\omega(X_I, \bullet)=0$.  Since  the exterior derivative $d$ commutes with pullbacks, if $\omega$ is closed, $\mathfrak{i}^*\omega=:\tilde \omega$ is also closed. Thus using the Cartan Magic formula relating Lie derivatives, contractions $i$ and the exterior derivative $\d $:   %(which is valid on general Banach manifolds), we get:
$$
 L_{X_I}\tilde\omega=(\d i_{X_I}+ i_{X_I}\d)\tilde\omega=0; 
$$
i.e. the first term also vanishes because $\tilde\omega(X_I,\bullet)=0$. So $\tilde\omega$ itself is invariant along $X_I$. Moreover, if we take the commutator of $X_I, X_J$, i.e. $[ X_I, X_J]= L_{X_I}X_J$,  contract it with $\tilde\omega$, and remember the formula:
$$L_{X_I}(\tilde\omega(X_J, \bullet))=\tilde\omega( L_{X_I}X_J, \bullet)+  (L_{X_I}\tilde\omega)(X_J, \bullet) \, ,
$$
we obtain that, since both $L_{X_I}(\tilde\omega(X_J, \bullet))=0$  and $L_{X_I}\tilde\omega=0$, it is also the case that $\tilde\omega( [{X_I}, X_J], \bullet)=0$. Thus, by the Frobenius theorem  the kernel of the pullback  $\mathfrak{i}^*(\omega)$ forms an integrable distribution which  integrates to give  the orbits of the symmetry transformation.  This means we can define a projection operator $\pi:\Gamma\rightarrow \Gamma/G$; and, ultimately the degeneracy of $\mathfrak{i}^*\omega$ allows one to define a \emph{reduced symplectic form}, $\bar\omega$,  on the space of orbits, given by $\pi^*\bar\omega=\mathfrak{i}^*\omega$. See \cite[Ch. 1]{Marsden2007}. This will be picked up in footnote \ref{ftnt:Aldo}, below.}
\be\label{eq:Gauss} \mathbb{i}_{\xi^\#}\Omega\approx 0,\ee
where $\approx$ means the equality holds after we impose the kinematical constraints, or conservation laws (see \cite[Ch. 1]{HenneauxTeitelboim} and \citep{Butterfield_symp, GomesButterfield_electro} for  philosophical introductions). % In the case of Yang-Mills theory, those are the Gauss constraints, $\Xi$,  and, in the absence of boundaries, we would obtain $\mathbb{i}_{\xi^\#}\Omega=\delta \Xi$, meaning gauge transformations are generated by the symplectic flow of the constraint . 
 For Yang-Mills theories, with a  general, non-Abelian algebra $\mathfrak{g}$,  \eqref{eq:Gauss} always obtains in the absence of boundaries. But in the presence of boundaries, it only obtains if $\xi_{|S}=0$ or $f=0$, which, again, are either severely limiting isolation conditions or do not respect downward consistency (Equation \eqref{eq:dwc}).\footnote{In more detail, let $\Omega=\int \mathrm{tr}(\delta A \wedge \delta E)$. Then we obtain: 
 $$\mathbb{i}_{\xi^\#}\Omega=\int \mathrm{tr}(\d \xi \dd E+[A, \delta E]+[\delta A, E])=\int \mathrm{tr}(\xi\, \delta(\D_A E))+\oint \mathrm{tr}(\xi \delta f), 
$$ where $\D_A$ is the gauge-covariant derivative (cf. footnote \ref{ftnt:D}). 
We can extract two important pieces of  information from this equation: (1) the flow  of gauge transformations is Hamiltonian, i.e. such that for each $\xi$ we have a generating function on phase space, $H_\xi$ such that: $\mathbb{i}_{\xi^\#}\Omega=\delta H_\xi$ iff $\delta \xi=0$ and either $\xi_{|S}=0$ or $\delta f_{|S}=0$. But, unless $f=0$,  $f$ is not gauge-invariant in the non-Abelian theory, and therefore we cannot fix $\delta f=0$ (unless $f=0$);   (2) $\xi^\#$ is in the kernel of the symplectic form iff  $\xi_{|S}=0$ or $f=0$. But using representational conventions, one can find gauge-invariant regional structures labeled by the  fluxes at the boundary; see \citep{Riello_symp} and footnote \ref{ftnt:Aldo}. }
 % and $\delta \xi=0$ (cf. \cite[Prop. 3.6 and Cor. 3.7]{GomesRiello_new}).\footnote{Here is a short proof, for the Abelian theory, in vacuum, where $\Omega=\int_M\delta A_i \wedge \delta E^i$ and  $\xi\in C^\infty(M)$, we obtain: $\mathbb{i}_{\xi^\#}\Omega=\int_M \pp_i \xi \delta E^i=-\int_M\xi \delta\pp_i E^i+\oint_S \xi \delta f.$}

  %That is, \eqref{eq:Gauss} only obtains if the transformations vanish at the boundary---meaning that they must be the identity there---and if $\xi$ does not depen on $A$. However, as we discussed in Section \ref{sec:rep_conv}, gauge transformations are defined with respect to a representational convention, or a section, and representational conventions \emph{do} depend on the gauge potential (cf. \eqref{eq:gf_bulk_Neu}).  In many contexts this dependence is  relatively unimportant; for unbounded regions for instance, it makes no difference to \eqref{eq:Gauss}. But when we want to glue subsystems, we have more than one convention at play and that dependence becomes crucial. Thus in the present context we must allow state-dependent gauge transformations at the boundary.  

In line with these considerations, the resolution pursued in \cite{GomesStudies, GomesRiello_new, Riello_new, GomesHopfRiello, GomesRiello2018,  GomesRiello2016}  is to consider all variations to be performed within the same representational convention. The dependence on the representational convention then appears explicitly in the variational procedure through the projection operator, $h_\sigma$, given in \eqref{eq:proj_h}. By taking into account the phase-space dependence of this projection, the  dynamical structures of the subsystem  become suitably gauge-invariant (cf. \cite[Section 3]{GomesRiello_new}). 

It is also interesting to note that, in a given convention, $h_\sigma(\mathbf{A})$ only captures the content of a principal connection, $\omega$, in directions that lie along the section, or representational convention,  $\sigma$. The vertical component of $\omega$---which is dynamically inert, since it is determined by gauge covariance---can be seen (in a suitable interpretation of differential forms, cf. \cite{Bonora1983}) as the BRST ghosts; see \cite{Thierry-MiegJMP}. When we have two regions, we have two sections, or two representational conventions. In the Thierry-Mieg interpretation, an infinitesimal relation between states $h_{\sigma_\pm}(\mathbf{A_\pm})$ is given by the vertical part of $\omega$; integrating this difference we obtain transition functions. We can think of that transition as our $g_{\sigma_+}(\mathbf{A}_-)$, defined in \eqref{eq:gauge-fixing}, defined at the boundary. So (1): there is an intimate relation between ghosts and the projection operators $h_\sigma$; and (2)  both mathematical objects are only dispensable in the classical domain with a single, unbounded region. Once we need to take into account multiple physical states---as we must in either the quantum regime or in the presence of boundaries---we need a representational convention.  The relationship between ghosts, representational conventions  and the gluing of regions is elaborated in \cite{GomesStudies, GomesRiello2016}. Thus we speculate:  the restoration of invariance of the regional dynamical structures is due to the  use of the classical BRST differential, that becomes manifest only upon either the gluing of regions or upon quantization; and that this is the main reason Gomes and Riello's functional connection-form works to restablish gauge-invariance.

 The resolution is pursued differently in \cite{DonnellyFreidel} and follow-up papers (see e.g. \cite{Geiller_edge2020} for a full list), which adds degrees of freedom at the boundary with appropriate gauge-covariance properties so as to cancel out the unwanted terms. The two approaches are related through a suitable interpretation of the new degrees of freedom as our $g_\sigma$ of \eqref{eq:gauge-fixing} (see e.g. \cite[Section 5]{Riello_symp}, \cite[Section 5]{ReggeTeitelboim1974} \cite[Section 4]{carrozza2021}). 

But let us focus on the Riello and Gomes's resolution through explicit representational convention. In the symplectic formalism, their choice of representational convention symplectically pairs  $h_\sigma(A)$  with the radiative content of the electric field. In more detail, the electric field can be split into  a radiative and Coulombic component, as discussed in \cite[Section  6.5]{GomesRiello_new}. The radiative component corresponds, roughly, to  radiation (also in the non-Abelian case), and it does not depend on the contemporaneous distribution of charges nor on the value of $f$ at the boundary; whereas the Coulombic component is entirely determined by these two pieces of information. The crucial mathematical property for the split in phase space is that the Coulombic component is symplectically orthogonal to the $h_\sigma(A)$.\footnote{ That representational convention is   a generalization of Coulomb gauge (see \cite{GomesButterfield_electro} for a philosophical/conceptual analysis of the relation between Coulomb gauge and the radiative/Coulombic split of the electric field). } The radiative/gauge-fixed regional phase space structure is fully gauge-invariant, but it leaves out a part of phase space that pairs up the group-valued $g_\sigma$ of \eqref{eq:gauge-fixing} with the electric flux.\footnote{ This is an alternative characterization of edge-modes: as the conjugate  to the electric flux. \cite{GomesRiello_new} make the symplectic pairing mathematically rigorous and find that this definition is imprecise: it is the entire Gauss constraint, together with the boundary flux, that becomes conjugate to the $g_\sigma$ in the entire region. }

When we take into consideration the full phase space, there are superselction sectors: i.e. different symplectic spaces attached to each gauge orbit of the electric flux, and these sectors are dynamically decoupled from each other. Although the full phase space structure of a regional subsystem  is therefore indexed by the   gauge-invariant class of $f$ at the boundary,  this superselection becomes redundant once we have at hand both the charged matter content and the radiative/gauge-fixed symplectic pair of each region \citep[p. 57]{GomesRiello_new}: 
\begin{quote}
Once both regional radiatives are known, even the regional Coulombic components
are completely determined---including the electric 
flux $f$ through $S$, which is thus no longer
an independent degree of freedom once the radiative modes are accessible in both regions.
Thus, in this case---when the larger (glued) region $M$ has no boundary---the regional
radiative modes encode the totality of the degrees of freedom in the joint system. In particular, the conclusion reached in section 3.4 from a regional viewpoint that $f$
through $S$ must be superselected is a mere artifact of excluding [radiative] observables
in the complement of that region.
The addition of charged matter does not change this conclusion.
\end{quote}

 In sum, using a representational convention and the appropriate variational principles, we can use a gauge-invariant characterization  of the regional phase space, in accord with downward consistency, given in \eqref{eq:dwc}. 
 Thus we can consider  the configuration space of the
theory over any spacetime as built out of the configuration spaces of the theory
over subregions of that spacetime.  The internalist's  splitting of the manifold naturally induces an identification of subsystems and regions, and ensuing identifications of their respective state spaces and symmetries.

\subsection{Preliminaries: from Yang-Mills to vacuum electromagnetism}\label{sec:premsYM}

Now we especialize to Yang-Mills theories. We do not need to explicitly exhibit the Lagrangian or the Hamiltonian, since in these cases the fundamental and the dynamical symmetries match, we just let $g\in\G$, where $\G:= C^\infty(M, G)$, for a  spatial manifold $M$.   In the vacuum Yang-Mills case,   $\varphi$ are identified as the field configurations, $A$; they are representatives of the equivalence classes, $[A]$.  Here $A\sim A'$ iff $A'=A^g$, and, in vacuum, 
\be\label{eq:Phi}\Phi\equiv\mathcal{A}:=\{A\in \Lambda^1(M, \mathfrak{g})\}\ee (the space of Lie-algebra-valued smooth  one-forms on $M$). The momentum variables conjugate to $A$ are the electric fields---Lie-algebra-valued vector fields, which we denote by $E\in \mathfrak{X}(M, \mathfrak{g})$. 

For the detailed exposition, in Appendix \ref{sec:gf_Lorentz}, we will  specialize to $A_i$ as the electromagnetic gauge field. The fundamental, or \textit{charge group}, of this theory is $G= \mathrm{U}(1)$, with an associated Lie-algebra $\mathfrak{g}=\RR$. When we include matter in the case of electromagnetism, we will assume it is of the Klein-Gordon type, , e.g. a map $\psi:M\rightarrow \bb C$. With an appropriate choice of units, the gauge transformations  are:
\be\label{eq:g_trans} 
\begin{cases}
A_i&\rightarrow (A^g)_i:= A_i+i\pp_i\ln g\\
 E^i&\rightarrow (E^g)^i:=E^i\\
 \psi&\rightarrow \psi^g:=g\psi
 \end{cases}
\ee
 for some  U(1)-valued function $g(x)$ (i.e. $g(x)$ is smooth complex-valued function satisfying $|g(x)|=1$).   That is, in the vacuum case the  $\varphi$ of the previous section would here be the electromagnetic potential, $A$, which changes non-trivially under the gauge transformation, whereas its conjugate variable, $E^i$ is invariant under it.

Given the embeddings $\iota_\pm: R_\pm\rightarrow M$, in the vacuum case, we get $\Phi_\pm\equiv\mathcal{A}_\pm:=\{A_\pm\in \Lambda^1(R_\pm, \mathfrak{g})\}$, where $\Lambda^1(R_\pm, \mathfrak{g})$ are the Lie-algebra-valued (i.e. here $\RR$-valued) smooth 1-forms on the spatial submanifolds $R_\pm$. Here we take the  surface $S$ to not impose any condition on the states, and thus it is specified gauge-invariant. As per the considerations of Section \ref{sec:subs}, the dynamics of the region then has an invariance group:   $\G_\pm:=\G\circ \iota_\pm= C^\infty(R_\pm, G)$ so that the the subsystem satisfies downward consistency.\footnote{More accurately, we would have to partition the phase spaces $T^*\F_\pm=\cup_{[f]}\F^{[f]}_\pm$, where $[f]$ is the equivalence class of electric fluxes (in the Abelian theory, since $f$ is gauge-invariant, no squre brackets are necessary). For each value of $[f]$ there is a well-defined gauge invariant regional symplectic form. Indeed,  we can abstractly define the projected, or reduced symplectic form: $\Omega^{[f]}_{\tiny\text{red}}$, as $\pi^*\Omega^{[f]\pm}_{\tiny\text{red}}=\mathfrak{i}_{[f]\pm}^*\Omega^{\pm}$, where $\mathfrak{i}_{[f]\pm}$ is the embedding of the Gauss constraint surfaces for $[f]$ (as in footnote \ref{ftnt:magic}. See \cite[Section 4]{Riello_symp} for the derivations of these facts and for the relation between the reduced symplectic form and the symplectic form in a representational convention.  \label{ftnt:Aldo}} We could also include matter in $\F$, as long as we implement boundary conditions that are gauge invariant according to  \eqref{eq:g_trans}; e.g. if matter  is absent from the boundary.

\subsection{Finding DES in gauge theories}\label{sec:DES}
In Section \ref{sec:gen_con},
  I summarize the main ideas involved in gluing or composing physical Yang-Mills states and articulate the matter of DES as a remaining physical variety after gluing. Section  \ref{sec:setup_gluing} summarizes the main technical achievements of the approach, which are considered in detail in Appendix \ref{sec:gf_Lorentz}, where I describe the procedure explicitly in the representational convention that corresponds to Coulomb gauge. 

\subsubsection{General considerations}\label{sec:gen_con}
The question of DES as I have construed it here amounts to whether there are \textit{physically distinct} ways that the composition of \textit{physically identical} subsystems states can go. We need to  assess the possibilities for satisfying \eqref{eq:phys_var}, finding $[\varphi]=[\varphi_+]\boxplus[\varphi_-]\neq [\varphi']=[\varphi'_+]\boxplus'[\varphi'_-]$ and non-trivially satisfying DES according to Section  \ref{sec:gluing}. Our main aim will be to unpack \eqref{eq:final_gluing}, which we reproduce here:
\be\label{eq:final_gluing1}\varphi_{\sigma_+}^{g_+}+\varphi_{\sigma_-}^{g_-}=:\varphi_\sigma\neq\varphi'_\sigma:=\varphi_{\sigma_+}^{g'_+}+\varphi_{\sigma_-}^{g'_-}.
\ee
%Gauge-fixings have simplified all questions of identity: $h(A)=h(A')$ iff $A\sim A'$ (or $[A]=[A']$). 
%The strategy now to find DES  is quite simple:  we are given two regional gauge-fixed states, and want to find out in how many universally physically distinct ways we can compose them (see Figure 1). 
We can translate \eqref{eq:final_gluing1} using the present notation as:
\be\label{eq:final_gluing2}h_+^{g_+}\Theta_++h_-^{g_-}\Theta_-=:h\neq h':=h_+^{g'_+}\Theta_-+h_-^{g'_-}\Theta_-.
\ee
 The $\Theta_\pm$ in \eqref{eq:final_gluing2} are the (Heaviside) characteristic functions of regions $R_\pm$.\footnote{The assumption of states as supported on the regions $R_\pm$ and adjacency of the regions fixes the embedding of the subsystems through the distributions $\Theta_\pm$. Then conditions for gluing become simply smoothness conditions.}          
%This equation says that whether two regional physical states are composable must be determined in terms of their representatives. And given the gauge-fixed representatives, the question is \textit{not} whether they smoothly join \textit{simpliciter}. 
We are given the physical content of the regional configurations, $[A_\pm]$ as input, and that is enough for our purposes of assessing DES; and  while the  $h_\pm$ that represent these physical states might not smoothly join,    they may still jointly correspond to a physically possible global state. 
As we saw in Section  \ref{sec:gluing}, whether two regional gauge-fixed  states $h_\pm$ are composable turns on whether there are  gauge transformations  on each region such that  
the transformed states---no longer of the form $h_\pm$---smoothly join, or glue. But if they are composable there will be many such transformations. Equation \eqref{eq:final_gluing2} selects only those transformations that lead to a global state in the chosen  representational convention, which thus allows us to infer physical differences from differences of the represented states. % But if we have already shown that $h(A)$ is gauge-invariant, how do we proceed?

% In \citep{GomesStudies, Gomes_new}, the construal of a gauge-fixing as a projection, mapping $\mathcal{A}\rightarrow{\mathcal {A}}$, and not as a quotienting, mapping $\mathcal{A}\rightarrow{[\mathcal {A}]}$, was argued as fundamental for the gluing of regions: both are gauge-invariant wrt gauge transformations of the domain $\mathcal{A}$, but only one allows a transformation in the range. In this respect, the external gauge-transformations will prove fundamental.

%That is, subsystem-intrinsic gauge transformations are defined as those acting on  the field configurations in the domain of the projection, whereas the subsystem extrinsic act on its range. Another way of putting it is that subsystem-intrinsic transformations acts between the members of the same equivalence class, whereas the extrinsic ones act as transformations between the representatives of these equivalence classes. 

\subsubsection{A summary of gluing the Abelian gauge potential in the Coulomb representational convention}\label{sec:setup_gluing}
%We are now finally ready to describe uniqueness conditions on the composition of physical states. 
%When I introduced subsystem-extrinsic transformations, in \eqref{eq:external_gt}, as the group action on the range of the projection map $h:\mathcal{A}\rightarrow{\mathcal{A}}$, I mentioned the crucial role they played in the gluing of regions. They are needed  because the only criterion for gluing quotients  employs representatives.\footnote{See   \cite[Sec 2]{GomesStudies} for more on the fundamental difference  between the sorts of projection implied by $h$ and the quotienting procedure.%: for the latter, no more gauge transformations are possible, and thus no possibility of gluing regions remains.} That is, there is no composition of physical states, $\boxplus$ of \eqref{eq:phys_var}, that is not formulated in terms of the composition of representatives $\oplus$.  Once we have eliminated redundancy and fixed a 1-1 correspondence with $[A_\pm]$ by the use of a gauge-fixing section, the image of $h$, i.e. $h[\mathcal{A}]\subset\mathcal{A}$, is invariant with respect to gauge transformations acting on its domain, but we can still change representatives by acting on its range, $\mathcal{A}$. 
Here we summarize, in a more perdestrian language, the  conclusions of Appendix \ref{sec:gf_Lorentz}. 

The existence of gauge transformations smoothening out the transition between $h_+$ and $h_-$ is a necessary and sufficient condition for their compatibility. 
%In sum, we have agreed to employ  $h_\pm$ and  global $h$'s  as representational conventions to assess physical differences.  Moreover, the $h_\pm$ are all we need as input to determine the compatibility of different regional physical states. That is: $h_\pm$ determine whether they can be joined by  gauge transformations (see footnote \ref{ftnt:subs_int_ext}). 
The condition is that there exist gauge transformations satisfying:\footnote{Note that we can still change  the representational convention itself. In footnote \ref{ftnt:subs_int_ext}, these types of transformations are labeled subsystem-extrinsic. This is how the smoothening gauge transformations need to be interpreted.}
\be 
(h_+-h_-)_{|S}=i\mathrm{grad}(\ln g_+-\ln g_-)_{|S};\label{eq:bdary_cont2}
\ee   (in spacetime index-free notation\footnote{Using indices, the equation is: $(h^\mu_+-h^\mu_-)_{|S}=i\partial^\mu(\ln g_+-\ln g_-)_{|S}$. }) which is the appropriate rewriting of the gluing condition \eqref{eq:bdary_cont}.

There could be many such possible ``adjustments'' of $h$; there are either none or an infinite amount of $g_\pm$ that will  satisfy \eqref{eq:bdary_cont2} and we need to partition all of these possibilities into physical equivalence classes. For the remaining question---whether the composition of regional states is physically unique---we employ a gauge-fixing of the global state, i.e. we demand that the global state is also given in some representational convention, $F$, or $\sigma$. Indeed, as discussed in Section \ref{sec:rep_conv} (see in particular the quote from \cite{Wallace2019}), that is the only way we can assess physical differences between alternative global states. %And this was how we originally obtained \eqref{eq:final_gluing}.  
   %To repeat: the question of DES amounts to whether there are \textit{physically distinct} ways the composition of \textit{physically identical} subsystems states can go.% Since we cannot work directly with the equivalence classes, we employ representational conventions for the regional and universal physical states. 

Thus we are given $h_\pm$  that are in  the regional representational convention (i.e. satisfy \eqref{eq:cov_gf}) %and \eqref{eq:bdary_cont2} (necessary if $h_\pm$ are to be composable),
 and want to glue them into a state that satisfies the global representational convention, $h$. That is:
\be\label{eq:gluing} h:=(h_++i\mathrm{grad}(\ln g_+))\Theta_++(h_-+i\mathrm{grad}(\ln g_-))\Theta_-,
\ee
must satisfy, for some $g_\pm$,  the unbounded gauge-fixing condition \eqref{eq:gf_M} of Appendix \ref{sec:gf_unbounded} (so that we uniquely determine the universal physical state). %Therefore the solution is obtained by finding the appropriate  gauge transformations $g_\pm$ that also satisfy \eqref{eq:bdary_cont2}.

\begin{figure}[t]
		\begin{center}
			\includegraphics[width=6cm]{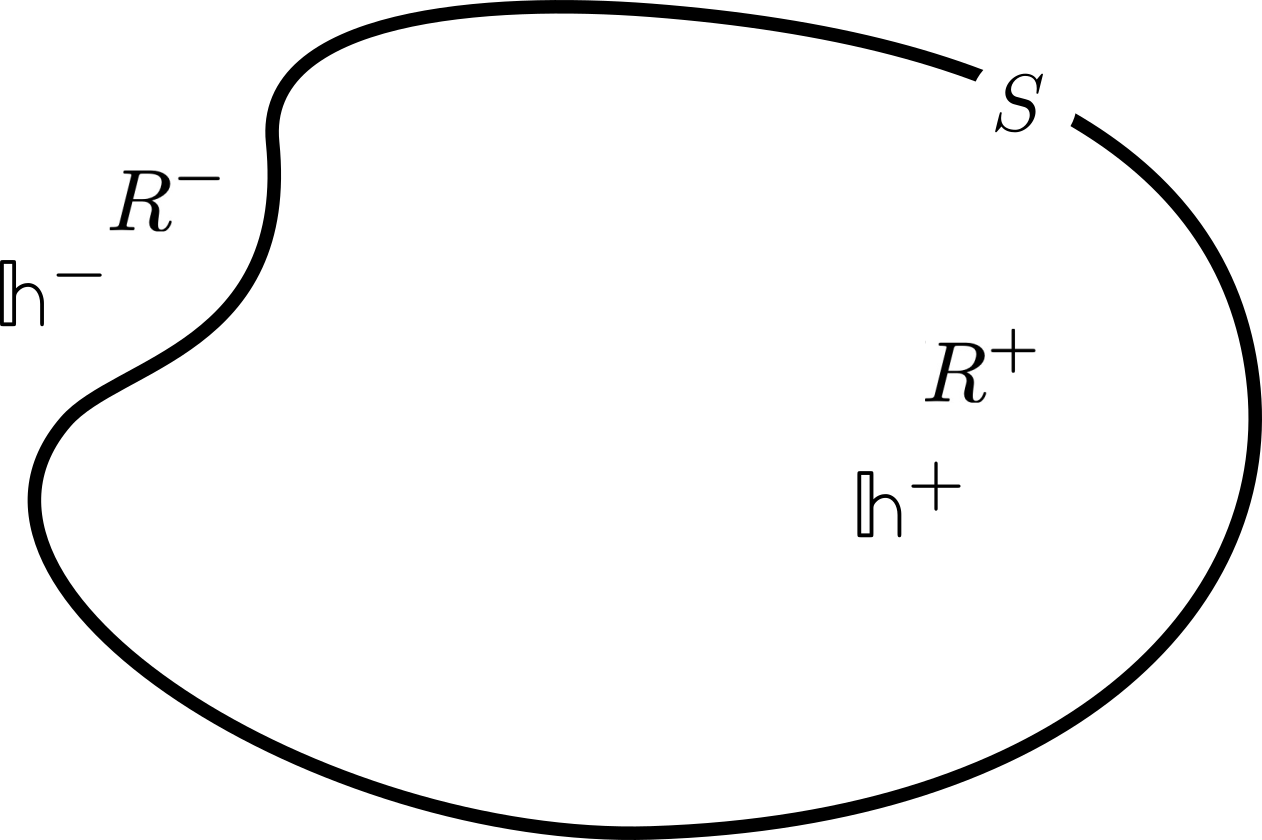}	
			\caption{ The two subregions of $M$, i.e. $R_\pm$, with the respective horizontal perturbations $ h^\pm$ on each side, along with the separating surface $S$. 
			}
			\label{fig4}
		\end{center}
\end{figure}

It is important to emphasize that the use of the gauge-fixed fields has eliminated all local redundancy. For instance, imposing that the regional states in their representational convention should match---$(h_+-h_-)_{|S}=0$---would  restrict our analysis to only a  subset of compatible physical configurations. In general, the universal $h$'s do not themselves restrict to $h_\pm$'s, as we saw in Section \ref{sec:gap} (see Equation \eqref{eq:mesh}).\footnote{Namely, $\iota_\pm^*h\neq h_\pm$. The generality of this inequality is a consequence of the non-locality of the gauge-fixing,  implicit in the inverse Laplacian. That is, the restrictions of universal $h$'s over $M$---satisfying \eqref{eq:gf_M}---to the regions $R_\pm$ are not necessarily themselves of the form of $h_\pm$, i.e. do not necesarily satisfy Neumann boundary conditions.  \label{ftnt:gluing} }  
% Demanding that the restrictions of $h$ are already in the form of some $h_\pm$  would restrict the space of compatible physical configurations to only a proper subset.
  Thus, as stated in Section \ref{sec:comparison}, if we are to use representational conventions, we cannot, when satisfying \eqref{eq:final_gluing2}, assume $g_-=\mathrm{Id}$ irrespective of the conventions used.\footnote{ Moreover, it would be impossible to meet in a  manifold that requires many charts (see Section \ref{sec:subs}).}    %\footnote{To get a bit ahead of ourselves:  this is precisely what will happen in \eqref{eq:overc}, when we apply a fundamental view of symmetries to the externalist's notion of boundary.\label{ftnt:gluing2}} 

Finally, since we have also partitioned the global state space, and identified 1-1 representatives of the physical equivalence classes,  whatever information  beyond the specification of the subsystem physical states that is required to determine $h$ will reveal a gap between $[\mathcal{A}]$ and the union of $[\mathcal{A}_\pm]$. % If we find underdetermination after gauge-fixing (or fixing the representational convention of) the global state, the underdetermination will no longer be descriptive. 
Thus any remaining underdetermination  will have physical significance, fulfilling the notion of DES. %This is easy to see: if there are multiple solutions $g_\pm$, they will correspond to different universal gauge-fixed configurations, e.g. $h$ and $h'$.
 % Since gauge-fixed configurations are physically identical if and only if they are identical, i.e. $[h]=[h']$ iff $h=h'$, further slack may realize the sought-after DES.\footnote{The slight complication to the statements above, is that, as we saw, the universal gauge-fixing \eqref{eq:h_div} doesn't completely fix the gauge (it only satisfies `Uniqueness'). The surviving group of DES would then have to quotient out the product of the two regional slacks by the universal degeneracy. In vacuum, this detail won't matter, since the action of these subgroups are stabilizers of the field.}
  In other words, the DES transformations---satisfying criteria Global Variance and Subsystem Invariance for DES, of Section  \ref{sec:debate}---\textit{will take the form of symmetries} on the subsystems, arising from the underdetermination of the global state by the gluing conditions that respect the global representational convention.\footnote{In the nomenclature of footnote \ref{ftnt:subs_int_ext}, these are subsystem-extrinsic symmetries, and as such conform to our intuition about symmetries with DES being applied from a perspective outside the subsystem and being undetectable from within.}

%The case of point-particles---which, in effect, formalizes the Galileo's ship thought-experiment---is different. %But before we turn to it, it is worthwhile illustrating how our analysis above plays out in the holonomy formulation of gauge theories since there we find a straightforward relation to the Aharonov-Bohm type of holism. 

\section{Conclusions}\label{sec:conclusions}
In gauge theories, empirical significance can be obscured by redundancy of representation.  Ultimately, that is why the direct empirical significance (DES), or the observability of symmetries  continues to be a debated question. Nonetheless, the standard treatment of DES is almost silent about fixing representation,\footnote{As noted in Section  \ref{sec:comparison} yielding equation  \eqref{eq:GW_DES}, if $\varphi_-$ is not kept fixed, one can always extend $g_+$. Such extensions have caused some confusion in the literature (see \citep{Friederich2014}).} with the exception of \citep{Gomes_new} and \cite{Wallace2019}, where the assumption {is} partially flagged, as noted in Section \ref{sec:rep_conv}, but not fully examined. Here I have paid due attention to this  issue.

In Section \ref{sec:summary}, I summarize the findings of this paper. In Section \ref{sec:asymptotic}, I discuss asymptotic idealisations in relation to the externalist view of subsystems. 

\subsection{Summary}\label{sec:summary}
The main question I have  investigated here is precisely how to establish a choice of representational convention  in the context of our search for DES.  This approach to DES reveals the inadequacy of the standard construction of Section   \ref{sec:comparison} and provides a straightforward alternative. And, while I agree with most aspects of \cite{GreavesWallace} and \cite{Wallace2019, Wallace2019b}'s analysis of symmetries,  I have recast the topic to focus on gauge-\textit{invariant} information about regions, by explicitly using representational conventions. This approach yielded a more precise formulation of the question of DES. 

The upshot is that gauge-fixing  disentangles the issue of redundant representation from DES, for \textit{all} types of systems and their subsystems.  Thus  we do not have to make extra assumptions (e.g. about the lack of bulk stabilizers): our construction is able to discern the existence of DES for any state. 
 
%For instance, no observable symmetry emerges  in the case of bounded Abelian gauge systems in a simply-connected, vacuum universe. In the presence of matter, the internalist's boundary only allows (relational) DES to be associated to rigid symmetries.\footnote{For non-Abelian theories, one can only single out rigid symmetries for configurations with stabilizers. In the nomenclature of \citep{Gomes_new}, these symmetries are dubbed `rigid'.  For a more general treatment including matter and non-trivial topology, where DES does emerge, see \citep{GomesRiello_new} and \citep{Gomes_new}.} 

The procedure identifies a type \emph{holism} lies at the core of DES: as articulated at length in \cite{Gomes_new}, there is a difference between $[\Phi]$ and $[\Phi_+]\cup[\Phi_-]$, even when $M=R_+\cup R_-$. If this is so, we should see the same sort of difference in other formalisms, that do not necessarily use gauge-fixing. We checked that this indeed occurs in the holonomy formalism for electromagnetism in Appendix \ref{sec:holonomy}.\footnote{In this we go beyond \cite{GreavesWallace, Wallace2019, Wallace2019b}, which explicitly open an exception to the use of holonomy variables (see e.g. \cite[p. 67]{GreavesWallace}). }

As another consistency-check, I then applied the gauge-fixed approach to  particle mechanics (Section  \ref{sec:particle}).  Thus, using precisely the same type of constructions as for gauge theories, I recovered the standard DES associated to Galileo's ship. In that context,  DES arises from the different ways to embed intrinsically identical subsystems into the universe.\footnote{No such possibilities exist in the field theoretic case, for the simply-connected manifold partitioned into regions; the embedding is determined by the partition. }%But DES arises for gauge theories in the same manner for the non-simply connected case, see \citep{GomesRiello_new}. }

%My  findings were of a similar character for both the internalist and the externalist notion of subsystem, as long as we restricted attention to dynamical symmetries. 

 The externalist case  requires  configuration space to be (non-covariantly) pared down. That is, we limit  not the set of physical possibilities, $[\varphi]$, but  the set of representatives, $\varphi$, at the boundary.  These boundary conditions would not be gauge-covariant under a fundamental view of symmetries. But by abandoning the requirement that gauge symmetries act equably on all configurations, the restriction does not break any symmetry.   This finding is entirely consistent with the idea that the environment state, whatever it is,  provides a representational convention. There is no gluing, and thus no requirement of meshing the global representational convention with the subsystem one, and no requirement that the subsystem symmetries must be compatible from the inside and the outside perspective on the boundary. 
 
 For many decades, the pared down asymptotic treatment of symmetries was assumed for the treatment of isolated subsystems. Thus, until recently, attempts to treat the finite, bounded subsystem of gauge theory were scarce (and mostly focused on computations of the entanglement entropy of black holes; the trailblazers were \cite{Carlip1997, Srednicki1993, Sorkin1983}). In our language, internal boundaries were not distinguished from external ones. 
  \citet[p. 11]{Wallace2019b} endorses the ensuing view of subsystems,   because he takes subsystems  as sufficiently isolated   to warrant an asymptotic-like treatment. One drawback is that such a  treatment would find no extension to spatially closed manifolds (as discussed in footnote \ref{ftnt:Wallace}). Another, is that the notion of isolation requires very strong boundary conditions, such as $F_{\mu\nu}{}_{|S}=0$. But recent developments in gauge theory have shown that we can have (non-asymptotic) bounded subsystems, in which e.g. $F_{\mu\nu}$ is non-vanishing everywhere, and which still enjoy the same set of fundamental symmetries for their intrinsic dynamics (even if they require time-varying boundary conditions). So clearly, there are good, weaker notions of subsystem recursivity that do not mimic the asymptotic ideal of  perfect isolation.  These recent developments were discussed in Section \ref{sec:subs}, and they warrant a treatment of internal  subsystems  in gauge theories that respects the downward consistency of symmetries.

    \paragraph{Classifying DES for gauge theory:}
  Using our  treatment based on representational conventions to assess DES, we can classify its occurrence for different types of systems (as computed in \cite{GomesRiello_new, Gomes_new} and summarized in the Appendices). As discussed in Appendix \ref{sec:stab}, stabilizers  represent certain degeneracies within any given  representational convention, and thus are crucial in articulating the results below. When they exist, stabilizers form a rigid (or subsystem-global) group of gauge transformations. \\
  
  First, assuming trivial topology, and an  internal boundary:
  
  \indent (i) in the Abelian case there is no physical variety in the absence of charged matter; or when  charged matter is present at the interface between the regions. That is because, to have underdetermination, the regional stabilizers of the gauge potential \emph{cannot} stabilize the state of all of the fields; but to preserve compatibility of the states at the boundary, it \emph{must} stabilize the boundary states (and Klein-Gordon matter fields have only the trivial stabilizer).  Thus the sector of the theory in which one has observable symmetries corresponds to regions that have charged matter in the bulk, but not in the interface of the regions. This sector contains the situation depicted by `t Hooft beam splitter experiment (see \cite[p.110]{thooft}  and  \cite[p. 651]{BradingBrown}); and likewise, the group of symmetries with DES is a rigid phase shift, given by $U(1)$.

\indent (ii) In the non-Abelian  case, we must distinguish a few possibilities. As in the Abelian case,  if the regions have the same set of stabilizers, and if a subgroup of stabilizers of the gauge potential act non-trivially on the regional states as a whole (e.g. by acting non-trivially on the matter fields), then there will be a physical variety, corresponding to the subgroup of the  (rigid) group of stabilizers. But such a condition is generically forbidden: generic states in non-Abelian Yang-Mills theory have only a trivial stabilizer. Moreover,  if the state at the interface of the region has stabilizers---meaning that there are gauge transformations that act as the identity only on the boundary states---then we also get one physical global state per choice of boundary stabilizer. These are what I take to be the \emph{physically relevant} notion of edge modes (see also \cite{carrozza2021}, for a similar argument).

    A comparison of these two cases  with the familiar Aharonov-Bohm phases in the Abelian theory is also worthwhile. There,   $M$ is taken to have a non-trivial topology, and the cohomology class of the gauge potential represents holistic physical information, that can nonetheless be represented at the boundary by suitable transition functions. Here too: there is a discrepancy between the tensor product of the regional physical  state  spaces and the physical state space of the union of the regions.  The discrepancy represents holistic physical information about the total system that is not contained in the individual subsystems. Nonetheless, we can represent this physical information through suitable mathematical operations either at the boundary between the two regions or on each region. \\

    Moving on to the external boundary, vacuum case:
    
\indent (iii) In this case, for either the Abelian or the non-Abelian case,  gauge-fixings formulated strictly  in terms of the gauge fields satisfy \textit{Uniqueness} and \textit{Universality} (as described at the end of Section  \ref{sec:gf}) \textit{only if} each transformation that  stabilizes the boundary is continuable to a transformation that stabilizes the universe. Otherwise,  gauge-fixings must also be indexed by the choice of boundary stabilizer.

    But these indices do not belong to the configuration space $\mathcal{A}$: they should be seen as additional degrees of freedom, which, ultimately, represent the externalist's version of  DES. They find a counterpart in the internalist scenario, in which the internal boundary state has stabilizers not shared by the bulk of the regions, as described in (ii) above.\footnote{The only way I know to make this construction kosher in the external boundary case, is that explored in \citep{Teh_abandon}, which endows these boundary `gauge' degrees of freedom with their own dynamics, following the work of \citep{DonnellyFreidel}. But one should not confuse these degrees of freedom with the generalized (non-Abelian) electric flux and its conjugate. These latter quantities  should not be interpreted as new degrees of freedom in the same way as edge-modes, and they do not contribute to the issue of DES (cf. \cite[Section  6]{GomesRiello_new} and Section \ref{sec:subs}). } That is, in the appropriate regime,  each choice of stabilizer intrinsic to the boundary corresponds to a different physical state,  matching the findings of \citep{GreavesWallace} (cf.  equation \eqref{eq:GW_DES}) and, in the asymptotic case of external boundaries, conforming to the intuition of Belot's `generalized shifts' \citep{Belot50}.%, as I describe in the next Section.

 \subsection{Externalist boundaries and asymptotics}\label{sec:asymptotic}

As a last remark,  I admit that the externalist's notion of boundaries  is ubiquitous in asymptotic treatments of symmetries.  In fact, we model the solar system in this way: the standard spatially asymptotically flat spacetime  imposes a particular form of the metric as one approaches the asymptotic boundary; it is not a diffeomorphism-invariant, geometric boundary condition. That is, the treatment of asymptotic symmetries cannot fall under the fundamental approach to symmetries, as discussed in Section  \ref{sec:fund_sym}. In this way, coordinates at the boundary acquire some physical meaning. And in this way, all the coordinates compatible with some given condition on the field acquire physical meaning: this variety is represented by the  non-trivial  boundary stabilizers, and they are what \cite{Wallace2019b} describes as observable symmetries; or what \cite{GreavesWallace} describe as symmetries with DES.

Note, moreover, that the non-trivial case of the asymptotic treatment arises precisely when there is a gap between the boundary and the bulk stabilizers. For instance, only the completely flat state (i.e. Minkowski space) extends to the bulk all the boundary stabilizers of a generic asymptotically flat spacetime. But as we find in the externalist case, we would only obtain DES if the bulk did not share the boundary stabilizers; e.g. if there is matter in the bulk or if the metric only asymptotes to the Minkowski metric. % A `generalized shift' (in Belot's nomenclature \citep{Belot50}), namely, a gauge transformation in the bulk that asymptotes to a non-trivial boundary stabilizer, is of no interest if applied to vacuum Minkowski: it does not change anything about the universe. 
In \cite{Wallace2019}, this intuition is preserved by a restriction of focus to cases without bulk stabilizers. But this restriction, as it stands there, is \emph{ad hoc}, while here our formalism includes all cases. 

The interesting examples are the ones for which different  stabilizers intrinsic to the boundary correspond to different physical states. We can find such cases in our formalism, and they \textit{agree} with \citep{GreavesWallace} and with the more general characterization of `physical symmetries' as corresponding to those of the quotient group $\G_S/\G_{\mathrm{Id}}$ (cf. \citep{Giulini_asymptotic}). 

Therefore I grant that even if, ideally, boundary conditions should be gauge-covariant so that the dynamical treatment of symmetries coincides with the fundamental treatment of symmetries, the externalist approach---which does not abide by that ideal---may work well for some purposes.

 But, to obtain a solid conceptual footing, the externalist's notion os subsystem requires further conceptual analysis (see \citep{Belot50} which partially lays the groundwork for such an analysis). Thus I believe we should not leave the underlying assumptions about asymptotic symmetries unexamined simply because they are useful; lest we  acquiesce to what amounts to a `shut up and calculate' mentality in the treatment of gauge and asymptotics. 

A more conceptually grounded approach is also a more conservative one: it goes  from small systems to big ones. We should first properly understand gauge systems in finite regions and then move to the asymptotic regime by progressive enlargement, keeping careful track of how  objects and relations maintain or lose their properties in the (singular) limit.\footnote{ This is how Riello relates the subsystem divisions used here to the asymptotic regime, in the case of Yang-Mills theory \citep{RielloSoft}. By doing so, he finds a singular limit for the asymptotic charges, recovering precisely the  results of  \citep{AshtekarStreubel}, who also treats boundary conditions in a  diffeomorphism-invariant manner in general relativity,  through Penrose compactification.}

This should be possible: the internalist case imposes only gauge-covariant boundary conditions, and thereby unifies the fundamental and the dynamical treatment of symmetries. Not only that: it  recovers qualitatively similar observable symmetries as we found in the externalist case.  

%Thus, in sum: this study has shown that previous literature on DES \citep{Healey2009, GreavesWallace,  Wallace2019a, Wallace2019b} errs in large part by conflating two conflicting scenarios for  boundaries: as boundaries of the entire universe---the externalist's notion---, or as an artefactual surface due to a choice of splitting. 

\subsection*{Acknowledgements}
I would like to thank, first and foremost, Aldo Riello, my collaborator on this topic, and Jeremy Butterfield, who helped me shape the argument and the language of the entire paper. I would also like to thank  Nic Teh for helpful discussions, Valeriya Chasova for copious corrections and remarks, and audiences at the ILMPS Meeting at Salzburg and at Bristol, where I gave a talk on this topic.

\begin{appendix}

\appendix

\section*{APPENDIX}

\section{Coulomb gauge}\label{sec:gf_Lorentz}

 In Section \ref{sec:sketch} I sketch a solution to this mathematical problem.  In Section 
 \ref{sec:DES_matter} I glimpse how this solution extends to other cases that were left out of this paper in the interest of simplicity: namely, I briefly discuss the necessary alterations and caveats incurred by the addition of matter, non-trivial topology, and non-Abelian gauge groups.  
 
\subsection{Coulomb gauge for the closed universe}\label{sec:gf_unbounded}
Now we will illustrate the previous definitions explicitly, by employing an explicit gauge-fixing functional $F$.  I will describe how this works when the manifold is closed but without boundary. Formally, this is simpler than the bounded case, which we will leave to Appendix \ref{sec:gf_bounded}. Nonetheless, the simpler case already suffices to illustrate many of the intricacies of  gauge-fixing. This Section  is more technically involved. Its main purpose is to illustrate: (i) how a representational convention can fail to fix all representational redundancy, due to a lack of  `wrinkles' of the represented state; and (ii) how a representational convention operates like a projection on configuration space, and is in that sense gauge-invariant.

I will start in Section \ref{sec:stab} by posing the intricacies of fixing the gauge in the presence of stabilizers of the gauge potential. And in Section \ref{sec:gf_Lorentz1}, I will lay out the details of the Coulomb gauge (and the corresponding projection operator, etc) in the bounded case. In Section \ref{sec:gf_bounded}, I do the same for the bounded case. 

\subsubsection{How to fix representational conventions: the problem of stabilizers}\label{sec:stab}

Physically,  fixing representational conventions uses features of the state to nail down representational redundancy of that state. 
 
 However, for certain states there may be group elements that have no grip on  representation.  In other words, there may be $\varphi$, and certain 
  $\tilde g$ for which $\varphi^{\tilde g}=\varphi$. Viz. there are  certain group elements that  act trivially on certain states. In these cases, the orbits formed by the action of the group, $\mathcal{O}_\varphi$,  will also be, in certain respects, singular.\footnote{The quotient $[\F]$ in these cases form `stratified manifolds', which are essentially a concatenation of bounded manifolds of different dimensions, with the manifolds of smaller dimension being obtained by the states with more symmetry and being the boundaries of the manifolds of higher dimension. See \cite{Fischer, Mitter:1979un, kondracki1983}.  }
    Thus we define the \emph{stabilizer}: 
\be\label{eq:stab}  \mathrm{Stab}(\varphi):=\{\tilde g\in\G~|~\varphi^{\tilde g}=\varphi\}.\ee
It is easy to see that $\mathrm{Stab}(\varphi^g)=g^{-1}\mathrm{Stab}(\varphi)g$,\footnote{A quick proof: for $\varphi':=\varphi^g$, we take $g'=g^{-1}\tilde g g$, where $\tilde g\in \mathrm{Stab}(\varphi)$. Then 
$\varphi'{}^{g'}=\varphi^{\tilde g}g=\varphi^g=\varphi'$. } and thus the conjugacy class of the stabilizer group is a property of the entire orbit (i.e. it is not dependent on the representative of the physical state). 

In field theory, the group $\G$ of local (or malleable) gauge transformations is infinite-dimensional, since it is a space of maps from a smooth manifold $M$ to some value space. 
In practice, the  features of the states used to fix the representation belong to the gauge potential, $A$, and not to the matter fields or the electric field. 

That is because, either configurations of the matter fields or of the electric fields, transforming via \eqref{eq:g_trans}, generically will possess stabilizers.\footnote{This is also true for the electric field in the non-Abelian case, transforming under conjugation by the group element.} For instance,  configurations in which the matter fields vanish \emph{anywhere}, will have stabilizers; if $\psi(x)=0$, then $\psi^g(x)=0$: the group cannot change representation on those points because it has no grip there. Thus, for example, if the matter fields vanish on an open set, a gauge transformation that is only non-zero on  that open set will stablize the state.\footnote{Of course, we could be interested in sectors of the theory in which the matter fields do not vanish anywhere; configurations in which they form a plenum. In these sectors, it is legitimate to use matter fields to fix representational conventions. And indeed, in these cases, as we will comment on in Section  \ref{sec:DES_matter},  there is no DES. That is, the gauge theory is separable. This is in line with  \citep{Wallace_deflating} (see \cite{Gomes_new}). But I disagree with  \cite{Wallace_deflating} that such a plenum is generic, in light of quantum de-localization. Specifying configuration or phase space is prior to quantization; and de-localization is a higher level consequence of quantization that has no bearing there. }

But to fix representational conventions we would like to choose those fields that generically have no stabilizer, i.e. that on an open and dense set of $\Phi$ have no stabilizer. This criterion selects the gauge potential as the field used to fix representational conventions. With that choice, a meagre set of states of the fields \emph{will} possess stabilizers, but the group of stabilizers will be rigid, or finite-dimensional, so that the values of a stabilizer on an open region determines that stabilizer everywhere.

The importance of these stabilizers for DES is that they are left unfixed by representational conventions, even if that convention fixes all local redundancy, i.e. completely fixes the representation of $A$. For instance, within fixed representational conventions for $A$, a disparity between stabilizers of two fixed subsystem states may give rise to DES as follows. Suppose two subsystem stabilizers $\tilde g_\pm$ do \emph{not} conjoin to form a global stabilizer of the joint region, $\tilde g$. Since we assume the representational convention has fixed all local redundancy, it will not allow a representational change that smoothens out the difference between the stabilizers. However, even if stabilizers do not change the representation of $A$, they can change the representation of the matter fields. In this case, incompatible stabilizers in each region will give rise to   a different global state, which, within a fixed representational conventions, implies a physical difference, that is,  \textit{Global Variance} will be satisfied.

 In \citet[p. 87]{Gomes_new}, it is argued that stabilizers are the only natural notion of global symmetries in a gauge theory, since they can, independently of any representational convention, pick out rigid subgroups from the   local groups of  gauge transformations. And indeed, the difference between stabilizers is an entirely gauge-invariant quantity. Thus finding observability criteria in terms of functionals of these stabilizers is consistent with gauge-invariance and a welcome development. 
 
 %\subsubsection{Representational conventions using the gauge potential, and stabilizers}\label{sec:rep_conv_Ab}

Here we will select the gauge potential as the field that orients the representational conventions. That is, representational conventions will be chosen by specifying particular forms for the gauge potential. 

 I will only consider gauge-fixings that completely fix the representation of $A$, but note that rising to this challenge does not require the solution $g(A)$ of \eqref{eq:gauge-fixing} to be unique. For, even if $F(A)=0$ underdetermines $g(A)$, as long as this underdetermination is only up to a stabilizer of $A$, as in \eqref{eq:stab},  the gauge-fixed representative of $[A]$ will not be underdetermined. In other words, suppose that $g(A)$ and $g'(A)$ both satisfy \eqref{eq:gauge-fixing}, as long as they differ by stabilizers of their argument, say $g'(A)=\tilde g(A)g(A)$, we will still obtain: 
 \be\label{eq:stab_diff}
 h(A)=A^{g(A)}=(A^{\tilde g(A)})^{g(A)}=A^{g'(A)}
 \ee
 and so the difference does not show up at the level of the projection operator.
 
 The presence of non-trivial stabilizers implies that  features of $[A]$ do not possess enough variety---``wrinkliness''---to completely fix the \textit{gauge transformations} that carry an $A\in [A]$ to an $h(A)$. In other words, a representational convention can fail to fix all representational redundancy, due to a lack of  `wrinkles' of the represented state. Nonetheless, the represented state cannot register any difference due to this remaining degeneracy, precisely because the state is not `wrinkly' enough for that remaining redundancy to get a grip on. In other words, if the only ``slack'' left in the determination of $g_\sigma(A)\in \G$  is due to stabilizers,  it is idle: there is no effect on the resulting gauge-fixed $h(A)$. That is because  that slack  has a trivial action  on the configuration.

For non-Abelian groups, $\mathcal{A}$ is generically stabilizer-free and so stabilizer groups are generically  trivial, i.e. just the identity. Nonetheless, particular physical states, such as the physical state of ``no $A$ field'', i.e.  $A\sim 0$, allow stabilizers, and thus do not allow gauge transformations to be uniquely fixed. In the Abelian case, all configurations share the same stabilizer, viz. the group of constant gauge transformations (cf. the discussion in Section  \ref{sec:gf_Lorentz}).

 Indeed,  for the purposes of this paper, this is the most important distinction between Abelian and non-Abelian theories. Namely: stabilizers are the same for all Abelian field configurations---they are the constant transformations---and,  on the other hand,  are trivial for generic non-Abelian field configurations.

\subsection{Details of Coulomb gauge}\label{sec:gf_Lorentz1}
Let us start by introducing a standard gauge-fixing for the entire manifold: Coulomb gauge.\footnote{Our specific findings do not depend on the particular gauge-fixing, as long as it adheres to the definitions in Section  \ref{sec:gf}; these definitions imply that some non-locality, or integration over the spatial region, will be involved in finding the particular gauge-fixed representation. And it will also necessarily satisfy `\textit{Uniqueness}', as explained at the end of Section  \ref{sec:gf}. All the constructions presented here have an exact analogue in the non-Abelian case. Essentially, the analogue  replaces $\pp_i$ by the gauge-covariant $\D_i=\pp_i+[A,\cdot]$, and constant gauge transformations are replaced by the more general concept of stabilizer \eqref{eq:stab}. 
See  \citep{GomesHopfRiello, GomesRiello_new, Gomes_new}. \label{ftnt:D} } Following the nomenclature of Section  \ref{sec:gf} for the gauge-fixing section $\sigma$, we define:
\be\label{eq:gf_M} F(A):= \mathrm{div}(A)=0.
\ee
 As we will now see, it is easy to see that this gauge-fixing satisfies \textit{Universality} and \textit{Uniqueness}. First,  \textit{Universality}: given a general $A$, not necessarily belonging to the given gauge-fixing section, i.e. $A$ which can be such that $h(A)\neq A$, we must ensure that there exists a gauge transformation that takes $A$ to that section. Second, \textit{Uniqueness}: we must ensure that whatever slack remains in the determination of this transformation cannot be ``detected'' by any $A$. 

As to  the first demand, equation \eqref{eq:g_trans} yields:
\begin{eqnarray}  \mathrm{div}(A^g)= \mathrm{div}(A)+i\nabla^2(\ln g)=0\label{eq:gf_bulk0}\\
\therefore\,\, g(A)=\exp(i\nabla^{-2}(\mathrm{div}(A)))\label{eq:gf_bulk}
\end{eqnarray}
where $\nabla^{-2}$ are the Green's functions associated to $\nabla^2$.\footnote{Roughly,  on the space of functions, for $f\in C^\infty(M)$ we define the Green's function as an operator inverse to the Laplacian, i.e. $\int_M \nabla^{-2}_{xy} (\nabla^2 f )(x)=f(y)$. For a closed manifold, the operator exists and is unique on the space of non-constant functions, see e.g. \citep{Trudinger}.   When the metric is sufficiently homogeneous, Green's functions can be obtained in explicit form.}  For all $A$, we can find a solution $g(A)$ to $\sigma(A^{g(A)})=0$ and thus a projection, $h(A)$. Therefore the gauge-fixing is \textit{Universal}.

Consistently,  the 1-form $h_i(A):=A^{g(A)}_i$, defined through \eqref{eq:gf_bulk} does satisfy \eqref{eq:gf_bulk0}. That is, it is easy to verify, since $\mathrm{div}(\mathrm{grad})=\nabla^2$, that 
\be\label{eq:h(A)}h(A):=A+i\,\mathrm{grad}(i\nabla^{-2}(\mathrm{div}(A))),\ee satisfies 
\be\label{eq:h_div} \mathrm{div}(h(A))=0.
\ee 

Moreover, the projection $h$ is invariant under  gauge transformations: $\forall g\in \G, h(A^{g})=h(A)$:
\begin{eqnarray}h(A^{g})&=& A+i\,\mathrm{grad}(\ln g)+i\,\mathrm{grad}((i\nabla^{-2}(\mathrm{div}(A+i\mathrm{grad}(\ln g)))\label{0}\\
&=&A+i\,\mathrm{grad}(i\nabla^{-2}(\mathrm{div}(A)))\\
&=&h(A)\label{1}
\end{eqnarray}
%where in passing from the first to the second  line, we used $\mathrm{div}(\mathrm{grad}(\ln g))=\nabla^2\ln g$.
 Moreover,  if $h(A)=h(A')$, then, putting all the gradients to one side of the equation, we obtain that  $A-A'=i\mathrm{grad} f$ ,\footnote{For $f=\nabla^{-2}(\mathrm{div}(A))-\nabla^{-2}(\mathrm{div}(A'))$.} and so $h(A)=h(A')$ iff $A\sim A'$. 
In this way, $h(A)$ captures the full gauge-invariant content of $A$. 

Lastly, notice something we glossed over when we found \eqref{eq:gf_bulk}:  $F$ doesn't determine $g(A)$ uniquely. However,   the underdetermination is solely due to stabilizers.  Here in the Abelian case,  $ g(A)$ and $ g'(A)$ are solutions to \eqref{eq:gf_bulk}, if and only if $ g'(A)=g(A)+c$, where $c$ is a constant.\footnote{For both to be solutions of \eqref{eq:gf_bulk0}, we must have $\nabla^2(g(A)-g'(A))=0$. But it is easy to show that $\nabla^2f=0$ iff $f$ is constant: 
$0=\int f\nabla^2f=\int |\mathrm{grad}f|^2\Leftrightarrow \mathrm{grad}f=0\Leftrightarrow f=$const (where we used integration by parts in the second equality). %The same proof would go through in the bounded case, with  Neumann conditions, since  $\pp_nf=0$
 \label{ftnt:19}}  Nonetheless, since here stabilizers have a trivial action  on the gauge potential (from \eqref{eq:g_trans}, since $\pp \ln c=0$),  the gauge-fixing still satisfies \textit{Uniqueness}; i.e. $A^{g(A)}=A^{g'(A)}$.

 It is important to note that fixing the representative requires finding something like $g(A)$, and this is always a non-local process. That is,  since $A$ is related to $g$ through a derivative, going the other way---tying $g$ to $A$---always requires integration.\footnote{A simple example: radial, or axial gauge, $A_r=0$. This is  not a complete gauge-fixing, but we still find $g(x,r)=\int^r_0 dr' A(r',x)$, where $r$ are radial coordinates, and $x$ are the remaining coordinates. \label{ftnt:nonlocal}}  This is manifested in  $g_\sigma$, and passed on to $h_\sigma$, by the presence of the Green's function.

\subsection{Coulomb gauge-fixing for the bounded case}\label{sec:gf_bounded}
Now we want to generalize \eqref{eq:gf_M} to the bounded case, i.e. for regions $R_\pm$ bounded by $S$, as in the introduction to Section  \ref{sec:gt}. 

%Although we want to define a gauge-fixing through the \textit{gauge potential}, $\sigma(A)$, we still aim to fix a particular value of the \textit{gauge transformations}: i.e. the transformations that take each representative to the gauge-fixing surface. 
Again, as in the discussion towards the end of Section  \ref{sec:gf}, we  want to fix the gauge using functions that exploit the `wrinkliness' of the states. %That is, we define gauge-fixings as $\sigma(A)=0$, with no extra parameters or explicit conditions on $g$ (which are, after all, entirely auxiliary quantities parametrizing gauge transformations). 

Since in the bulk of the manifold we have a scalar second-order differential equation for $g$, viz. equation \eqref{eq:gf_bulk0}, we want to impose scalar conditions for $g$ on  the boundary: either Dirichlet or Neumann boundary conditions, which are, respectively, of  zeroth and first order in derivatives. However, as per the definition of the gauge-fixing section $\sigma$, these conditions should descend to $g$ from $F(A)$; we do not  simply impose Neumann or Dirichlet boundary conditions on  $g$. Apart from being conceptually uncouth and falling outside our previous definition of gauge-fixing sections, such imposed conditions  on the boundary values of $g$ would not be gauge-covariant, a rather unwelcome side-effect. That is, the gauge-fixing projection would depend on the initial, entirely arbitrary choice of $A$; we would thus have $g_{\sigma, A_{|R}}$, where $S$ is the boundary.\footnote{ Unless there is a fixed choice for the boundary $A$, as in the externalist scenario. Then, of course, we may omit this dependence.}  In this case, if I were to choose one $A'$ to start with, and you chose another, $A''$, and we use the same boundary conditions \textit{on} $g$, we would find distinct gauge-fixed fields for the same $A$. 

% Appropriate boundary conditions for  second order elliptic differential equation with sources for $g$, as in equation \eqref{eq:gf_bulk0} must be a single scalar condition on $g$: a mixture of first-order normal derivatives of $g$ at the boundary and its (undifferentiated) value at the boundary. But $g$ is ``unphysical fluff'': as remarked above, to introduce covariant boundary conditions, we must pose them in terms of the gauge-potential, $A$, not $g$.

   Dirichlet or Neumann (scalar) boundary conditions imply that we  must fix one, and only one,  scalar degree of freedom of $A$ at the boundary ($A$ has dim$(M)$-many such degrees of freedom, of course). Since $A$ can only constrain the gradients of $g$ through their relation in equation  \eqref{eq:g_trans}, this rules out Dirichlet conditions. And since only the spacetime direction normal to the boundary is singled out by the introduction  of a boundary, we can only naturally introduce Neumman  boundary conditions  by fixing  the normal component $A_n$ of the gauge potential (see Section  \ref{sec:comparison} for more on this). There are mathematically and physically more upright ways to introduce these boundary conditions (cf. \cite{GomesButterfield_electro, GomesRiello_new}), but this simple argument here suffices.    

Lastly, if we do not want to introduce further arbitrary parameters in our gauge-fixing, the simplest choice for $F(A_\pm)$ as given by: 
 \be\label{eq:cov_gf}
 F(A_\pm)\equiv \begin{cases}
\mathrm{div}(A^\pm)&=0\\
  A^\pm_n &=0.
 \end{cases}
 \ee
 Given  arbitrary regional configurations $A_\pm$, we solve the following second-order set of differential equations with field-dependent, covariant boundary conditions:
 \begin{eqnarray}\nabla^2(\ln g_\pm)&=&i\mathrm{div}(A_\pm)\label{eq:bdary_lap}\\
 \pp_n (\ln g_\pm)&=&\pm i A_n\label{eq:Neu}
 \end{eqnarray}
where the $\pm$ signs on the right hand side of \eqref{eq:Neu} come from the opposite directions of the normal at $S$. The solution is as in \eqref{eq:gf_bulk}, namely, 
 \be g_\pm(A_\pm)=\exp(\pm i\nabla_{\text{Neu}(_\pm A_n)}^{-2}(\mathrm{div}(A_\pm)))\label{eq:gf_bulk_Neu}\ee
  with the difference that now the Green's functions (inverse Laplacian), $\nabla_{\text{Neu}(A_n)}^{-2}$ are defined for the field-dependent, Neumann boundary conditions \eqref{eq:Neu}; therefore $ \pp_n \ln g_\pm(A_\pm)=\pm i A_n$ holds automatically.%\footnote{I.e. a basis of harmonic functions $\nabla^2_{N_\pm(A_n)} f_\pm=0$ is such that $\pp_n f_\pm=\pm i A_n$. With this diagonal basis, the Laplacian is itself a diagonal matrix, and therefore one obtains the inverse $\nabla_{N_\pm(A_n)}^{-2}$ } 
  
  In precise analogy to \eqref{eq:h(A)}, we obtain 
  \be h_\pm(A_\pm):=A_\pm+i\,\mathrm{grad}(i\nabla_{\text{Neu}(_\pm A_n)}^{-2}(\mathrm{div}(A_\pm))),\ee  
  The projected field $h_\pm$  also satisfies \eqref{eq:cov_gf}. That is,  $h(A_\pm)_n=0$, even if $A_n^\pm\neq0$. In other words, even though we have not restricted the set of $A_\pm$'s, independently of its behavior at the boundary, any $A_\pm$ can be brought to satisfy equations \eqref{eq:cov_gf} through a gauge transformation---also generically non-trivial at the boundary. The reason is simple:  the system of equations \eqref{eq:bdary_lap} and \eqref{eq:Neu} always has solutions (existence).\footnote{There is also the added benefit, in the dynamical 3+1 setting of Yang-Mills gauge theories, that such gauge-fixings correspond to \textit{Helmholtz decompositions} separating the Coulombic from the radiative degrees of freedom of the region. Radiative degrees of freedom are those that are intrinsic to a region; they do not depend on further incoming  information at the boundary. See \cite[Sec. 3]{GomesRiello_new} for more on this point. }   This shows that we are respecting the `\textit{internalist}'s mantra for the internal boundary: no truncation of gauge transformations or of configuration variables is needed at this internal boundary. The projection works just fine without such truncations.

  Moreover,  as expected,: $h(A_\pm)=h(A'_\pm)$ iff $A'=A^{g_\pm}$ for some $g_\pm\in \G_\pm$ (the proof is a little more complicated than  the unbounded case, due to the field-dependent boundary conditions, but it proceeds in much the same way as \eqref{0}-\eqref{1}).

  As in the previous, unbounded case, the only ambiguity in solutions is due to stabilizers. Again, in the  Abelian case, this means $g_\pm(A_\pm)$ and $g'_\pm(A_\pm)$ are solutions to \eqref{eq:bdary_lap} and \eqref{eq:Neu} iff $g'_\pm(A_\pm)=g_\pm(A_\pm)+c$; as  is easy to check.\footnote{As in footnote \ref{ftnt:19}, to  be solutions to \eqref{eq:gf_bulk0}, we must have $\nabla^2(g-g')=0$ and $\pp_n( g-g')=0$. Again, calling $f=(g-g')$, we have
$0=\int f\nabla^2f=\int |\mathrm{grad}f|^2+2\oint f\pp_nf\Leftrightarrow \mathrm{grad}f=0$, and since $\pp_n f=0$,  $f=$const. In the non-Abelian case, we can have stabilizers of the boundary that are not shared by the bulk: each such stabilizer will likewise contribute to a degeneracy in the gluing.  \label{ftnt:20}}
  
  And again, for the same reasons, this ambiguity will have no effect on the representative. In other words, the associated projected potentials, $h(A_\pm):=A_\pm^{g(A_\pm)}$ and $h'(A_\pm):=A_\pm^{g'(A_\pm)}$, are identical.  Thus $\sigma$ satisfies both \textit{Universality} and \textit{Uniqueness} and  provides a bona-fide \textit{gauge-fixing. }

% In sum: we now have our set of physical representatives; we have eliminated redundancy. Connecting back to the notation of Section  \ref{sec:DES_gi}: $h(A)$ and $h(A_\pm)$ carry the same information, respectively, as $[\varphi]\equiv[A]$  and $[\varphi_\pm]\equiv[A_\pm]$. 
 So, not only is the configuration space $\mathcal{A}$ defined by the spaces $\mathcal{A}_\pm$, but each such space has its own principal fiber bundle structure.\footnote{What we have just shown for each space is essentially equivalent to  the existence of a local \textit{slice}, which is the mathematical jargon for a gauge-fixing section (local on field-space) on infinite-dimensional configuration spaces. The existence of a local slice is \textit{the} characterizing feature for (the closest analogues of a)  principal fiber bundle structure in this context. See, e.g. \citep{Mitter:1979un, kondracki1983, YangMillsSlice}.} %But this says nothing about how $\mathcal{A}/\mathcal{G}$ is related to $\mathcal{A}_\pm/\mathcal{G}_\pm$, the question surrounding DES.

\subsection{A sketch of the solution}\label{sec:sketch}
After this stage setting, we sketch the solution to our original problem: in the type of systems we have focussed on---vacuum, simply connected Universe---do regional physical states uniquely determine the entire physical state? %Gauge-fixing has disentangled the issue of redundant representation from empirical significance, and thus cleared the way to a full resolution. 

Essentially, to find $g_\pm$ as above, we obtain, from \eqref{eq:h_div}, i.e. from $\mathrm{div}(h)=0$ (and $\mathrm{div}(h_\pm)=0$) that $\nabla^2(g_\pm)=0$; and the action of the divergence operator on the Heaviside functions  on \eqref{eq:gluing} (and the Neumann conditions $h^\pm_n=0$) enforces a continuity equation for $g_\pm$ in terms of $h_\pm$ (the gluing condition, \eqref{eq:bdary_cont2}). This gives us enough information to fix the appropriate boundary conditions for the solutions $g_\pm$  (see \cite[Sec.4, p.30-33]{GomesRiello_new}).   

When all the chips have fallen, one can prove existence and almost uniqueness for the $g_\pm$ of \eqref{eq:gluing}. Unsurprisingly, the only degeneracy left  is again made up of regional stabilizers; they form the only well-defined rigid (or global) subgroup of the local gauge symmetries. 

To finish an assessement of DES, we now need to give more information about sectors of the theory.  In the case studied here---the Abelian case, in the absence of charged matter fields---any degeneracy in the stabilizers is idle: it is not felt by the gauge fields. This result is irrespective of the boundary conditions on the fields $E$ and $A$, as long as these conditions are posed gauge-invariantly (i.e. respect downward consistency, as described in Section  \ref{sec:two_bdaries}). Moreover, since in the present case $E$ is gauge-invariant, there are no further considerations that impinge on the gluing of subsystems ($E_\pm$ is just required to match at $S$). 

In other words, we  find \textit{unique} $g_\pm$  \textit{strictly as functionals of} the values of $h_\pm$ pulled-back to the boundary, $i^*h_\pm=:h^S_\pm$, where $i:S\rightarrow M$ (no derivatives of $h_\pm$ at the boundary are necessary, cf. footnote \ref{ftnt:C0}) and of regional stabilizers $c_\pm$:\footnote{Also note that we are using $S$ as a superscript to denote the intrinsic---pulled-back---quantity, that is different from the $S$ subscript that denotes mere restriction of the base point of vector quantities (cf. footnote \ref{ftnt:C0}).}  
\be\label{eq:rigid_var} g_\pm=g_\pm(h^S_\pm{}, c_\pm) \quad\text{with}\quad g_\pm(h^S_\pm{}, c_\pm)=g_\pm(h^S_\pm{}, 0)+c_\pm.\ee
Thus the difference between two solutions is entirely due to stabilizers.\footnote{ For illustration purposes, I display the solution here: 
$$\ln g_\pm = \zeta_{(\pm)}^{ \pm\Pi}\quad\text{with}\quad \Pi=\Big(\mathcal{R}^{-1}_+  + \mathcal{R}^{-1}_-\Big)^{-1}\left(  (\nabla^2_S)^{-1}\mathrm{div}_S( h_+-h_- )_{S}\right), $$
where the subscript $S$ denotes operators and quantities intrinsic (i.e. pulled-back)  to the interface surface $S$; $\zeta_{(\pm)}^u$ is a harmonic function on (respectively) $R_\pm$ with Neumann boundary condition $ \pp_n \zeta^u_{(\pm)}=u$, and $\mathcal{R}$ is the Dirichlet-to-Neumann operator. For the meaning of these operators, and also the analogous solution for the general non-Abelian Yang-Mills gauge theories, see \cite[Sec. 4]{GomesRiello_new}, and  \cite[Appendix D]{Gomes_new}.} 

As before,  since we are in vacuum, stabilizers---for electromagnetism, constant gauge transformations---do not affect the gauge potential. That is, some internal directions are not fixed by gluing, but they also do not change the vacuum states, as we saw in \eqref{eq:stab_diff}. Thus the underdetermination of  $g_\pm$ \textit{cannot} be converted  into a \textit{physical} variety \citep{Gomes_new}. Therefore, given $h_\pm$, there is a unique $h$ which can be obtained from their union. In this particular case, we are left \textit{without} DES for local gauge theory.

\subsection{Matter, non-Abelian, and  non simply-connected $M$:  the observability of symmetries in other theories and other sectors, glimpsed}\label{sec:DES_matter}
%In the non-Abelian case, the slack is not made up  of constant gauge transformations, but of  regional \textit{stabilizers}. Particular field configurations are isotropic: they admit internal directions that leave the configurations invariant, just like the Minkowski metric admits Poincar\'e transformations leaving it invariant. 
In contrast to the vacuum sector studied in the previous Section, in the presence of matter, both for Abelian and non-Abelian,   the stabilizer redundancy \textit{can} lead to real physical difference. It can do this because it may act non-trivially on the matter and electric fields. That is, as we saw in \eqref{eq:rigid_var}, our gluing procedure left a redundancy, corresponding to  certain rigid---more commonly known as global---symmetries acting on each region. 

Now we must more carefully consider what kind of boundary conditions defining our subsystems would allow DES. In the non-Abelian case,   gauge symmetries act on the electric field, and so \eqref{eq:g_trans} is no longer valid. The gluing condition \eqref{eq:bdary_cont}, acquires two more sets of equations beyond (the non-Abelian analogue of) \eqref{eq:bdary_cont2}.\footnote{ Reinstating $\sigma$ for the choice of convention---that is a functional of the gauge potential alone, we define $h_\pm^\psi:=g^\sigma_\pm(A)\psi_\pm$ and $h^E_\pm:=\mathrm{Ad}_{g^\sigma_\pm(A)}E_\pm$. We then have:  in the Abelian case, the states $\psi_\pm{}_{|S}$ the externally applied gauge transformations $g_\pm$ would have to satisfy $(g_+h_+^\psi-g_-h^\psi_-)_{|S}=0$ and, in the non-Abelian case,  $(\mathrm{Ad}_{g_+}h^E_+-\mathrm{Ad}_{g_-}h^E_-)_{|S}=0$. If $g_\pm$  are defined up to some degeneracy $\tilde g_\pm$ (e.g. the stabilizers of $A_\pm$). Composing all the group transformations (and since both the  $\G_\pm$ overlap on $S$) it is easy to see that this will only occur if both of the following conditions are satisfied: $(\mathrm{Ad}_{\tilde g_-^{-1}\tilde g_+}E_+-E_+)_{|S}=0$ and  $({\tilde g_-^{-1}\tilde g_+}\psi_+-\psi_+)_{|S}=0$. And so the combination of stabilizers must preserve the boundary value of the other fields. \label{ftnt:further_bdary}}

And as before, as required by downward consistency,  sectors should be defined so that they  cannot discern between gauge-related boundary values of these fields either, as discussed in Section  \ref{sec:int_bdary} (but I will not discuss the non-Abelian case at length).

% If there is no charged matter residing in the boundary between the regions, the remaining stabilizers of $A$  (if there are any)  can act separately on each region. Such  actions will not change the regional states, but, if there is matter in the bulk, they will change the global state, as long as the stabilizers do not also stabilize the matter states. 
 
 In the Abelian case, the representation of  a Klein-Gordon charged scalar  is nowhere stabilized by a non-trivial action of the gauge transformations,  as can be seen from \eqref{eq:g_trans}. %That is, the regional  rigid symmetries are fixed separately and independently, and, even after fixing the global state, we are still left with a degeneracy.
 So,  in the simple Abelian case of $U(1)$ symmetry, if the initial state has  matter fields on $S$, no mismatch of stabilizers can maintain the composition of the states in their representational convention (cf. footnote \ref{ftnt:further_bdary}). But if there are no matter fields on $S$, \emph{prima facie} we would have  an initial variety corresponding to the action of $\U(1)\times \U(1)$. This would have dynamical significance for as long as matter did not wander into the boundary $S$. And the  sector such that $S$ has no matter fields and which gives some spatial partition of the manifold still respects downward consistency, since it is a gauge invariant specification. In other words, were we to write down an action for the subsystem, the boundary contribution from $S$ would be gauge-invariant, and, for an interval $I$ for which matter does not cross $S$, we would have observable rigid symmetries corresponding to a physical variety of joint states.
   
To find out precisely what the physical variety here is, we also need to reinstate the action of the global stabilizer: the global representational convention was still left ambiguous up to a global stabilizer,  as we saw in Section \ref{sec:gf_unbounded}. And, as per the \emph{unobserability thesis} of Section \ref{sec:unobs},  a global symmetry is unobservable (i.e. not empirically significant). 

From \eqref{eq:rigid_var},  any  $c_+$ in $R_+$ has a unique--e.g. subsystem-global, cf. \S\ref{sec:subs_loc}---extension to $c_+$ acting on $R_-$. Thus applying a global $-c_+$ symmetry, we find that for any choice of $c_\pm$, the regional states can always be seen as transformed by:
\be\label{eq:reg_id} g_+(h^S_\pm{}, 0)\quad \text{and}\,\,g_-(h^S_\pm{}, c_--c_+),
\ee
for a given choice of $\bar c=c_+-c_-$.
Thus we obtain a remaining $\U(1)$ variety of observationally distinct global states.  This is precisely what is expected from e.g. `t Hooft's beam splitter thought-experiment (cf. \cite[p.110]{thooft} and and  \cite[p. 651]{BradingBrown}).

We can phrase this result in \citet[p. 13]{Wallace2019}'s notation (cf. Equation 11):  since the rigid  symmetry $\varphi_+\mapsto \varphi_+^{c_+}$ is subsystem-global, and thus has a unique extension to $\Phi_-$ which does not alter either the representational convention or the gluing, we can write this global action as: 
\be ([\varphi_+], g_+;  [\varphi_-], g_-)_\sigma\mapsto  ([\varphi_+], c^+g_+;  [\varphi_-], c^+g_-)_\sigma
\ee
where we have reinstated the subscript-$\sigma$ notation of Section \ref{sec:intro_gf} (used to designate the use of representational conventions to link the equivalence classes to the states; see \eqref{eq:doublet}). Then, as remarked by \cite[p. 13]{Wallace2019}: ``Importantly, since the symmetry acts simultaneously on the two systems, the
symmetry-invariant information about the combined system is not exhausted by
O and O' but also includes the relational quantity $g'^{-1}g$.''\footnote{ In our notation: $O\equiv [\varphi_+], O'\equiv [\varphi'], g'\equiv g_-, g\equiv g_+$.}  %and coincides with the existence and properties of conserved charges in each region. 
%There is one more important thing to notice here: \eqref{eq:reg_id} involves setting a symmetry to the identity in one of the subsystems, and so we may appear to be in tension with our warnings of  Section \ref{sec:rep_states}. But this is not so. As remarked there, and vindicated in  Section \ref{sec:setup_gluing}, a problem only arises if  the regional restriction of the global state (in its representational convention) does not fall into the  regional representational conventions. In that case, we may need to `adjust' the regional state in its representational convention and we cannot afford to limit these ajustments with pre-given boundary conditions, such as $g_\pm{}|_{S}=\mathrm{Id}$. In contrast, in this Section , we have kept all representational conventions intact, so that a straightforward comparison between cases is possible and physically significant, as advocated in \S\ref{sec:rep_conv} (see Wallace's quote in particular). 
\\

In the non-Abelian case,  we can only articulate the analogue of \eqref{eq:gluing} perturbatively,  for reasons mainly to do with the Gribov problem \citep{Gribov:1977wm} (the Gribov problem says there is no gauge-fixing section that covers the entire configuration space, cf footnote \ref{ftnt:Gribov}). But again we could find the same type of variety of global physical states, or same variety of observable symmetries if we have shared regional stabilizers, i.e such that the stabilizer of one region can be uniquely extended to act on the other region. Thus, for instance, for $G=SU(N)$, a configuration that   is in the orbit of $A=0$,  has  $SU(N)$ stabilizers of the gauge potential: infinitesimally, the constant generators of the Lie-algebra. The existence of observable symmetries would then depend on the sector of the theory we are in. According to  footnote \ref{ftnt:further_bdary}, we would only have DES for those boundary conditions that were also stabilized by the constant generators. For instance, if $\tau^I$ is an element of the Lie-algebra basis $\mathfrak{g}$, we would require, at the boundary: $[\tau^I, E_+]_{|S}=0$. Note that such conditions are gauge-invariant, since both the electric field and the stabilizers transform in the adjoint representation and thus they respect downward consistency. We would only get a full set of $SU(N)$ symmetries with DES if the sector was defined with vanishing electric field at the boundary.\footnote{ Moreover, in the non-Abelian case, it is possible to have a stabilizer of the boundary that is not shared by the bulk of the region. In that case, we will also have non-uniqueness of the composition (see also footnote \ref{ftnt:20}). We will come back to  this last point in Section  \ref{sec:GW_ext}. Of course, as mentioned in Section \ref{sec:premsYM}, below \eqref{eq:g_trans}, stabilizers are trivial for generic non-Abelian field configurations, in both bulk and boundary.} 

Thus the question of DES is not dependent on the detail of the boundary contributions to the dynamics: it depends only on the compatibility between the boundary values of the fields and the stabilizers of $A$.  Again, for the time interval in which these conditions hold---namely, such  that  $A$ maintains  the stabilizers in time, and the boundary values of $E$ and $\psi$ are also stabilized throughout evolution, in the sense above---we will have the corresponding observable symmetries. \\

    In case $M$ is not simply-connected,  there is more freedom in how one embeds, or puts together, the regions. This topological redundancy  produces physical variety even in the absence of matter. Such a variety will be equivalent to Aharonov-Bohm phases (cf. \citep{GomesRiello_new, Gomes_new}).\footnote{ Such topological variety is more akin to the standard Galileo ship case, as we will see in Section  \ref{sec:DES_particles}. \label{ftnt:DES_gt}} 

\section{Using gauge-fixings for the externalist's subsystem}\label{sec:GW_ext}\label{sec:ext_sub}

 Let us now see in more detail how our analysis through gauge-fixing, when applied to  the dynamical view of symmetries in  the externalist's notion of subsystem  recovers the results of \citep{GreavesWallace, Wallace2019b}. 
 
 First, it should be clear that there is, \emph{prima facie}, a tension between a fundamental approach to symmetries (as discussed in Section \ref{sec:fund_sym}) and  assigning a fixed boundary value to the states. It is in fact, not hard to show that only the dynamical approach works in these cases, and we will do so below.  At least, that is, if the externalist is saddled with providing a specification of the state at the boundary as in \eqref{eq:ext_gf}---an assumption that I am making.\footnote{Were we able to provide a gauge-invariant specification of $A$ at the boundary, it wouldn't  help fix the gauge: it would then be underdetermined.  }

Thus suppose that instead of the covariant boundary conditions used in the internalist boundary case, \eqref{eq:cov_gf}, we  implement $A{}_{|S}=\lambda$ for some fixed boundary 1-form, $\lambda$. That is (omitting the subscript $\sigma$ on $g_\sigma$):
  \be\label{eq:ext_gf}
 F(A^{g})\equiv \begin{cases}
\mathrm{div}(A^{g})&=0\\
  A^{g}{}_{|S}&=\lambda 
 \end{cases}
 \ee
%The conclusion is that the boundary condition $A{}_{|S}=\lambda$ is only compatible with the `\textit{dynamical}' view on symmetries, and with a boundary taken as external to the whole universe, as discussed in Section  \ref{sec:bdary_disc}.

 To require  $A^g_{|S}= \lambda$ as a boundary condition, we must appropriately pare down configuration space,   so that only $A{}_{|S}\equiv \lambda$ are allowed, i.e. $\mathcal{A}'=\{A_i\in \Lambda^1(M, \mathfrak{g})~,~A_i^I{}_{|S}\equiv \lambda^I_i\}$.\footnote{Also recall the notation $|_S$ denotes equality of all derivatives at the boundary: cf. \eqref{eq:bdary_cont} and footnote \ref{ftnt:C0}.} This is the space where the projection $h:\cal A'\rightarrow \cal A'$ will be taken to operate. Here $\lambda$ is functioning as the fixed environment state, and this boundary condition is analogous to fixing the representation of the environment  in equation \eqref{eq:states2}---one of the dubious suppositions at stake in Section  \ref{sec:comparison}.
 
 The reason we must pare down configuration space is the same reason that we cannot take a fundamental view of symmetry with the boundary conditions of \eqref{eq:ext_gf}. The obstruction is that the boundary-value problem \eqref{eq:ext_gf} is over-determined for $g_\sigma$ if $\cal A$ and $\G$ are not constrained at the boundary (where I reinsteated the subscript, for clarity). Namely, knowing the normal component of $A$ at the boundary suffices for a complete solution, since it determines a  boundary-value problem for $g_\sigma$ in terms of $\mathrm{div}(A)$ and $A_n$, but the boundary state also gives two more boundary conditions (given by the other components of $A$). In more detail, given \emph{any} $A$, the $g_\sigma$ must satisfy  \eqref{eq:bdary_lap_ext} with $\pp_n \ln (g_\sigma)=A_n-\lambda_n$. This fixes $g_\sigma$. But the remaining gradients of $\ln (g_\sigma)$ will not in general coincide with the remaining components of $A_n-\lambda_n$. Even if we pare down the space $\cal A$ where the gauge-fixing projection is operating to $\cal A'$, such that $A{}_{|S}\equiv\lambda$, we now have a Neumann boundary problem for $g_\sigma$, but the remaining gradients of $g_\sigma$ at the boundary are also constrained to vanish.
 
 Thus, instead of \eqref{eq:Neu}, solving these equations for $g(A)$ in $F(A^{g})=0$, for consistency we must simultaneously pare down the configuration space $\cal A$ and  require the boundary condition $\pp_i(\ln g)_{S}\equiv0$. More generally, the same argument would apply in the non-Abelian case, where preserving the boundary condition implies that the gauge transformations must be boundary-stabilizers, called $\mathcal{G}_S(A)$ in Section  \ref{sec:comparison}. %As presaged, this is precisely the perspective on subsystems dubbed `externalist',  that is only compatible with the dynamical view on symmetries. 

We can then choose one of these stabilizers, as a non-covariant---there is no need for covariance, since $A$ is fixed at the boundary---Dirichlet boundary condition for the gauge transformations. Namely,  $g_{|S}=\tilde g_{|S}=:\kappa$, for some arbitrary boundary-stabilizing $\tilde g\in \G_S$. So we have, in analogy to \eqref{eq:bdary_lap} and \eqref{eq:Neu}, the system:
 \begin{eqnarray}\nabla^2(\ln g)&=&i\mathrm{div}(A)\label{eq:bdary_lap_ext}\\
 g_{|S}&=&\kappa\label{eq:Neu_ext}
 \end{eqnarray}
Differente choices of $\kappa$ can be thought of as related by the action of the stabilizer group at the boundary (even if the action on $A$ there is trivial). Such changes correspond what Belot calls `generalized shifts' \citep{Belot50}: these are `transformations' that don't change the fixed state at the boundary. 
 
 We can only  claim this choice satisfies   \textit{`Uniqueness'}, thereby yielding a bona-fide gauge-fixing as seen in Section  \ref{sec:gf} if different choices of $\kappa$ produce the same $h(A)$. Otherswise, the surface in (the pared down)  $\mathcal{A}'$ defined by \eqref{eq:ext_gf} may depend on the choice of $\kappa$ (which does not appear in the defining equation, \eqref{eq:ext_gf}). 
 The above conditions demand $g$ stabilizes the boundary state, but that is it; each  choice $g_{|S}=\kappa$ can in principle yield a different gauge-fixed $A$. 
 
 That is, for $\kappa\neq \kappa'$, we may have  substantially different solutions. Augmenting the notation to include $\kappa$ as a subscript, and understanding $\sigma$ as implicit, we may have $g_{\kappa}(A)\neq g_{\kappa'}(A)$, and perhaps even such that their difference is not due to stabilizers, and therefore $h_{\kappa}(A)\neq h_{\kappa'}(A)$. 

There are three possibilities: (i)  $A$'s boundary state  has only the trivial stabilizer; or (ii) every $\kappa$ can be extended to a universal stabilizer; or (iii) some boundary stabilizers are not so extendible. Let us examine these in turn.

Suppose first (i), that  $A$ has only the trivial boundary stabilizer: then  there is no DES, for $\kappa=$Id (the same conclusion holds from \eqref{eq:GW_DES}). This matches \cite{Wallace2019}'s conclusion about what he defines as subsystem-local symmetries, since these obligatorily go to the identity at the boundary.  
 
Now suppose  that $A$ has some  stabilizers intrinsic to the boundary. 
 The system \eqref{eq:bdary_lap_ext} and \eqref{eq:Neu_ext} has a different unique solution $g_\kappa(A)$ for each $\kappa$. If we are in possibility (ii) and these solutions were related by a universal stabilizer, the difference between $g_\kappa(A)$ and $g_{\kappa'}(A)$ would not affect $h$. So in vacuum, the difference would be immaterial; and if there is matter in the bulk, the difference would again be physically relevant. This case describes electromagnetism, since there $g_\kappa(A)=g_{\kappa'}(A)+(\kappa-\kappa')$ as in \eqref{eq:rigid_var}. 
 
 But if we are in possibility (iii) and they \emph{are not} related by a universal stabilizer, that is, if the boundary stabilizer does not extend throughout the bulk, different $\kappa$ \textit{will} produce different physical states even in vacuum. Since in this case $A$ is assumed not to have a universal stabilizer,  the gauge-fixed, or projected states  $h_\kappa$ (cf. \eqref{eq:h(A)}), will differ, depending on the boundary value of the gauge-group: $h_\kappa(A)\neq h_{\kappa'}(A)$.
 In this case, , each $A$ corresponds to a collection of $h_\kappa(A)$'s, parametrized by a choice of stabilizer intrinsic to the boundary, $\kappa$.
  % Thus, to find a bona-fide gauge-fixing, $\sigma(A, \kappa)=0$, we must consider $\kappa$ to possess the same status as the gauge fields $A$. 
 In vacuum, this  can only occur in the   non-Abelian case.  And although the equations would no longer be \eqref{eq:bdary_lap_ext} and \eqref{eq:Neu_ext}, the general manipulations still apply.\footnote{We cannot proceed in precise analogy to footnotes \ref{ftnt:19} and \ref{ftnt:20} here. Writing $g$ as the (path)exponential of an infinitesimal $\xi$ for simplification, we have $\D^2(\xi_\kappa(A)-\xi'_\kappa(A))=0$, in the non-Abelian analogue. But now the integration by parts trick of footnote \ref{ftnt:20} no longer works, because we are using Dirichlet, not Neumann conditions. }

Since in possibility (iii) the group of boundary-intrinsic $\kappa$ that are not extendible is isomorphic to the quotient \eqref{eq:GW_DES}, it is then true that we have leftover physically inequivalent configurations in that same amount, even in vacuum. They can be taken to possess `non-relational DES' if you will,  because these inequivalent possibilities are related solely by `gauge transformations' of the boundary conditions: $\kappa$ and $\kappa'$ \textit{would be} symmetry-related under a fundamental view after all, and these transformations do not change the state of $A$ at the boundary.\footnote{In the treatment of `non-relational' DES of \citep{Teh_abandon}, $\kappa$'s are treated in a  \textit{dynamical} fashion. See also \citep{Teh_stack} for the categorical geometrical treatment. See also \cite{DonnellyFreidel}.}  More importantly, such degeneracy has no representation as the action of a rigid group on the bulk of the region, as it does when the stabilizer of the boundary is shared by a bulk infused with charges. The only plausible view on $\kappa$ is that it represents degrees of freedom intrinsc to the boundary.\\

%Finally, what happens when all of $A$'s  stabilizers intrinsic to the boundary come from universal stabilizers?  This was the situation we posed as a counter-example to the standard derivation of DES in Section  \ref{sec:standard_GW}. In this case, we \textit{can} relate the gauge-fixings for different choices of $\kappa$ by \textit{universal} stabilizers. E.g. for electromagnetism: $g_\kappa[A]=g_{\kappa'}[A]+(\kappa-\kappa')$. Here, the different constants $\kappa$ and $\kappa'$ at the boundary are immaterial: we still  have a gauge-fixing, as defined at the end of Section  \ref{sec:gf}; each $A$ corresponds to a unique $h(A)$.

We can now summarize our findings: in either the externalist or the internalist scenario, in vaccuum and in the simply-connected case,   we find that stabilizers intrinsic to the boundary that do not correspond to either regional or universal stabilizers give observable boundary-intrinsic symmetries; and the physical difference between these has no immediate realization through the action of a symmetry group in the bulk of the region; but this scenario can only occur in the non-Abelian theory. If both kinds of stabilizers---bulk and boundary---match-up (trivially or not), neither the internalist nor the externalist obtains DES in vacuum.  Moreover, in this case, the internalist and the externalist also agree about DES  in the presence of charged matter within the region(s) (cf. footnote \ref{ftnt:DES_gt} and Section  \ref{sec:hol_charge}): they exist only when bulk charges are present and the stabilizer is non-trivial.

\section{Comparison with the holonomy formalism}\label{sec:holonomy}

The holonomy interpretation of electromagnetism takes as its basic elements assignments of unit complex numbers to loops in spacetime. A loop is the image of a smooth  embedding of the oriented circle,  $\gamma:S^1\rightarrow \Sigma$; the image is therefore a closed, oriented, non-intersecting curve. One can form a basis of gauge-invariant quantities for the holonomies (cf. \citep{Barrett_hol} and \cite[Ch.4.4]{Healey_book} and references therein),\footnote{Of course, any discussion of matter charges and normalization of action functionals would require $e$ and $\hbar$ to appear. However, I am not treating matter, so these questions of  choice of unit do not become paramount. As before, if needed, I set my units  to $e=\hbar=1$; as is the standard choice in quantum chromodynamics (or as in the so-called Hartree convention for atomic units).} 
\be\label{eq:hol} hol(\gamma):=\exp{(i\int_\gamma A)}.\ee

\subsection{The basic formalism}
Let us look at this in more detail.  By exponentiation (path-ordered in the non-Abelian case), we can assign a complex number  (matrix element in the non-Abelian case) $hol(C)$ to the oriented embedding of the unit interval: $C:[0,1]\mapsto M$. This makes it easier to see how composition works: if the endpoint of $C_1$ coincides with the starting point of $C_2$, we define the composition $C_1\circ C_2$ as, again, a map from $[0,1]$ into $M$, which takes $[0,1/2]$ to traverse $C_1$ and $[1/2, 1]$ to traverse $C_2$.  The inverse $C^{-1}$ traces out the same curve with the opposite orientation, and therefore $C\circ C^{-1}=C(0)$.\footnote{It is rather intuitive that we don't want to consider curves that trace the same path back and forth, i.e.  \textit{thin} curves. Therefore we  define a closed curve as \textit{thin} if it is possible to shrink it down to a point while remaining within its image. Quotienting the space of curves by those that are thin, we obtain the space of \textit{hoops}, and this is the actual space considered in the treatment of holonomies.   I will not call attention to this finer point, since it follows from a rather intuitive understanding of the composition of curves.} 
Following this composition law, it is easy to see from \eqref{eq:hol} that 
\be\label{eq:loop_com} hol(C_1\circ C_2)=hol(C_1)hol(C_2),\ee with the right hand side understood as complex multiplication in the Abelian case, and as composition of linear transformations, or  multiplication of matrices, in the non-Abelian case.

For both Abelian and non-Abelian groups, given the above notion of composition, holonomies are conceived of as smooth homomorphisms from the space of loops into a suitable Lie group. One obtains a representation of these abstractly defined holonomies as
holonomies of a connection on a principal fiber bundle with that Lie group as structure group; the collection of such holonomies carries the same amount of information as the gauge-field $A$. However, only for an Abelian theory can we cash this relation out in terms of gauge-invariant functionals. That is, while \eqref{eq:hol} is gauge-invariant, the non-Abelian counterpart (with a path-ordered exponential), is not.\footnote{For non-Abelian theories the gauge-invariant counterparts of \eqref{eq:hol} are Wilson loops, see e.g. \citep{Barrett_hol}, 
$ W(\gamma):=\text{Tr}\, \mathcal{P}\exp{(i\int_\gamma A)}
$,
where one must take the trace of the (path-ordered) exponential of the gauge-potential. It is true that all the gauge-invariant content of the theory can be reconstructed from Wilson loops. But,  importantly for our purposes, it is no longer true that there is a homomorphism from the composition of loops to the composition of Wilson loops. That is, it is no longer true that the counterpart \eqref{eq:loop_com} holds,  $W(\gamma_1\circ\gamma_2)\neq W(\gamma_1)W(\gamma_2)$. This is  due solely to the presence of the trace. The general composition constraints---named after Mandelstam---come from generalizations of the Jacobi identity for Lie algebras, and depend on $N$ for SU($N$)-theories; e.g. for $N=2$, they apply to three paths and are: $W(\gamma_1)W(\gamma_2)W(\gamma_3)-\frac12(W(\gamma_1\gamma_2)W(\gamma_3)+W(\gamma_2\gamma_3)W(\gamma_1)+W(\gamma_1\gamma_3)W(\gamma_2))+\frac14(W(\gamma_1\gamma_2\gamma_3) + W(\gamma_1\gamma_3\gamma_2) = 0$.  \label{ftnt:tr_hol}}

\subsection{DES and separability}\label{sec:hol_charge}
As both Healey \cite[Ch. 4.4]{Healey_book} and Belot (\cite[Sec.12]{Belot2003} and \cite[Sec.3]{Belot1998}) have pointed out, even classical electromagnetism, in the holonomy interpretation, evinces a form of non-locality, which one might otherwise have thought was a hallmark of non-classical physics. But is it still the case that  the state of a region supervenes on assignments of intrinsic properties to patches of the region (where the patches may be
taken to be arbitrarily small)? This is essentially the question of \textit{separability} of the theory (see \cite[Ch.2.4]{Healey_book}, \cite[Sec.12]{Belot2003}, \cite[Sec.3]{Belot1998}, and \citep{Myrvold2010}). 
\begin{figure}[t]
		\begin{center}
			\includegraphics[width=8cm]{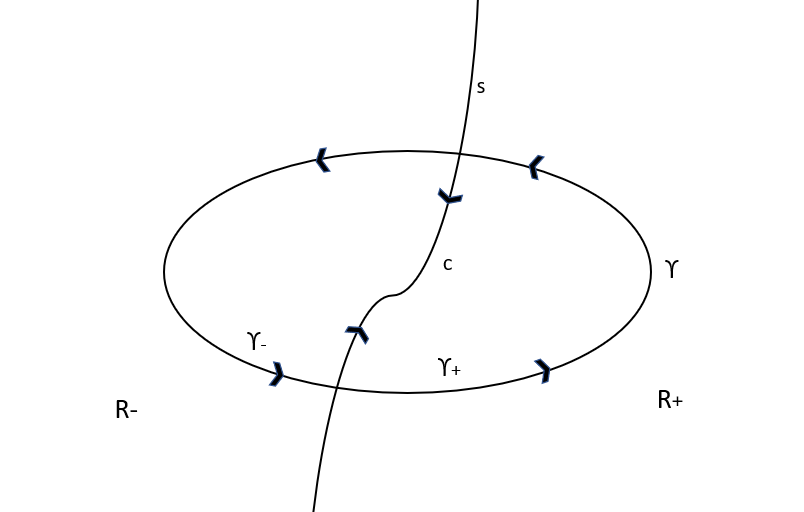}	
			\caption{ Two subregions, i.e. $R_\pm$,  with the separating surface $S$. A larger loop $\gamma$ crosses both regions. But, since $\gamma_1$ and $\gamma_2$ traverse $S$ along $C$ in opposite directions, $\gamma=\gamma_1\circ\gamma_2$.
			}
			\label{fig4}
		\end{center}
\end{figure}

 Clearly, the question of DES asked in the present paper is intimately related to the one of separability. For DES, in many of its incarnations, e.g. \citep{BradingBrown, GreavesWallace, Teh_emp, Friederich2014}, is conditional on the existence of universal gauge-invariant quantities that are \textit{not} specified by the regional gauge-invariant content. But we are not  interested here in cases of ``topological holism'', as related to the Aharonov-Bohm effect. We are asking whether a vacuum, simply-connected universe still displays non-separability. For this topic, we can directly follow Myrvold's definition \cite[p.427]{Myrvold2010} (which builds on Healey's notion of Weak Separability \cite[p. 46]{Healey_book} and on Belot's notion of Synchronic
Locality \cite[p. 540]{Belot1998}):\\ \smallskip

\noindent$\bullet$ \textit{Patchy Separability for Simply-Connected Regions}. For any simply-connected spacetime region $R$, there are arbitrarily fine open coverings $\mathcal{N}=\{R_i\}$ of $R$ such that the state of $R$ supervenes on an assignment of qualitative intrinsic physical properties to elements of $\mathcal{N}$.\\

If \textit{Patchy Separability for Simply-Connected Regions} holds, there will be no room for DES. And indeed, in vacuum,  it is easy to show that it \textit{does} hold. In Figure 2, we see a loop $\gamma$ not contained in either $R_+$ or $R_-$. However, we can decompose it as $\gamma=\gamma_+\circ\gamma_-$, where each regional loop $\gamma_\pm$ does not enter the complementary region ($R_\mp$, respectively). Following \eqref{eq:loop_com}, it is then true that, since holonomies form a basis of gauge-invariant quantities,  the universal gauge-invariant content of the theory supervenes on the regional gauge-invariant content of the theory.\smallskip

It is also easy to see how \textit{Patchy Separability for Simply-Connected Regions} fails when charges are present within the regions but absent from the boundary $S$  (see in particular  \cite[Sec. 4.3.2]{GomesRiello_new}, and footnote 70 in \citep{Gomes_new}). For, in the presence of charges, we can form \textit{gauge-invariant} functions from a non-closed curve $C'$  that crosses $S$ and has one positive and one negative charge, $\psi_\pm(x_\pm)$, at each end of $C'$, at $x_\pm\in R_\pm$. That is, the following quantity is a gauge-invariant function:
$$Q(C', \psi_\pm)= \psi_-(x_-)hol(C')\psi_+(x_+)
$$
for $C'(0)=x_-, C'(1)=x_+$. It is easy to check from the transformation property $\psi\mapsto g\psi$, that $Q$ is gauge-invariant. Moreover, we cannot break this invariant up into the two regions, since we have assumed no charges lie at the boundary. This is just a  translation of the results mentioned at the end of section \ref{sec:sketch} into the holonomy formalism. %See  \cite[Sec. 4.3.2]{GomesRiello_new}, and  \citep{Gomes_new} for a complete analysis for how DES \textit{does} emerge in these cases.

Unfortunately,  this holonomy-based analysis cannot be reproduced for non-Abelian theories (see footnote \ref{ftnt:tr_hol}); and it does not apply to an externalist's notion  of boundaries; and it cannot be translated to the point particle language. Since we will have to analyse point-particles and the externalist's notion of boundaries,  and since we want our formalism to apply also to the non-Abelian case, a treatment with holonomies---even if good for illustration---will not do.

\section{Point-particle systems}\label{sec:particle}
To compare the local gauge theory discussed above to the case that originally motivated the notion of DES---Galileo's ship---we introduce representational conventions to the study of point particles in Euclidean space.\footnote{This discussion echoes \citep{RovelliGauge2013}, which considers precisely the question of matching physical information about point-particle subsystems. The thought-experiment is made more explicit in the context we are exploring here in \cite[Sec 2]{GomesStudies}. For an enlightening discussion of the topic, see also \citep{Teh_emp}. }

Adopting subsystem recursivity, as discussed in Section  \ref{sec:sub_rec}, and in particular, downward consistency, we assume the sectors of the theory are defined in symmetry-invariant way (with respect to the global symmetries). In particular, this implies that the subsystem inherits the full group of symmetries of the universe. Of course, these limited assumptions will only allow us to discuss observable symmetries for the initial state. How these symmetries extend in time will depend on the details of how we embed the subsystem in the rest of the universe, and if they are to extend at all one requires this embedding to respect some condition of dynamical isolation.

In sum, we would like here to gauge-fix the Galileian symmetries, for two subsystems, replacing ship and shore respectively. After some prescription for composing the system, we would still like to evaluate whether different compositions are physically distinguishable or not, and therefore we must again choose a representational convention for the global state. 

In Section \ref{sec:gf_particles} I introduce natural representational conventions in the particle case, and in Section \ref{sec:DES_particles} I use these conventions to find the standard notion of DES for Galilean symmetries. 

\subsection{Gauge-fixing}\label{sec:gf_particles}
For particle systems, it is straightforward to fix translations by the center of mass and rotations by diagonalizing the moment of inertia tensor around the center of mass. It is again true that these choices of gauge-fixing/representational conventions  may not satisfy `\textit{uniqueness}'. In the case of translations, this can happen for infinite, homogeneous mass distributions; there just is no unique center of mass to speak about.  For rotations, the lack of uniqueness will obtain when the configuration has some rotational symmetry along an axis.  We will only consider a finite number of pointlike mass particles, leaving only the degeneracy of rotations as relevant. 

To be more explicit, the total system is given by $N$ particles of mass $m_\alpha$, $\alpha\in I=\{1,\cdots N\}$, with position vectors $\mathbf{q}_\alpha$ in some arbitrary inertial frame of $\bb{R}^3$, constituting the configuration space $\mathcal{Q}=\RR^{3N}$; with conjugate momentum variables $\mathbf{p}^\alpha$. The subsystems here are defined by selecting two subsets of these particles, $I_\pm\subset I$, so that $I_+\cap I_-=\emptyset$ and $I_+\cup I_-=I$; that is, they are mutually exclusive and jointly exhaustive. The subsets define sectors that satisfy downward consistency, i.e. are symmetry-invariant, and are the analogues of $R_\pm$, whereas the relevant configuration space, $\mathcal{Q}$, is analogous to $\mathcal{A}$, and $\mathcal{Q}_\pm$ to $\mathcal{A}_\pm$. Thus we assume that the same global symmetries that act on the entire universe act on these subsystem configuration spaces. 

 The translations act as $T:\mathbf{q}_\alpha\mapsto \mathbf{q}_\alpha +\mathbf{t}$, for a given vector $\mathbf{t}$. The rotations act as $R:\mathbf{q}_\alpha\mapsto \mathbf{R}\mathbf{q}_\alpha $, where $\mathbf{R}\in SO(3)$, acting in coordinates as $R:{q}^i_\alpha\mapsto {R}^i_j{q}^j_\alpha$. A Galilei boost is just the  translation in momentum space (i.e. an infinitesimal translation of the type $\mathbf{t}(t):=\mathbf{v}t$, where $t$ is time), acting as $B:\mathbf{q}_\alpha\mapsto \mathbf{q}_\alpha$ and $B:\mathbf{p}^\alpha\mapsto \mathbf{p}^\alpha+\mathbf{v}$. %\footnote{ There are no boundaries in systems of point particles, and therefore we should allow no distinction between `fundamental' and `dynamical' views of symmetries (Section  \ref{sec:fund_sym}). Given the usual Newtonian dynamics, we should therefore consider the full dynamical  group of Galilean translations or boosts. This extension simply amounts to the usual one, whereby a symmetry group acting on configuration space $\mathcal{Q}$ is naturally extended  to act on phase space, $T\mathcal{Q}$, as long as the Hamiltonian, or the Lagrangian, are invariant under the time-dependent transformations.\label{ftnt:Gal} } 
The action of the group on configuration space is a semi-direct product of the two groups  $\mathcal{G}=SO(3)\ltimes \RR^3$, with group action $\G\times\G\rightarrow \G$ denoted by `$\cdot$', i.e. $(g,g')\mapsto g\cdot g'$.  We will focus on this action, and denote  $g_\pm=(R_\pm, t_\pm)\in \mathcal{G}_\pm$, with $\circ$  the action of rotations and translations on the configurations, e.g. $\circ: (g, \mathbf{q})\mapsto g\circ \mathbf{q}$.

 For each (sub)system, $J=I, I_+$ or $I_-$, we first fix center of mass coordinates through the gauge-fixing $F_t(q)=0$, as:
\begin{eqnarray}
F_t(q)=\sum_{\alpha\in J} m_\alpha \mathbf{q}_\alpha+\mathbf{t}=0
\end{eqnarray}
and so define $\mathbf{t}_\sigma(q)=\sum_{\alpha\in J} m_\alpha \mathbf{q}_\alpha$. 
Fixing the rotations is slightly more complicated. We first define the translationally fixed positions through the translationally fixed coordinates, as $\bar{\mathbf{q}}_\alpha:=\mathbf{q}_\alpha+\mathbf{t}_\sigma(q)$. Now we can define the moment of inertia tensor as $\mathbf{L}$ with components:
$$L^{ij}:=\sum _{\alpha\in J}m_{\alpha}(\|\bar{\mathbf {q}}_\alpha \|^{2}\delta ^{ij}-\bar q_\alpha^{i}\bar q_\alpha^{j}\,)$$
$L^{ij}$ is a real symmetric matrix. A real symmetric matrix has an almost unique eigendecomposition into the product of a rotation matrix and a diagonal matrix. We therefore fix rotations through $F_R(q)=0$, as:
\be F_R(q)=\mathbf{R}^T\mathbf{L}\mathbf{R}-\mathbf{\Lambda}=0,
\ee
where $\Lambda=\mathrm{diag}(\Lambda_1, \Lambda_2, \Lambda_3)$ is a diagonal matrix whose non-zero elements are called the principal moments of inertia. When all principal moments of inertia are distinct, the principal axes through the center of mass are uniquely specified. If two principal moments are the same, there is no unique choice for the two corresponding principal axes. If all three principal moments are the same, the moment of inertia is the same about any axis. These constitute the possible degeneracies in the determination of $\mathbf{\Lambda}$. And so we find the configuration-dependent rotation matrix $\mathbf{R}_\sigma(q)$. As with the translation element $t_\sigma(q)$, this matrix depends on the positions of all the particles, $\{\mathbf{q}_\alpha\}$, a dependence we denote simply by $(q)$.

We have thus completely fixed the coordinate system for the particles, and therefore a complete representational convention of the configurations is given by the n-tuples:
\be \mathbf{h}(q)_\alpha=\mathbf{R}_\sigma(q)(\mathbf{q}_\alpha+\mathbf{t}_\sigma(q))=g_\sigma(q)\circ \mathbf{q}_\alpha\ee
in perfect analogy with our definition of $h(A)$ in \eqref{eq:h(A)}; e.g. 
\be\label{eq:g_inv}
\mathbf{h}(g\circ q)_\alpha=\mathbf{h}(q)_\alpha\ee 
where $g_\sigma(q)=(\mathbf{R}_\sigma(q),\mathbf{t}_\sigma(q))$ is the necessary translation and rotation to bring the configurations to the  frame chosen by $\sigma$, i.e. so that $F_t(h)=F_R(h)=0$. This configuration-dependent group element obeys:
 \be\label{eq:cov_g}g_\sigma(g'\circ{q})=g_\sigma({q})\cdot g'{}^{-1} \ee which is what guarantees \eqref{eq:g_inv}. Again we can see $\mathbf{h}:\mathcal{Q}\rightarrow \mathcal{Q}$ as a projection from configuration space to configuration space, and such that the image of $\mathbf{h}$ is the gauge-fixing surface and this image is invariant under gauge transformations on the domain.\footnote{ But, as before, in equation \eqref{eq:external_gt}, we can also apply subsystem-extrinsic gauge transformations to the range of $\mathbf{h}$; cf. footnote \ref{ftnt:subs_int_ext}.}

\subsection{Finding DES}\label{sec:DES_particles}

Again, the idea  is:---assume each subsystem employs these representational conventions. Then we ask: in how many physically distinct ways can we compose given physical states of the subsystems? 

In the beginning of Section  \ref{sec:DES} we saw that a universal gauge-fixed field, $h$, did not necessarily restrict to the corresponding regional gauge-fixed fields, $h_\pm$, because of non-locality---this is why subsystem-extrinsic gauge transformations were required (cf. footnote \ref{ftnt:gluing}). Again it is true that, given a universal configuration in the preferred coordinates, $\mathbf{h}_\alpha(q)$, restriction to subsystems---to the analogue of $R_\pm$---will not be in their center of mass and diagonalized moment of inertia. Therefore, again: in order to relate an $h$ to the $h_\pm$, we must allow some adjustments (or subsystem-extrinsic transformations) so that  we find an expresion of the glued states in the global representational convention (omitting the particle indices):
\be\label{eq:DES_part}\mathbf{h}=(g_+\circ \mathbf{h}_+)\oplus (g_-\circ \mathbf{h}_-).\ee
And in particular, we cannot assume one of the subsystems will remain in the center of mass coordinates (so that $g_-=$Id).  

But the main difference to the previous, field theoretic case is the following: there  we nailed down the composition $\oplus$ in terms of the embeddings of the manifolds: it amounted to smooth composition along a shared boundary. 
 For fields, the splitting of the universe into adjacent regions nails down the embedding of the regions supporting the subsystems into the larger spacetime manifold. Consider: were the two regions $R_\pm$ not adjacent, and had their placement been left free,   we would have had a further freedom of composition given by the possible embeddings of one submanifold with respect to the other.\footnote{And indeed the topological ambiguity related to the Aharonov-Bohm effect is the effect of such an added  freedom, since then adjancy still leaves some features of the embedding undetermined.} Of course, the two regional states  then would not have determined the global state, and so adjency is implied by completeness. In the field theory example, by stipulating that the two regional subsystems descended from a splitting of the universe and were to jointly determine the global state, we topologically fixed the embeddings of the regions.\footnote{Indeed, for non-simply connected manifolds, adjacency does \textit{not} fix the topological embedding uniquely, giving rise to a DES for gauge systems associated to the Aharonov-Bohm effect; see \cite[Sec.4.5]{GomesRiello_new}.}

 In contrast, here in the particle case   an analogue of the gluing condition, \eqref{eq:bdary_cont2}, is missing. So even if we hold that the two subsystems should jointly describe the state of the universe,  we have the extra step of stipulating how to embed the subsystems. It is \textit{this} freedom that gives rise to Galileo's ship and Einstein elevator realizations of DES. For it is still possible to respect downward consistency, and have the subsystem symmetries be Gallilean, by embedding that system into the universe with a force that acts equally on all its components  (see e.g. \citep{saunders2013} for a thorough analysis of the constant but non-zero force, and its relation to  Newton's Corollary VI). Of course, this will only be possible for certain embeddings: those that satisfy downward consistency and, for an arbitrary time-dependent acceleration, it may well be near impossible to find an environment for which the embedding satisfies the equations of motion of the universe. 
 
Thus,  instead of finding  explicit $g_\pm$ in \eqref{eq:DES_part}, we divide the process into two parts: we first arbitrarily embed the subsystems into the same Euclidean space, and then we find a transformation that brings the newly defined composite system to its gauge-fixed frame.  At the end, we want to find out what information is  required to determine $\mathbf{h}$ beyond that provided by the subsystem-intrinsic physical information, given by  $\mathbf{h}_\pm$.

Here $\oplus$ firstly designates an embedding of the two frames into the larger universe. We embed them by defining a new frame, which is related to the ones used in $\mathcal{Q}_\pm$ by  arbitrary transformations $ g^{\text{emb}}_\pm\in SO(3)\ltimes \RR^3(t)$, where $\RR^3(t)$ denotes translations that have an arbitrary time-dependence. We thus obtain a universal configuration, 
\be\mathbf{q}_{\alpha}
=\begin{cases}
g^{\text{emb}}_+\circ \mathbf{h}^+_\alpha\qquad \text{for}\qquad \alpha\in I_+\\
g^{\text{emb}}_-\circ\mathbf{h}^-_\alpha\qquad \text{for}\qquad \alpha\in I_-\\
\end{cases};\label{a}
\ee
with the understanding that $\alpha$ runs through the appropriate domains for $I_\pm$, we can replace those indices by $\pm$. The positions of the particles are now all seen to inhabit the same Euclidean 3-space, and $\oplus$ becomes simple vector addition. 

Of course, this $\mathbf{q}_{\alpha}$ is not yet in the form of $\mathbf{h}_{\alpha}$; that is, it is not in a universal center of mass and eigenframe of the moment of inertia coordinate system. As above, a gauge-fixing yields $g({q}_{\alpha})$, and therefore, by linearity (omitting particle indices):
\be\mathbf{h}:=g(q)\circ\mathbf{q}=(g(q)\circ( g^{\text{emb}}_+\circ \mathbf{h}_+))+ (g(q)\circ( g^{\text{emb}}_-\circ \mathbf{h}_-)).\label{37}\ee
But we can put \eqref{37} in a slightly more concise form. Since here the symmetries act universally (i.e. they are subsystem-global, cf. Section \ref{sec:subs_loc}) and  we know the  covariance property \eqref{eq:cov_g} holds (this is what guarantees  \eqref{eq:g_inv}),   there is no loss of generality if we replace \eqref{a} by:
\be\mathbf{q}'_{\alpha}
=\begin{cases}
g_+^{\text{emb}'}\circ \mathbf{h}^+_\alpha\qquad &\text{for}\qquad \alpha\in I_+\\
\mathbf{h}^-_\alpha\qquad &\text{for}\qquad \alpha\in I_-\\
\end{cases}\label{b}
\ee
where $g_+^{\text{emb}'}:=(g^{\text{emb}}_-)^{-1}\cdot g^{\text{emb}}_+$ (we can compose them since they all act on the same Euclidean space). %Equation \eqref{b} is merely a justification for fixing the frame of one of the subsystems of the original embeddings as coincident with the frame of the universe, before universal gauge-fixing.
 Thus, finally, we can write our solution (again omitting the index $\alpha$) as:
\be \mathbf{h}= (g(q')\circ( g_+^{\text{emb}'}\circ \mathbf{h}_+))+ (g(q')\circ \mathbf{h}_-),
\ee
where `+' is now simply vector addition in the center of mass frame. 

We can write  $g(q')=g(\mathbf{h}_\pm, g_+^{\text{emb}'})$. Therefore the solution is uniquely defined, in terms of $g_+^{\text{emb}'}$ and $\mathbf{h}_\pm$. Now $g_+^{\text{emb}'}$ is universally gauge invariant: it is a quotient of two rigid symmetries, as we obtained in Section \ref{sec:DES_matter}; we can no longer get rid of it by a universal change of coordinates. But $g_+^{\text{emb}'}$ is \textit{not} solely determined by $\mathbf{h}_\pm$. This is in contrast to what we found for the field theory, in equation \eqref{eq:rigid_var}, where, for a simply-connected, vacuum universe, up to stabilizers, the transformations were uniquely determined by $h_\pm$. Here in the particle case, there is no way to associate $ g_+^{\text{emb}'}$---the information required beyond $\mathbf{h}_\pm$---with stabilizers of the configurations. 

 The physical variety, i.e. the variety of ways to compose physical states of subsystems, is therefore given by $g_+^{\text{emb}'}$: namely,  by \textit{how we embed one of the subsytems with respect to the other}. Everything else is uniquely determined by $\mathbf{h}_\pm$.\footnote{Namely, for two ships, these $\mathbf{h}_\pm$ would be the description of all the particles of each ship with respect to its own gauge-fixed coordinates (center of mass and diagonal moment of inertia).}%\footnote{ I should  mention once more that dynamical considerations are secondary: we can apply this scheme to whatever symmetries of the N-particle system emerge from the dynamics, such as Gallilean symmetries (see footnote \ref{ftnt:Gal}).}
 
Again, dynamical considerations would come into play once we take into account the time interval in which the subsystems remain (approximately) isolated, and, more importantly, in determining the type of environment for which $g_+^{\text{emb}'}$ is a dynamically allowed embedding. This needs to analysed in a case by case basis.  % But here there is an added complication in defining the sectors of the theory: we may take a sector to be defined by some condition on the embedding, e.g. that the subsystem is `far enough away' from other masses or charges (see \citep[Section 6]{Wallace2019a}). But such a definition of a sector is not fully invariant under the symmetries of the universe, and so does not satisfy downward consistency. Such a sector will only be approximately invariant, or invariant only under idealizations about asymptotic infinity. And there is a physical reason for why this should be so: given a sector, for any $I>0$, there are large enough boosts and/or translations of the subsystems for which the isolation condition is not maintained. And so, we  define a sector relative to the separation of the embedded subsystems, and consistently, given a sector, the symmetries that do satisfy downward consistency will depend on that relative separation. 

\end{appendix}

%\bibliographystyle{apacite} 
%\bibliography{references2}

\begin{thebibliography}{}

\bibitem [\protect \citeauthoryear {%
Ashtekar%
}{%
Ashtekar%
}{%
{\protect \APACyear {1987}}%
}]{%
AshtekarNapoli}
\APACinsertmetastar {%
AshtekarNapoli}%
\begin{APACrefauthors}%
Ashtekar, A.%
\end{APACrefauthors}%
\unskip\
\newblock
\APACrefYear{1987}.
\newblock
\APACrefbtitle {{Asymptotic Quantization: Based on 1984 Naples Lectures}}
  {{Asymptotic Quantization: Based on 1984 Naples Lectures}}.
\newblock
\APACaddressPublisher{}{(Monographs and Textbooks in Physical Science Lecture
  Notes, Vol 2)}.
\PrintBackRefs{\CurrentBib}

\bibitem [\protect \citeauthoryear {%
Ashtekar~A.%
}{%
Ashtekar~A.%
}{%
{\protect \APACyear {1981}}%
}]{%
AshtekarStreubel}
\APACinsertmetastar {%
AshtekarStreubel}%
\begin{APACrefauthors}%
Ashtekar~A., S\BPBI M.%
\end{APACrefauthors}%
\unskip\
\newblock
\APACrefYearMonthDay{1981}{}{}.
\newblock
{\BBOQ}\APACrefatitle {Symplectic geometry of radiative modes and conserved
  quantities at null infinity} {Symplectic geometry of radiative modes and
  conserved quantities at null infinity}.{\BBCQ}
\newblock
\APACjournalVolNumPages{Proc. R. Soc. Lond. A, 376}{}{}{}.
\newblock
\begin{APACrefDOI} \doi{10.1098/rspa.1981.0109} \end{APACrefDOI}
\PrintBackRefs{\CurrentBib}

\bibitem [\protect \citeauthoryear {%
Barrett%
}{%
Barrett%
}{%
{\protect \APACyear {1991}}%
}]{%
Barrett_hol}
\APACinsertmetastar {%
Barrett_hol}%
\begin{APACrefauthors}%
Barrett, J\BPBI W.%
\end{APACrefauthors}%
\unskip\
\newblock
\APACrefYearMonthDay{1991}{Sep}{01}.
\newblock
{\BBOQ}\APACrefatitle {Holonomy and path structures in general relativity and
  Yang-Mills theory} {Holonomy and path structures in general relativity and
  yang-mills theory}.{\BBCQ}
\newblock
\APACjournalVolNumPages{International Journal of Theoretical
  Physics}{30}{9}{1171--1215}.
\newblock
\begin{APACrefURL} \url{https://doi.org/10.1007/BF00671007} \end{APACrefURL}
\newblock
\begin{APACrefDOI} \doi{10.1007/BF00671007} \end{APACrefDOI}
\PrintBackRefs{\CurrentBib}

\bibitem [\protect \citeauthoryear {%
Belot%
}{%
Belot%
}{%
{\protect \APACyear {1998}}%
}]{%
Belot1998}
\APACinsertmetastar {%
Belot1998}%
\begin{APACrefauthors}%
Belot, G.%
\end{APACrefauthors}%
\unskip\
\newblock
\APACrefYearMonthDay{1998}{12}{}.
\newblock
{\BBOQ}\APACrefatitle {{Understanding Electromagnetism}} {{Understanding
  Electromagnetism}}.{\BBCQ}
\newblock
\APACjournalVolNumPages{The British Journal for the Philosophy of
  Science}{49}{4}{531-555}.
\newblock
\begin{APACrefURL} \url{https://doi.org/10.1093/bjps/49.4.531} \end{APACrefURL}
\newblock
\begin{APACrefDOI} \doi{10.1093/bjps/49.4.531} \end{APACrefDOI}
\PrintBackRefs{\CurrentBib}

\bibitem [\protect \citeauthoryear {%
Belot%
}{%
Belot%
}{%
{\protect \APACyear {2003}}%
}]{%
Belot2003}
\APACinsertmetastar {%
Belot2003}%
\begin{APACrefauthors}%
Belot, G.%
\end{APACrefauthors}%
\unskip\
\newblock
\APACrefYearMonthDay{2003}{}{}.
\newblock
{\BBOQ}\APACrefatitle {Symmetry and gauge freedom} {Symmetry and gauge
  freedom}.{\BBCQ}
\newblock
\APACjournalVolNumPages{Studies in History and Philosophy of Science Part B:
  Studies in History and Philosophy of Modern Physics}{34}{2}{189 - 225}.
\newblock
\begin{APACrefURL}
  \url{http://www.sciencedirect.com/science/article/pii/S1355219803000042}
  \end{APACrefURL}
\newblock
\begin{APACrefDOI} \doi{https://doi.org/10.1016/S1355-2198(03)00004-2}
  \end{APACrefDOI}
\PrintBackRefs{\CurrentBib}

\bibitem [\protect \citeauthoryear {%
Belot%
}{%
Belot%
}{%
{\protect \APACyear {2013}}%
}]{%
Belot_sym}
\APACinsertmetastar {%
Belot_sym}%
\begin{APACrefauthors}%
Belot, G.%
\end{APACrefauthors}%
\unskip\
\newblock
\APACrefYearMonthDay{2013}{}{}.
\newblock
{\BBOQ}\APACrefatitle {{Symmetry and Equivalence}} {{Symmetry and
  Equivalence}}.{\BBCQ}
\newblock
\BIn{} \APACrefbtitle {The Oxford Handbook of Philosophy of Physics.} {The
  oxford handbook of philosophy of physics.}
\newblock
\APACaddressPublisher{}{Oxford University Press. Edited by Batterman, R.}
\PrintBackRefs{\CurrentBib}

\bibitem [\protect \citeauthoryear {%
Belot%
}{%
Belot%
}{%
{\protect \APACyear {2018}}%
}]{%
Belot50}
\APACinsertmetastar {%
Belot50}%
\begin{APACrefauthors}%
Belot, G.%
\end{APACrefauthors}%
\unskip\
\newblock
\APACrefYearMonthDay{2018}{}{}.
\newblock
{\BBOQ}\APACrefatitle {Fifty Million Elvis Fans Can't be Wrong} {Fifty million
  elvis fans can't be wrong}.{\BBCQ}
\newblock
\APACjournalVolNumPages{Nous}{52}{4}{946-981}.
\newblock
\begin{APACrefURL}
  \url{https://onlinelibrary.wiley.com/doi/abs/10.1111/nous.12200}
  \end{APACrefURL}
\newblock
\begin{APACrefDOI} \doi{10.1111/nous.12200} \end{APACrefDOI}
\PrintBackRefs{\CurrentBib}

\bibitem [\protect \citeauthoryear {%
Berghofer%
\ \protect \BOthers {.}}{%
Berghofer%
\ \protect \BOthers {.}}{%
{\protect \APACyear {2021}}%
}]{%
Elements_gauge}
\APACinsertmetastar {%
Elements_gauge}%
\begin{APACrefauthors}%
Berghofer, P.%
, François, J.%
, Friederich, S.%
, Gomes, H.%
, Hetzroni, G.%
, Maas, A.%
\BCBL {}\ \BBA {} Sondenheimer, R.%
\end{APACrefauthors}%
\unskip\
\newblock
\APACrefYearMonthDay{2021}{}{}.
\newblock
\APACrefbtitle {Gauge Symmetries, Symmetry Breaking, and Gauge-Invariant
  Approaches.} {Gauge symmetries, symmetry breaking, and gauge-invariant
  approaches.}
\PrintBackRefs{\CurrentBib}

\bibitem [\protect \citeauthoryear {%
Bergmann%
\ \BBA {} Komar%
}{%
Bergmann%
\ \BBA {} Komar%
}{%
{\protect \APACyear {1960}}%
}]{%
Bergmann_Komar}
\APACinsertmetastar {%
Bergmann_Komar}%
\begin{APACrefauthors}%
Bergmann, P\BPBI G.%
\BCBT {}\ \BBA {} Komar, A\BPBI B.%
\end{APACrefauthors}%
\unskip\
\newblock
\APACrefYearMonthDay{1960}{Apr}{}.
\newblock
{\BBOQ}\APACrefatitle {{Poisson Brackets Between Locally Defined Observables in
  General Relativity}} {{Poisson Brackets Between Locally Defined Observables
  in General Relativity}}.{\BBCQ}
\newblock
\APACjournalVolNumPages{Physical Review Letters}{4}{}{432--433}.
\newblock
\begin{APACrefURL} \url{https://link.aps.org/doi/10.1103/PhysRevLett.4.432}
  \end{APACrefURL}
\newblock
\begin{APACrefDOI} \doi{10.1103/PhysRevLett.4.432} \end{APACrefDOI}
\PrintBackRefs{\CurrentBib}

\bibitem [\protect \citeauthoryear {%
Bonora%
\ \BBA {} Cotta-Ramusino%
}{%
Bonora%
\ \BBA {} Cotta-Ramusino%
}{%
{\protect \APACyear {1983}}%
}]{%
Bonora1983}
\APACinsertmetastar {%
Bonora1983}%
\begin{APACrefauthors}%
Bonora, L.%
\BCBT {}\ \BBA {} Cotta-Ramusino, P.%
\end{APACrefauthors}%
\unskip\
\newblock
\APACrefYearMonthDay{1983}{dec}{}.
\newblock
{\BBOQ}\APACrefatitle {{Some remarks on BRS transformations, anomalies and the
  cohomology of the Lie algebra of the group of gauge transformations}} {{Some
  remarks on BRS transformations, anomalies and the cohomology of the Lie
  algebra of the group of gauge transformations}}.{\BBCQ}
\newblock
\APACjournalVolNumPages{Communications in Mathematical
  Physics}{87}{4}{589--603}.
\newblock
\begin{APACrefURL} \url{http://link.springer.com/10.1007/BF01208267}
  \end{APACrefURL}
\newblock
\begin{APACrefDOI} \doi{10.1007/BF01208267} \end{APACrefDOI}
\PrintBackRefs{\CurrentBib}

\bibitem [\protect \citeauthoryear {%
Brading%
\ \BBA {} Brown%
}{%
Brading%
\ \BBA {} Brown%
}{%
{\protect \APACyear {2004}}%
}]{%
BradingBrown}
\APACinsertmetastar {%
BradingBrown}%
\begin{APACrefauthors}%
Brading, K.%
\BCBT {}\ \BBA {} Brown, H\BPBI R.%
\end{APACrefauthors}%
\unskip\
\newblock
\APACrefYearMonthDay{2004}{}{}.
\newblock
{\BBOQ}\APACrefatitle {Are Gauge Symmetry Transformations Observable?} {Are
  gauge symmetry transformations observable?}{\BBCQ}
\newblock
\APACjournalVolNumPages{The British Journal for the Philosophy of
  Science}{55}{4}{645--665}.
\newblock
\begin{APACrefURL} \url{http://www.jstor.org/stable/3541620} \end{APACrefURL}
\PrintBackRefs{\CurrentBib}

\bibitem [\protect \citeauthoryear {%
Buividovich%
\ \BBA {} Polikarpov%
}{%
Buividovich%
\ \BBA {} Polikarpov%
}{%
{\protect \APACyear {2008}}%
}]{%
Buividovich_2008}
\APACinsertmetastar {%
Buividovich_2008}%
\begin{APACrefauthors}%
Buividovich, P.%
\BCBT {}\ \BBA {} Polikarpov, M.%
\end{APACrefauthors}%
\unskip\
\newblock
\APACrefYearMonthDay{2008}{Dec}{}.
\newblock
{\BBOQ}\APACrefatitle {Entanglement entropy in gauge theories and the
  holographic principle for electric strings} {Entanglement entropy in gauge
  theories and the holographic principle for electric strings}.{\BBCQ}
\newblock
\APACjournalVolNumPages{Physics Letters B}{670}{2}{141–145}.
\newblock
\begin{APACrefURL} \url{http://dx.doi.org/10.1016/j.physletb.2008.10.032}
  \end{APACrefURL}
\newblock
\begin{APACrefDOI} \doi{10.1016/j.physletb.2008.10.032} \end{APACrefDOI}
\PrintBackRefs{\CurrentBib}

\bibitem [\protect \citeauthoryear {%
Butterfield%
}{%
Butterfield%
}{%
{\protect \APACyear {2007}}%
}]{%
Butterfield_symp}
\APACinsertmetastar {%
Butterfield_symp}%
\begin{APACrefauthors}%
Butterfield, J.%
\end{APACrefauthors}%
\unskip\
\newblock
\APACrefYearMonthDay{2007}{}{}.
\newblock
{\BBOQ}\APACrefatitle {ON SYMPLECTIC REDUCTION IN CLASSICAL MECHANICS} {On
  symplectic reduction in classical mechanics}.{\BBCQ}
\newblock
\BIn{} J.~Butterfield\ \BBA {} J.~Earman\ (\BEDS), \APACrefbtitle {Philosophy
  of Physics} {Philosophy of physics}\ (\BPG~1 - 131).
\newblock
\APACaddressPublisher{Amsterdam}{North-Holland}.
\newblock
\begin{APACrefURL}
  \url{http://www.sciencedirect.com/science/article/pii/B978044451560550004X}
  \end{APACrefURL}
\newblock
\begin{APACrefDOI} \doi{https://doi.org/10.1016/B978-044451560-5/50004-X}
  \end{APACrefDOI}
\PrintBackRefs{\CurrentBib}

\bibitem [\protect \citeauthoryear {%
Carlip%
}{%
Carlip%
}{%
{\protect \APACyear {1997}}%
}]{%
Carlip1997}
\APACinsertmetastar {%
Carlip1997}%
\begin{APACrefauthors}%
Carlip, S.%
\end{APACrefauthors}%
\unskip\
\newblock
\APACrefYearMonthDay{1997}{}{}.
\newblock
{\BBOQ}\APACrefatitle {Statistical mechanics and black hole thermodynamics}
  {Statistical mechanics and black hole thermodynamics}.{\BBCQ}
\newblock
\APACjournalVolNumPages{Nuclear Physics B - Proceedings
  Supplements}{57}{1}{8-12}.
\newblock
\begin{APACrefURL}
  \url{https://www.sciencedirect.com/science/article/pii/S0920563297003484}
  \end{APACrefURL}
\newblock
\APACrefnote{Constrained Dynamics and Quantum Gravity 1996}
\newblock
\begin{APACrefDOI} \doi{https://doi.org/10.1016/S0920-5632(97)00348-4}
  \end{APACrefDOI}
\PrintBackRefs{\CurrentBib}

\bibitem [\protect \citeauthoryear {%
Carrozza%
\ \BBA {} Hoehn%
}{%
Carrozza%
\ \BBA {} Hoehn%
}{%
{\protect \APACyear {2021}}%
}]{%
carrozza2021}
\APACinsertmetastar {%
carrozza2021}%
\begin{APACrefauthors}%
Carrozza, S.%
\BCBT {}\ \BBA {} Hoehn, P\BPBI A.%
\end{APACrefauthors}%
\unskip\
\newblock
\APACrefYearMonthDay{2021}{}{}.
\newblock
\APACrefbtitle {Edge modes as reference frames and boundary actions from
  post-selection.} {Edge modes as reference frames and boundary actions from
  post-selection.}
\PrintBackRefs{\CurrentBib}

\bibitem [\protect \citeauthoryear {%
Chasova%
}{%
Chasova%
}{%
{\protect \APACyear {2019}}%
}]{%
Chasova}
\APACinsertmetastar {%
Chasova}%
\begin{APACrefauthors}%
Chasova, V.%
\end{APACrefauthors}%
\unskip\
\newblock
\APACrefYearMonthDay{2019}{}{}.
\newblock
{\BBOQ}\APACrefatitle {Direct Empirical Status of Theoretical Symmetries in
  Physics} {Direct empirical status of theoretical symmetries in
  physics}.{\BBCQ}
\newblock
\APACjournalVolNumPages{PhD thesis}{}{}{}.
\PrintBackRefs{\CurrentBib}

\bibitem [\protect \citeauthoryear {%
Dasgupta%
}{%
Dasgupta%
}{%
{\protect \APACyear {2016}}%
}]{%
Dasgupta_sym}
\APACinsertmetastar {%
Dasgupta_sym}%
\begin{APACrefauthors}%
Dasgupta, S.%
\end{APACrefauthors}%
\unskip\
\newblock
\APACrefYearMonthDay{2016}{}{}.
\newblock
{\BBOQ}\APACrefatitle {{Symmetry as an Epistemic Notion (Twice Over)}}
  {{Symmetry as an Epistemic Notion (Twice Over)}}.{\BBCQ}
\newblock
\APACjournalVolNumPages{The British Journal for the Philosophy of
  Science}{67}{3}{837-878}.
\newblock
\begin{APACrefURL} \url{https://doi.org/10.1093/bjps/axu049} \end{APACrefURL}
\newblock
\begin{APACrefDOI} \doi{10.1093/bjps/axu049} \end{APACrefDOI}
\PrintBackRefs{\CurrentBib}

\bibitem [\protect \citeauthoryear {%
Dewar%
}{%
Dewar%
}{%
{\protect \APACyear {2017}}%
}]{%
Dewar2017}
\APACinsertmetastar {%
Dewar2017}%
\begin{APACrefauthors}%
Dewar, N.%
\end{APACrefauthors}%
\unskip\
\newblock
\APACrefYearMonthDay{2017}{09}{}.
\newblock
{\BBOQ}\APACrefatitle {{Sophistication about Symmetries}} {{Sophistication
  about Symmetries}}.{\BBCQ}
\newblock
\APACjournalVolNumPages{The British Journal for the Philosophy of
  Science}{70}{2}{485-521}.
\newblock
\begin{APACrefURL} \url{https://doi.org/10.1093/bjps/axx021} \end{APACrefURL}
\newblock
\begin{APACrefDOI} \doi{10.1093/bjps/axx021} \end{APACrefDOI}
\PrintBackRefs{\CurrentBib}

\bibitem [\protect \citeauthoryear {%
Donnelly%
\ \BBA {} Freidel%
}{%
Donnelly%
\ \BBA {} Freidel%
}{%
{\protect \APACyear {2016}}%
}]{%
DonnellyFreidel}
\APACinsertmetastar {%
DonnellyFreidel}%
\begin{APACrefauthors}%
Donnelly, W.%
\BCBT {}\ \BBA {} Freidel, L.%
\end{APACrefauthors}%
\unskip\
\newblock
\APACrefYearMonthDay{2016}{}{}.
\newblock
{\BBOQ}\APACrefatitle {{Local subsystems in gauge theory and gravity}} {{Local
  subsystems in gauge theory and gravity}}.{\BBCQ}
\newblock
\APACjournalVolNumPages{JHEP}{09}{}{102}.
\newblock
\begin{APACrefDOI} \doi{10.1007/JHEP09(2016)102} \end{APACrefDOI}
\PrintBackRefs{\CurrentBib}

\bibitem [\protect \citeauthoryear {%
Donnelly%
\ \BBA {} Giddings%
}{%
Donnelly%
\ \BBA {} Giddings%
}{%
{\protect \APACyear {2016}}%
}]{%
Donnelly_Giddings}
\APACinsertmetastar {%
Donnelly_Giddings}%
\begin{APACrefauthors}%
Donnelly, W.%
\BCBT {}\ \BBA {} Giddings, S\BPBI B.%
\end{APACrefauthors}%
\unskip\
\newblock
\APACrefYearMonthDay{2016}{Jan}{}.
\newblock
{\BBOQ}\APACrefatitle {Diffeomorphism-invariant observables and their nonlocal
  algebra} {Diffeomorphism-invariant observables and their nonlocal
  algebra}.{\BBCQ}
\newblock
\APACjournalVolNumPages{Physical Review D}{93}{2}{}.
\newblock
\begin{APACrefURL} \url{http://dx.doi.org/10.1103/PhysRevD.93.024030}
  \end{APACrefURL}
\newblock
\begin{APACrefDOI} \doi{10.1103/physrevd.93.024030} \end{APACrefDOI}
\PrintBackRefs{\CurrentBib}

\bibitem [\protect \citeauthoryear {%
Dougherty%
}{%
Dougherty%
}{%
{\protect \APACyear {2017}}%
}]{%
Dougherty2017}
\APACinsertmetastar {%
Dougherty2017}%
\begin{APACrefauthors}%
Dougherty, J.%
\end{APACrefauthors}%
\unskip\
\newblock
\APACrefYearMonthDay{2017}{}{}.
\newblock
{\BBOQ}\APACrefatitle {Sameness and Separability in Gauge Theories} {Sameness
  and separability in gauge theories}.{\BBCQ}
\newblock
\APACjournalVolNumPages{Philosophy of Science}{84}{5}{1189-1201}.
\newblock
\begin{APACrefURL} \url{https://doi.org/10.1086/694083} \end{APACrefURL}
\newblock
\begin{APACrefDOI} \doi{10.1086/694083} \end{APACrefDOI}
\PrintBackRefs{\CurrentBib}

\bibitem [\protect \citeauthoryear {%
Earman%
}{%
Earman%
}{%
{\protect \APACyear {1987}}%
}]{%
Earman_local}
\APACinsertmetastar {%
Earman_local}%
\begin{APACrefauthors}%
Earman, J.%
\end{APACrefauthors}%
\unskip\
\newblock
\APACrefYearMonthDay{1987}{September}{}.
\newblock
{\BBOQ}\APACrefatitle {Locality, Nonlocality and Action at a Distance: A
  Skeptical Review of Some Philosophical Dogmas} {Locality, nonlocality and
  action at a distance: A skeptical review of some philosophical
  dogmas}.{\BBCQ}
\newblock
\BIn{} R.~Kargon, P.~Achinstein\BCBL {}\ \BBA {} W\BPBI T.~Kelvin\ (\BEDS),
  \APACrefbtitle {Kelvin's Baltimore lectures and modern theoretical physics :
  historical and philosophical perspectives} {Kelvin's baltimore lectures and
  modern theoretical physics : historical and philosophical perspectives}\
  (\BPGS\ 449 -- 490).
\newblock
\APACaddressPublisher{Cambridge}{MIT Press}.
\newblock
\begin{APACrefURL} \url{http://d-scholarship.pitt.edu/12972/} \end{APACrefURL}
\PrintBackRefs{\CurrentBib}

\bibitem [\protect \citeauthoryear {%
Fischer%
}{%
Fischer%
}{%
{\protect \APACyear {1970}}%
}]{%
Fischer}
\APACinsertmetastar {%
Fischer}%
\begin{APACrefauthors}%
Fischer, A.%
\end{APACrefauthors}%
\unskip\
\newblock
\APACrefYearMonthDay{1970}{}{}.
\newblock
{\BBOQ}\APACrefatitle {{The Theory of Superspace}} {{The Theory of
  Superspace}}.{\BBCQ}
\newblock
\BIn{} \APACrefbtitle {Proceedings of the Relativity Conference held 2-6 June,
  1969 in Cincinnati, OH. Edited by Moshe Carmeli, Stuart I. Fickler, and Louis
  Witten. New York: Plenum Press, 1970., p.303.} {Proceedings of the relativity
  conference held 2-6 june, 1969 in cincinnati, oh. edited by moshe carmeli,
  stuart i. fickler, and louis witten. new york: Plenum press, 1970., p.303.}
\PrintBackRefs{\CurrentBib}

\bibitem [\protect \citeauthoryear {%
Friederich%
}{%
Friederich%
}{%
{\protect \APACyear {2014}}%
}]{%
Friederich2014}
\APACinsertmetastar {%
Friederich2014}%
\begin{APACrefauthors}%
Friederich, S.%
\end{APACrefauthors}%
\unskip\
\newblock
\APACrefYearMonthDay{2014}{04}{}.
\newblock
{\BBOQ}\APACrefatitle {{Symmetry, Empirical Equivalence, and Identity}}
  {{Symmetry, Empirical Equivalence, and Identity}}.{\BBCQ}
\newblock
\APACjournalVolNumPages{The British Journal for the Philosophy of
  Science}{66}{3}{537-559}.
\newblock
\begin{APACrefURL} \url{https://doi.org/10.1093/bjps/axt046} \end{APACrefURL}
\newblock
\begin{APACrefDOI} \doi{10.1093/bjps/axt046} \end{APACrefDOI}
\PrintBackRefs{\CurrentBib}

\bibitem [\protect \citeauthoryear {%
Friederich%
}{%
Friederich%
}{%
{\protect \APACyear {2017}}%
}]{%
Friederich2017}
\APACinsertmetastar {%
Friederich2017}%
\begin{APACrefauthors}%
Friederich, S.%
\end{APACrefauthors}%
\unskip\
\newblock
\APACrefYearMonthDay{2017}{}{}.
\newblock
{\BBOQ}\APACrefatitle {Symmetries and the Identity of Physical States}
  {Symmetries and the identity of physical states}.{\BBCQ}
\newblock
\BIn{} M.~Massimi, J\BHBI W.~Romeijn\BCBL {}\ \BBA {} G.~Schurz\ (\BEDS),
  \APACrefbtitle {EPSA15 Selected Papers} {Epsa15 selected papers}\ (\BPGS\
  153--165).
\newblock
\APACaddressPublisher{Cham}{Springer International Publishing}.
\PrintBackRefs{\CurrentBib}

\bibitem [\protect \citeauthoryear {%
Geiller%
}{%
Geiller%
}{%
{\protect \APACyear {2017}}%
}]{%
Geiller:2017xad}
\APACinsertmetastar {%
Geiller:2017xad}%
\begin{APACrefauthors}%
Geiller, M.%
\end{APACrefauthors}%
\unskip\
\newblock
\APACrefYearMonthDay{2017}{}{}.
\newblock
{\BBOQ}\APACrefatitle {{Edge modes and corner ambiguities in 3d Chern–Simons
  theory and gravity}} {{Edge modes and corner ambiguities in 3d Chern–Simons
  theory and gravity}}.{\BBCQ}
\newblock
\APACjournalVolNumPages{Nucl. Phys.}{B924}{}{312-365}.
\newblock
\begin{APACrefDOI} \doi{10.1016/j.nuclphysb.2017.09.010} \end{APACrefDOI}
\PrintBackRefs{\CurrentBib}

\bibitem [\protect \citeauthoryear {%
Geiller%
\ \BBA {} Jai-akson%
}{%
Geiller%
\ \BBA {} Jai-akson%
}{%
{\protect \APACyear {2020}}%
}]{%
Geiller_edge2020}
\APACinsertmetastar {%
Geiller_edge2020}%
\begin{APACrefauthors}%
Geiller, M.%
\BCBT {}\ \BBA {} Jai-akson, P.%
\end{APACrefauthors}%
\unskip\
\newblock
\APACrefYearMonthDay{2020}{Sep}{}.
\newblock
{\BBOQ}\APACrefatitle {Extended actions, dynamics of edge modes, and
  entanglement entropy} {Extended actions, dynamics of edge modes, and
  entanglement entropy}.{\BBCQ}
\newblock
\APACjournalVolNumPages{Journal of High Energy Physics}{2020}{9}{}.
\newblock
\begin{APACrefURL} \url{http://dx.doi.org/10.1007/JHEP09(2020)134}
  \end{APACrefURL}
\newblock
\begin{APACrefDOI} \doi{10.1007/jhep09(2020)134} \end{APACrefDOI}
\PrintBackRefs{\CurrentBib}

\bibitem [\protect \citeauthoryear {%
Gilbarg%
\ \BBA {} Trudinger%
}{%
Gilbarg%
\ \BBA {} Trudinger%
}{%
{\protect \APACyear {2001}}%
}]{%
Trudinger}
\APACinsertmetastar {%
Trudinger}%
\begin{APACrefauthors}%
Gilbarg, D.%
\BCBT {}\ \BBA {} Trudinger, N.%
\end{APACrefauthors}%
\unskip\
\newblock
\APACrefYear{2001}.
\newblock
\APACrefbtitle {Elliptic Partial Differential Equations of Second Order}
  {Elliptic partial differential equations of second order}.
\newblock
\APACaddressPublisher{}{Springer}.
\PrintBackRefs{\CurrentBib}

\bibitem [\protect \citeauthoryear {%
Giulini%
}{%
Giulini%
}{%
{\protect \APACyear {1995}}%
}]{%
Giulini_asymptotic}
\APACinsertmetastar {%
Giulini_asymptotic}%
\begin{APACrefauthors}%
Giulini, D.%
\end{APACrefauthors}%
\unskip\
\newblock
\APACrefYearMonthDay{1995}{}{}.
\newblock
{\BBOQ}\APACrefatitle {ASYMPTOTIC SYMMETRY GROUPS OF LONG-RANGED GAUGE
  CONFIGURATIONS} {Asymptotic symmetry groups of long-ranged gauge
  configurations}.{\BBCQ}
\newblock
\APACjournalVolNumPages{Modern Physics Letters A}{10}{28}{2059-2070}.
\newblock
\begin{APACrefURL} \url{https://doi.org/10.1142/S0217732395002210}
  \end{APACrefURL}
\newblock
\begin{APACrefDOI} \doi{10.1142/S0217732395002210} \end{APACrefDOI}
\PrintBackRefs{\CurrentBib}

\bibitem [\protect \citeauthoryear {%
Gomes%
}{%
Gomes%
}{%
{\protect \APACyear {2019}}%
}]{%
GomesStudies}
\APACinsertmetastar {%
GomesStudies}%
\begin{APACrefauthors}%
Gomes, H.%
\end{APACrefauthors}%
\unskip\
\newblock
\APACrefYearMonthDay{2019}{}{}.
\newblock
{\BBOQ}\APACrefatitle {Gauging the boundary in field-space} {Gauging the
  boundary in field-space}.{\BBCQ}
\newblock
\APACjournalVolNumPages{Studies in History and Philosophy of Science Part B:
  Studies in History and Philosophy of Modern Physics}{}{}{}.
\newblock
\begin{APACrefURL}
  \url{http://www.sciencedirect.com/science/article/pii/S1355219818302144}
  \end{APACrefURL}
\newblock
\begin{APACrefDOI} \doi{https://doi.org/10.1016/j.shpsb.2019.04.002}
  \end{APACrefDOI}
\PrintBackRefs{\CurrentBib}

\bibitem [\protect \citeauthoryear {%
Gomes%
}{%
Gomes%
}{%
{\protect \APACyear {2021}}%
{\protect \APACexlab {{\protect \BCnt {1}}}}}]{%
Gomes_new}
\APACinsertmetastar {%
Gomes_new}%
\begin{APACrefauthors}%
Gomes, H.%
\end{APACrefauthors}%
\unskip\
\newblock
\APACrefYearMonthDay{2021{\protect \BCnt {1}}}{}{}.
\newblock
{\BBOQ}\APACrefatitle {Holism as the significance of gauge symmetries} {Holism
  as the significance of gauge symmetries}.{\BBCQ}
\newblock
\APACjournalVolNumPages{European Journal of Philosophy of Science, vol 11,
  87}{}{}{}.
\PrintBackRefs{\CurrentBib}

\bibitem [\protect \citeauthoryear {%
Gomes%
}{%
Gomes%
}{%
{\protect \APACyear {2021}}%
{\protect \APACexlab {{\protect \BCnt {2}}}}}]{%
Samediff_1}
\APACinsertmetastar {%
Samediff_1}%
\begin{APACrefauthors}%
Gomes, H.%
\end{APACrefauthors}%
\unskip\
\newblock
\APACrefYearMonthDay{2021{\protect \BCnt {2}}}{}{}.
\newblock
{\BBOQ}\APACrefatitle {{Same-diff? Part I: Conceptual similarities (and one
  difference) between gauge transformations and diffeomorphisms}} {{Same-diff?
  Part I: Conceptual similarities (and one difference) between gauge
  transformations and diffeomorphisms}}.{\BBCQ}
\newblock
\APACjournalVolNumPages{Arxiv: 2110.07203. Submitted.}{}{}{}.
\PrintBackRefs{\CurrentBib}

\bibitem [\protect \citeauthoryear {%
Gomes%
}{%
Gomes%
}{%
{\protect \APACyear {2021}}%
{\protect \APACexlab {{\protect \BCnt {3}}}}}]{%
Samediff_2}
\APACinsertmetastar {%
Samediff_2}%
\begin{APACrefauthors}%
Gomes, H.%
\end{APACrefauthors}%
\unskip\
\newblock
\APACrefYearMonthDay{2021{\protect \BCnt {3}}}{}{}.
\newblock
{\BBOQ}\APACrefatitle {{Same-diff? Part II: A compendium of similarities
  between gauge transformations and diffeomorphisms}} {{Same-diff? Part II: A
  compendium of similarities between gauge transformations and
  diffeomorphisms}}.{\BBCQ}
\newblock
\APACjournalVolNumPages{Arxiv: 2110.07204. Submitted.}{}{}{}.
\PrintBackRefs{\CurrentBib}

\bibitem [\protect \citeauthoryear {%
Gomes%
\ \BBA {} Butterfield%
}{%
Gomes%
\ \BBA {} Butterfield%
}{%
{\protect \APACyear {2021}}%
}]{%
GomesButterfield_electro}
\APACinsertmetastar {%
GomesButterfield_electro}%
\begin{APACrefauthors}%
Gomes, H.%
\BCBT {}\ \BBA {} Butterfield, J.%
\end{APACrefauthors}%
\unskip\
\newblock
\APACrefYearMonthDay{2021}{}{}.
\newblock
{\BBOQ}\APACrefatitle {How to choose a gauge: the example of electromagnetism}
  {How to choose a gauge: the example of electromagnetism}.{\BBCQ}
\newblock
\APACjournalVolNumPages{In preparation}{}{}{}.
\PrintBackRefs{\CurrentBib}

\bibitem [\protect \citeauthoryear {%
Gomes%
, Hopfmüller%
\BCBL {}\ \BBA {} Riello%
}{%
Gomes%
\ \protect \BOthers {.}}{%
{\protect \APACyear {2019}}%
}]{%
GomesHopfRiello}
\APACinsertmetastar {%
GomesHopfRiello}%
\begin{APACrefauthors}%
Gomes, H.%
, Hopfmüller, F.%
\BCBL {}\ \BBA {} Riello, A.%
\end{APACrefauthors}%
\unskip\
\newblock
\APACrefYearMonthDay{2019}{}{}.
\newblock
{\BBOQ}\APACrefatitle {A unified geometric framework for boundary charges and
  dressings: Non-Abelian theory and matter} {A unified geometric framework for
  boundary charges and dressings: Non-abelian theory and matter}.{\BBCQ}
\newblock
\APACjournalVolNumPages{Nuclear Physics B}{941}{}{249 - 315}.
\newblock
\begin{APACrefURL}
  \url{http://www.sciencedirect.com/science/article/pii/S0550321319300483}
  \end{APACrefURL}
\newblock
\begin{APACrefDOI} \doi{https://doi.org/10.1016/j.nuclphysb.2019.02.020}
  \end{APACrefDOI}
\PrintBackRefs{\CurrentBib}

\bibitem [\protect \citeauthoryear {%
Gomes%
\ \BBA {} Riello%
}{%
Gomes%
\ \BBA {} Riello%
}{%
{\protect \APACyear {2017}}%
}]{%
GomesRiello2016}
\APACinsertmetastar {%
GomesRiello2016}%
\begin{APACrefauthors}%
Gomes, H.%
\BCBT {}\ \BBA {} Riello, A.%
\end{APACrefauthors}%
\unskip\
\newblock
\APACrefYearMonthDay{2017}{}{}.
\newblock
{\BBOQ}\APACrefatitle {{The observer’s ghost: notes on a field space
  connection}} {{The observer’s ghost: notes on a field space
  connection}}.{\BBCQ}
\newblock
\APACjournalVolNumPages{Journal of High Energy Physics (JHEP)}{05}{}{017}.
\newblock
\begin{APACrefURL}
  \url{https://link.springer.com/article/10.1007%2FJHEP05%282017%29017}
  \end{APACrefURL}
\newblock
\begin{APACrefDOI} \doi{10.1007/JHEP05(2017)017} \end{APACrefDOI}
\PrintBackRefs{\CurrentBib}

\bibitem [\protect \citeauthoryear {%
Gomes%
\ \BBA {} Riello%
}{%
Gomes%
\ \BBA {} Riello%
}{%
{\protect \APACyear {2018}}%
}]{%
GomesRiello2018}
\APACinsertmetastar {%
GomesRiello2018}%
\begin{APACrefauthors}%
Gomes, H.%
\BCBT {}\ \BBA {} Riello, A.%
\end{APACrefauthors}%
\unskip\
\newblock
\APACrefYearMonthDay{2018}{Jul}{}.
\newblock
{\BBOQ}\APACrefatitle {Unified geometric framework for boundary charges and
  particle dressings} {Unified geometric framework for boundary charges and
  particle dressings}.{\BBCQ}
\newblock
\APACjournalVolNumPages{Physical Review D}{98}{}{025013}.
\newblock
\begin{APACrefURL} \url{https://link.aps.org/doi/10.1103/PhysRevD.98.025013}
  \end{APACrefURL}
\newblock
\begin{APACrefDOI} \doi{10.1103/PhysRevD.98.025013} \end{APACrefDOI}
\PrintBackRefs{\CurrentBib}

\bibitem [\protect \citeauthoryear {%
Gomes%
\ \BBA {} Riello%
}{%
Gomes%
\ \BBA {} Riello%
}{%
{\protect \APACyear {2021}}%
}]{%
GomesRiello_new}
\APACinsertmetastar {%
GomesRiello_new}%
\begin{APACrefauthors}%
Gomes, H.%
\BCBT {}\ \BBA {} Riello, A.%
\end{APACrefauthors}%
\unskip\
\newblock
\APACrefYearMonthDay{2021}{}{}.
\newblock
{\BBOQ}\APACrefatitle {{The quasilocal degrees of freedom of Yang-Mills
  theory}} {{The quasilocal degrees of freedom of Yang-Mills theory}}.{\BBCQ}
\newblock
\APACjournalVolNumPages{SciPost Phys.}{10}{}{130}.
\newblock
\begin{APACrefURL} \url{https://scipost.org/10.21468/SciPostPhys.10.6.130}
  \end{APACrefURL}
\newblock
\begin{APACrefDOI} \doi{10.21468/SciPostPhys.10.6.130} \end{APACrefDOI}
\PrintBackRefs{\CurrentBib}

\bibitem [\protect \citeauthoryear {%
Gourgoulhon%
}{%
Gourgoulhon%
}{%
{\protect \APACyear {2007}}%
}]{%
3+1_book}
\APACinsertmetastar {%
3+1_book}%
\begin{APACrefauthors}%
Gourgoulhon, E.%
\end{APACrefauthors}%
\unskip\
\newblock
\APACrefYearMonthDay{2007}{}{}.
\newblock
{\BBOQ}\APACrefatitle {{3+1 formalism and bases of numerical relativity}} {{3+1
  formalism and bases of numerical relativity}}.{\BBCQ}
\newblock
\APACjournalVolNumPages{Lecture notes in Physics 846, Springer}{}{}{}.
\PrintBackRefs{\CurrentBib}

\bibitem [\protect \citeauthoryear {%
Greaves%
\ \BBA {} Wallace%
}{%
Greaves%
\ \BBA {} Wallace%
}{%
{\protect \APACyear {2014}}%
}]{%
GreavesWallace}
\APACinsertmetastar {%
GreavesWallace}%
\begin{APACrefauthors}%
Greaves, H.%
\BCBT {}\ \BBA {} Wallace, D.%
\end{APACrefauthors}%
\unskip\
\newblock
\APACrefYearMonthDay{2014}{}{}.
\newblock
{\BBOQ}\APACrefatitle {Empirical Consequences of Symmetries} {Empirical
  consequences of symmetries}.{\BBCQ}
\newblock
\APACjournalVolNumPages{British Journal for the Philosophy of
  Science}{65}{1}{59--89}.
\PrintBackRefs{\CurrentBib}

\bibitem [\protect \citeauthoryear {%
Gribov%
}{%
Gribov%
}{%
{\protect \APACyear {1978}}%
}]{%
Gribov:1977wm}
\APACinsertmetastar {%
Gribov:1977wm}%
\begin{APACrefauthors}%
Gribov, V\BPBI N.%
\end{APACrefauthors}%
\unskip\
\newblock
\APACrefYearMonthDay{1978}{}{}.
\newblock
{\BBOQ}\APACrefatitle {{Quantization of Nonabelian Gauge Theories}}
  {{Quantization of Nonabelian Gauge Theories}}.{\BBCQ}
\newblock
\APACjournalVolNumPages{Nucl. Phys.}{B139}{}{1}.
\newblock
\APACrefnote{[,1(1977)]}
\newblock
\begin{APACrefDOI} \doi{10.1016/0550-3213(78)90175-X} \end{APACrefDOI}
\PrintBackRefs{\CurrentBib}

\bibitem [\protect \citeauthoryear {%
Harlow%
\ \BBA {} Wu%
}{%
Harlow%
\ \BBA {} Wu%
}{%
{\protect \APACyear {2019}}%
}]{%
Harlow_cov}
\APACinsertmetastar {%
Harlow_cov}%
\begin{APACrefauthors}%
Harlow, D.%
\BCBT {}\ \BBA {} Wu, J\BHBI Q.%
\end{APACrefauthors}%
\unskip\
\newblock
\APACrefYearMonthDay{2019}{}{}.
\newblock
{\BBOQ}\APACrefatitle {{Covariant phase space with boundaries}} {{Covariant
  phase space with boundaries}}.{\BBCQ}
\newblock

\PrintBackRefs{\CurrentBib}

\bibitem [\protect \citeauthoryear {%
Haro%
\ \BBA {} Butterfield%
}{%
Haro%
\ \BBA {} Butterfield%
}{%
{\protect \APACyear {2021}}%
}]{%
ButterfieldHaro_sym}
\APACinsertmetastar {%
ButterfieldHaro_sym}%
\begin{APACrefauthors}%
Haro, S\BPBI D.%
\BCBT {}\ \BBA {} Butterfield, J.%
\end{APACrefauthors}%
\unskip\
\newblock
\APACrefYearMonthDay{2021}{}{}.
\newblock
{\BBOQ}\APACrefatitle {{Symmetry and Duality}} {{Symmetry and Duality}}.{\BBCQ}
\newblock
\APACjournalVolNumPages{Synthese, 198}{}{}{}.
\PrintBackRefs{\CurrentBib}

\bibitem [\protect \citeauthoryear {%
Hayward%
}{%
Hayward%
}{%
{\protect \APACyear {2013}}%
}]{%
Hayward_book}
\APACinsertmetastar {%
Hayward_book}%
\begin{APACrefauthors}%
Hayward, S\BPBI A.%
\end{APACrefauthors}%
\unskip\
\newblock
\APACrefYear{2013}.
\newblock
\APACrefbtitle {{Black Holes}} {{Black Holes}}.
\newblock
\APACaddressPublisher{}{WORLD SCIENTIFIC}.
\newblock
\begin{APACrefURL} \url{https://www.worldscientific.com/doi/abs/10.1142/8604}
  \end{APACrefURL}
\newblock
\begin{APACrefDOI} \doi{10.1142/8604} \end{APACrefDOI}
\PrintBackRefs{\CurrentBib}

\bibitem [\protect \citeauthoryear {%
Healey%
}{%
Healey%
}{%
{\protect \APACyear {2007}}%
}]{%
Healey_book}
\APACinsertmetastar {%
Healey_book}%
\begin{APACrefauthors}%
Healey, R.%
\end{APACrefauthors}%
\unskip\
\newblock
\APACrefYear{2007}.
\newblock
\APACrefbtitle {{Gauging What's Real: The Conceptual Foundations of Gauge
  Theories}} {{Gauging What's Real: The Conceptual Foundations of Gauge
  Theories}}.
\newblock
\APACaddressPublisher{}{Oxford University Press}.
\PrintBackRefs{\CurrentBib}

\bibitem [\protect \citeauthoryear {%
Healey%
}{%
Healey%
}{%
{\protect \APACyear {2009}}%
}]{%
Healey2009}
\APACinsertmetastar {%
Healey2009}%
\begin{APACrefauthors}%
Healey, R.%
\end{APACrefauthors}%
\unskip\
\newblock
\APACrefYearMonthDay{2009}{08}{}.
\newblock
{\BBOQ}\APACrefatitle {{Perfect Symmetries}} {{Perfect Symmetries}}.{\BBCQ}
\newblock
\APACjournalVolNumPages{The British Journal for the Philosophy of
  Science}{60}{4}{697-720}.
\newblock
\begin{APACrefURL} \url{https://doi.org/10.1093/bjps/axp033} \end{APACrefURL}
\newblock
\begin{APACrefDOI} \doi{10.1093/bjps/axp033} \end{APACrefDOI}
\PrintBackRefs{\CurrentBib}

\bibitem [\protect \citeauthoryear {%
Henneaux%
\ \BBA {} Teitelboim%
}{%
Henneaux%
\ \BBA {} Teitelboim%
}{%
{\protect \APACyear {1992}}%
}]{%
HenneauxTeitelboim}
\APACinsertmetastar {%
HenneauxTeitelboim}%
\begin{APACrefauthors}%
Henneaux, M.%
\BCBT {}\ \BBA {} Teitelboim, C.%
\end{APACrefauthors}%
\unskip\
\newblock
\APACrefYear{1992}.
\newblock
\APACrefbtitle {Quantization of Gauge Systems} {Quantization of gauge systems}.
\newblock
\APACaddressPublisher{}{Princeton University Press}.
\PrintBackRefs{\CurrentBib}

\bibitem [\protect \citeauthoryear {%
Jacobson%
\ \BBA {} Nguyen%
}{%
Jacobson%
\ \BBA {} Nguyen%
}{%
{\protect \APACyear {2019}}%
}]{%
Jacobson_2019}
\APACinsertmetastar {%
Jacobson_2019}%
\begin{APACrefauthors}%
Jacobson, T.%
\BCBT {}\ \BBA {} Nguyen, P.%
\end{APACrefauthors}%
\unskip\
\newblock
\APACrefYearMonthDay{2019}{Aug}{}.
\newblock
{\BBOQ}\APACrefatitle {Diffeomorphism invariance and the black hole information
  paradox} {Diffeomorphism invariance and the black hole information
  paradox}.{\BBCQ}
\newblock
\APACjournalVolNumPages{Physical Review D}{100}{4}{}.
\newblock
\begin{APACrefURL} \url{http://dx.doi.org/10.1103/PhysRevD.100.046002}
  \end{APACrefURL}
\newblock
\begin{APACrefDOI} \doi{10.1103/physrevd.100.046002} \end{APACrefDOI}
\PrintBackRefs{\CurrentBib}

\bibitem [\protect \citeauthoryear {%
Kondracki%
\ \BBA {} Rogulski%
}{%
Kondracki%
\ \BBA {} Rogulski%
}{%
{\protect \APACyear {1983}}%
}]{%
kondracki1983}
\APACinsertmetastar {%
kondracki1983}%
\begin{APACrefauthors}%
Kondracki, W.%
\BCBT {}\ \BBA {} Rogulski, J.%
\end{APACrefauthors}%
\unskip\
\newblock
\APACrefYear{1983}.
\newblock
\APACrefbtitle {On the Stratification of the Orbit Space for the Action of
  Automorphisms on Connections. On Conjugacy Classes of Closed Subgroups. On
  the Notion of Stratification} {On the stratification of the orbit space for
  the action of automorphisms on connections. on conjugacy classes of closed
  subgroups. on the notion of stratification}.
\newblock
\APACaddressPublisher{}{Inst., Acad.}
\newblock
\begin{APACrefURL} \url{https://books.google.co.uk/books?id=LK0JrgEACAAJ}
  \end{APACrefURL}
\PrintBackRefs{\CurrentBib}

\bibitem [\protect \citeauthoryear {%
Kosso%
}{%
Kosso%
}{%
{\protect \APACyear {2000}}%
}]{%
Kosso}
\APACinsertmetastar {%
Kosso}%
\begin{APACrefauthors}%
Kosso, P.%
\end{APACrefauthors}%
\unskip\
\newblock
\APACrefYearMonthDay{2000}{}{}.
\newblock
{\BBOQ}\APACrefatitle {The Empirical Status of Symmetries in Physics} {The
  empirical status of symmetries in physics}.{\BBCQ}
\newblock
\APACjournalVolNumPages{The British Journal for the Philosophy of
  Science}{51}{1}{81--98}.
\newblock
\begin{APACrefURL} \url{http://www.jstor.org/stable/3541749} \end{APACrefURL}
\PrintBackRefs{\CurrentBib}

\bibitem [\protect \citeauthoryear {%
Ladyman%
}{%
Ladyman%
}{%
{\protect \APACyear {2015}}%
}]{%
Ladyman_DES}
\APACinsertmetastar {%
Ladyman_DES}%
\begin{APACrefauthors}%
Ladyman, J.%
\end{APACrefauthors}%
\unskip\
\newblock
\APACrefYearMonthDay{2015}{}{}.
\newblock
{\BBOQ}\APACrefatitle {Representation and Symmetry in Physics} {Representation
  and symmetry in physics}.{\BBCQ}
\newblock
\APACjournalVolNumPages{unpublished}{}{}{}.
\PrintBackRefs{\CurrentBib}

\bibitem [\protect \citeauthoryear {%
Lange%
}{%
Lange%
}{%
{\protect \APACyear {2007}}%
}]{%
Lange_meta}
\APACinsertmetastar {%
Lange_meta}%
\begin{APACrefauthors}%
Lange, M.%
\end{APACrefauthors}%
\unskip\
\newblock
\APACrefYearMonthDay{2007}{}{}.
\newblock
{\BBOQ}\APACrefatitle {Laws and meta-laws of nature: Conservation laws and
  symmetries} {Laws and meta-laws of nature: Conservation laws and
  symmetries}.{\BBCQ}
\newblock
\APACjournalVolNumPages{Studies in History and Philosophy of Science Part B:
  Studies in History and Philosophy of Modern Physics}{38}{3}{457-481}.
\newblock
\begin{APACrefURL}
  \url{https://www.sciencedirect.com/science/article/pii/S1355219806000943}
  \end{APACrefURL}
\newblock
\begin{APACrefDOI} \doi{https://doi.org/10.1016/j.shpsb.2006.08.003}
  \end{APACrefDOI}
\PrintBackRefs{\CurrentBib}

\bibitem [\protect \citeauthoryear {%
Marsden%
}{%
Marsden%
}{%
{\protect \APACyear {2007}}%
}]{%
Marsden2007}
\APACinsertmetastar {%
Marsden2007}%
\begin{APACrefauthors}%
Marsden, J.%
\end{APACrefauthors}%
\unskip\
\newblock
\APACrefYearMonthDay{2007}{}{}.
\newblock
{\BBOQ}\APACrefatitle {{Symplectic Reduction}} {{Symplectic Reduction}}.{\BBCQ}
\newblock
\BIn{} \APACrefbtitle {Hamiltonian Reduction by Stages} {Hamiltonian reduction
  by stages}\ (\BPGS\ 3--42).
\newblock
\APACaddressPublisher{Berlin, Heidelberg}{Springer Berlin Heidelberg}.
\newblock
\begin{APACrefURL} \url{https://doi.org/10.1007/978-3-540-72470-4_1}
  \end{APACrefURL}
\newblock
\begin{APACrefDOI} \doi{10.1007/978-3-540-72470-4_1} \end{APACrefDOI}
\PrintBackRefs{\CurrentBib}

\bibitem [\protect \citeauthoryear {%
Martens%
\ \BBA {} Read%
}{%
Martens%
\ \BBA {} Read%
}{%
{\protect \APACyear {2020}}%
}]{%
ReadMartens}
\APACinsertmetastar {%
ReadMartens}%
\begin{APACrefauthors}%
Martens, N\BPBI C.%
\BCBT {}\ \BBA {} Read, J.%
\end{APACrefauthors}%
\unskip\
\newblock
\APACrefYearMonthDay{2020}{}{}.
\newblock
{\BBOQ}\APACrefatitle {Sophistry about symmetries?} {Sophistry about
  symmetries?}{\BBCQ}
\newblock
\APACjournalVolNumPages{Synthese}{}{}{}.
\newblock
\begin{APACrefURL} \url{http://philsci-archive.pitt.edu/17184/}
  \end{APACrefURL}
\PrintBackRefs{\CurrentBib}

\bibitem [\protect \citeauthoryear {%
Mathieu%
}{%
Mathieu%
}{%
{\protect \APACyear {2020}}%
}]{%
Teh_stack}
\APACinsertmetastar {%
Teh_stack}%
\begin{APACrefauthors}%
Mathieu, M\BPBI L\BPBI S\BPBI A\BPBI T\BPBI N., P.%
\end{APACrefauthors}%
\unskip\
\newblock
\APACrefYearMonthDay{2020}{}{}.
\newblock
{\BBOQ}\APACrefatitle {Homological perspective on edge modes in linear
  Yang–Mills and Chern–Simons theory.} {Homological perspective on edge
  modes in linear yang–mills and chern–simons theory.}{\BBCQ}
\newblock
\APACjournalVolNumPages{Lett Math Phys}{}{}{}.
\PrintBackRefs{\CurrentBib}

\bibitem [\protect \citeauthoryear {%
Mitter%
\ \BBA {} Viallet%
}{%
Mitter%
\ \BBA {} Viallet%
}{%
{\protect \APACyear {1981}}%
}]{%
Mitter:1979un}
\APACinsertmetastar {%
Mitter:1979un}%
\begin{APACrefauthors}%
Mitter, P\BPBI K.%
\BCBT {}\ \BBA {} Viallet, C\BPBI M.%
\end{APACrefauthors}%
\unskip\
\newblock
\APACrefYearMonthDay{1981}{}{}.
\newblock
{\BBOQ}\APACrefatitle {{On the Bundle of Connections and the Gauge Orbit
  Manifold in {Yang-Mills} Theory}} {{On the Bundle of Connections and the
  Gauge Orbit Manifold in {Yang-Mills} Theory}}.{\BBCQ}
\newblock
\APACjournalVolNumPages{Commun. Math. Phys.}{79}{}{457}.
\newblock
\begin{APACrefDOI} \doi{10.1007/BF01209307} \end{APACrefDOI}
\PrintBackRefs{\CurrentBib}

\bibitem [\protect \citeauthoryear {%
Myrvold%
}{%
Myrvold%
}{%
{\protect \APACyear {2010}}%
}]{%
Myrvold2010}
\APACinsertmetastar {%
Myrvold2010}%
\begin{APACrefauthors}%
Myrvold, W\BPBI C.%
\end{APACrefauthors}%
\unskip\
\newblock
\APACrefYearMonthDay{2010}{03}{}.
\newblock
{\BBOQ}\APACrefatitle {{Nonseparability, Classical, and Quantum}}
  {{Nonseparability, Classical, and Quantum}}.{\BBCQ}
\newblock
\APACjournalVolNumPages{The British Journal for the Philosophy of
  Science}{62}{2}{417-432}.
\newblock
\begin{APACrefURL} \url{https://doi.org/10.1093/bjps/axq036} \end{APACrefURL}
\newblock
\begin{APACrefDOI} \doi{10.1093/bjps/axq036} \end{APACrefDOI}
\PrintBackRefs{\CurrentBib}

\bibitem [\protect \citeauthoryear {%
Nguyen%
, Teh%
\BCBL {}\ \BBA {} Wells%
}{%
Nguyen%
\ \protect \BOthers {.}}{%
{\protect \APACyear {2018}}%
}]{%
Teh_surplus}
\APACinsertmetastar {%
Teh_surplus}%
\begin{APACrefauthors}%
Nguyen, J.%
, Teh, N\BPBI J.%
\BCBL {}\ \BBA {} Wells, L.%
\end{APACrefauthors}%
\unskip\
\newblock
\APACrefYearMonthDay{2018}{03}{}.
\newblock
{\BBOQ}\APACrefatitle {{Why Surplus Structure Is Not Superfluous}} {{Why
  Surplus Structure Is Not Superfluous}}.{\BBCQ}
\newblock
\APACjournalVolNumPages{The British Journal for the Philosophy of
  Science}{}{}{}.
\newblock
\begin{APACrefURL} \url{https://doi.org/10.1093/bjps/axy026} \end{APACrefURL}
\newblock
\APACrefnote{axy026}
\newblock
\begin{APACrefDOI} \doi{10.1093/bjps/axy026} \end{APACrefDOI}
\PrintBackRefs{\CurrentBib}

\bibitem [\protect \citeauthoryear {%
Putnam%
}{%
Putnam%
}{%
{\protect \APACyear {1975}}%
}]{%
Putnam_analytic}
\APACinsertmetastar {%
Putnam_analytic}%
\begin{APACrefauthors}%
Putnam, H.%
\end{APACrefauthors}%
\unskip\
\newblock
\APACrefYearMonthDay{1975}{}{}.
\newblock
{\BBOQ}\APACrefatitle {{The Analytic and Synthetic}} {{The Analytic and
  Synthetic}}.{\BBCQ}
\newblock
\BIn{} \APACrefbtitle {Mind, Language and Reality: Philosophical Papers} {Mind,
  language and reality: Philosophical papers}\ (\BPGS\ 33--69).
\newblock
\APACaddressPublisher{}{Cambridge University Press}.
\PrintBackRefs{\CurrentBib}

\bibitem [\protect \citeauthoryear {%
S.~Ramirez%
\ \BBA {} Teh%
}{%
S.~Ramirez%
\ \BBA {} Teh%
}{%
{\protect \APACyear {2019}}%
}]{%
Teh_abandon}
\APACinsertmetastar {%
Teh_abandon}%
\begin{APACrefauthors}%
Ramirez, S.%
\BCBT {}\ \BBA {} Teh, N.%
\end{APACrefauthors}%
\unskip\
\newblock
\APACrefYearMonthDay{2019}{}{}.
\newblock
{\BBOQ}\APACrefatitle {Abandoning Galileo's Ship: The quest for non-relational
  empirical signicance} {Abandoning galileo's ship: The quest for
  non-relational empirical signicance}.{\BBCQ}
\newblock
\APACjournalVolNumPages{preprint}{}{}{}.
\PrintBackRefs{\CurrentBib}

\bibitem [\protect \citeauthoryear {%
S\BPBI M.~Ramirez%
}{%
S\BPBI M.~Ramirez%
}{%
{\protect \APACyear {2019}}%
}]{%
Seb_ramirez}
\APACinsertmetastar {%
Seb_ramirez}%
\begin{APACrefauthors}%
Ramirez, S\BPBI M.%
\end{APACrefauthors}%
\unskip\
\newblock
\APACrefYearMonthDay{2019}{October}{}.
\newblock
{\BBOQ}\APACrefatitle {A puzzle concerning local symmetries and their empirical
  significance} {A puzzle concerning local symmetries and their empirical
  significance}.{\BBCQ}
\newblock
\begin{APACrefURL} \url{http://philsci-archive.pitt.edu/16509/}
  \end{APACrefURL}
\PrintBackRefs{\CurrentBib}

\bibitem [\protect \citeauthoryear {%
Regge%
\ \BBA {} Teitelboim%
}{%
Regge%
\ \BBA {} Teitelboim%
}{%
{\protect \APACyear {1974}}%
}]{%
ReggeTeitelboim1974}
\APACinsertmetastar {%
ReggeTeitelboim1974}%
\begin{APACrefauthors}%
Regge, T.%
\BCBT {}\ \BBA {} Teitelboim, C.%
\end{APACrefauthors}%
\unskip\
\newblock
\APACrefYearMonthDay{1974}{}{}.
\newblock
{\BBOQ}\APACrefatitle {{Role of Surface Integrals in the Hamiltonian
  Formulation of General Relativity}} {{Role of Surface Integrals in the
  Hamiltonian Formulation of General Relativity}}.{\BBCQ}
\newblock
\APACjournalVolNumPages{Annals Phys.}{88}{}{286}.
\newblock
\begin{APACrefDOI} \doi{10.1016/0003-4916(74)90404-7} \end{APACrefDOI}
\PrintBackRefs{\CurrentBib}

\bibitem [\protect \citeauthoryear {%
Riello%
}{%
Riello%
}{%
{\protect \APACyear {2020}}%
}]{%
RielloSoft}
\APACinsertmetastar {%
RielloSoft}%
\begin{APACrefauthors}%
Riello, A.%
\end{APACrefauthors}%
\unskip\
\newblock
\APACrefYearMonthDay{2020}{}{}.
\newblock
{\BBOQ}\APACrefatitle {{Soft charges from the geometry of field space}} {{Soft
  charges from the geometry of field space}}.{\BBCQ}
\newblock
\APACjournalVolNumPages{JHEP}{}{}{}.
\PrintBackRefs{\CurrentBib}

\bibitem [\protect \citeauthoryear {%
Riello%
}{%
Riello%
}{%
{\protect \APACyear {2021}}%
{\protect \APACexlab {{\protect \BCnt {1}}}}}]{%
Riello_new}
\APACinsertmetastar {%
Riello_new}%
\begin{APACrefauthors}%
Riello, A.%
\end{APACrefauthors}%
\unskip\
\newblock
\APACrefYearMonthDay{2021{\protect \BCnt {1}}}{}{}.
\newblock
{\BBOQ}\APACrefatitle {Edge modes without edge modes} {Edge modes without edge
  modes}.{\BBCQ}
\newblock
\APACjournalVolNumPages{forthcoming}{}{}{}.
\PrintBackRefs{\CurrentBib}

\bibitem [\protect \citeauthoryear {%
Riello%
}{%
Riello%
}{%
{\protect \APACyear {2021}}%
{\protect \APACexlab {{\protect \BCnt {2}}}}}]{%
Riello_symp}
\APACinsertmetastar {%
Riello_symp}%
\begin{APACrefauthors}%
Riello, A.%
\end{APACrefauthors}%
\unskip\
\newblock
\APACrefYearMonthDay{2021{\protect \BCnt {2}}}{}{}.
\newblock
{\BBOQ}\APACrefatitle {{Symplectic reduction of Yang-Mills theory with
  boundaries: from superselection sectors to edge modes, and back}}
  {{Symplectic reduction of Yang-Mills theory with boundaries: from
  superselection sectors to edge modes, and back}}.{\BBCQ}
\newblock
\APACjournalVolNumPages{SciPost Phys.}{10}{}{125}.
\newblock
\begin{APACrefURL} \url{https://scipost.org/10.21468/SciPostPhys.10.6.125}
  \end{APACrefURL}
\newblock
\begin{APACrefDOI} \doi{10.21468/SciPostPhys.10.6.125} \end{APACrefDOI}
\PrintBackRefs{\CurrentBib}

\bibitem [\protect \citeauthoryear {%
Rovelli%
}{%
Rovelli%
}{%
{\protect \APACyear {2014}}%
}]{%
RovelliGauge2013}
\APACinsertmetastar {%
RovelliGauge2013}%
\begin{APACrefauthors}%
Rovelli, C.%
\end{APACrefauthors}%
\unskip\
\newblock
\APACrefYearMonthDay{2014}{}{}.
\newblock
{\BBOQ}\APACrefatitle {{Why Gauge?}} {{Why Gauge?}}{\BBCQ}
\newblock
\APACjournalVolNumPages{Found. Phys.}{44}{1}{91-104}.
\newblock
\begin{APACrefDOI} \doi{10.1007/s10701-013-9768-7} \end{APACrefDOI}
\PrintBackRefs{\CurrentBib}

\bibitem [\protect \citeauthoryear {%
Saunders%
}{%
Saunders%
}{%
{\protect \APACyear {2013}}%
}]{%
saunders2013}
\APACinsertmetastar {%
saunders2013}%
\begin{APACrefauthors}%
Saunders, S.%
\end{APACrefauthors}%
\unskip\
\newblock
\APACrefYearMonthDay{2013}{}{}.
\newblock
{\BBOQ}\APACrefatitle {{Rethinking Newton’s Principia}} {{Rethinking
  Newton’s Principia}}.{\BBCQ}
\newblock
\APACjournalVolNumPages{Philosophy of Science}{80}{1}{22--48}.
\PrintBackRefs{\CurrentBib}

\bibitem [\protect \citeauthoryear {%
Singer%
}{%
Singer%
}{%
{\protect \APACyear {1978}}%
}]{%
Singer:1978dk}
\APACinsertmetastar {%
Singer:1978dk}%
\begin{APACrefauthors}%
Singer, I\BPBI M.%
\end{APACrefauthors}%
\unskip\
\newblock
\APACrefYearMonthDay{1978}{}{}.
\newblock
{\BBOQ}\APACrefatitle {{Some Remarks on the Gribov Ambiguity}} {{Some Remarks
  on the Gribov Ambiguity}}.{\BBCQ}
\newblock
\APACjournalVolNumPages{Commun. Math. Phys.}{60}{}{7-12}.
\newblock
\begin{APACrefDOI} \doi{10.1007/BF01609471} \end{APACrefDOI}
\PrintBackRefs{\CurrentBib}

\bibitem [\protect \citeauthoryear {%
Sorkin%
}{%
Sorkin%
}{%
{\protect \APACyear {1983}}%
}]{%
Sorkin1983}
\APACinsertmetastar {%
Sorkin1983}%
\begin{APACrefauthors}%
Sorkin, R.%
\end{APACrefauthors}%
\unskip\
\newblock
\APACrefYearMonthDay{1983}{}{}.
\newblock
{\BBOQ}\APACrefatitle {On the entropy of the vacuum outside a horizon} {On the
  entropy of the vacuum outside a horizon}.{\BBCQ}
\newblock
\BIn{} \APACrefbtitle {Tenth International Conference on General Relativity and
  Gravitation (held Padova, 4-9 July, 1983), Contributed Papers} {Tenth
  international conference on general relativity and gravitation (held padova,
  4-9 july, 1983), contributed papers}\ (\BVOL~2, \BPGS\ 734--736).
\PrintBackRefs{\CurrentBib}

\bibitem [\protect \citeauthoryear {%
Srednicki%
}{%
Srednicki%
}{%
{\protect \APACyear {1993}}%
}]{%
Srednicki1993}
\APACinsertmetastar {%
Srednicki1993}%
\begin{APACrefauthors}%
Srednicki, M.%
\end{APACrefauthors}%
\unskip\
\newblock
\APACrefYearMonthDay{1993}{Aug}{}.
\newblock
{\BBOQ}\APACrefatitle {Entropy and area} {Entropy and area}.{\BBCQ}
\newblock
\APACjournalVolNumPages{Physical Review Letters}{71}{}{666--669}.
\newblock
\begin{APACrefURL} \url{https://link.aps.org/doi/10.1103/PhysRevLett.71.666}
  \end{APACrefURL}
\newblock
\begin{APACrefDOI} \doi{10.1103/PhysRevLett.71.666} \end{APACrefDOI}
\PrintBackRefs{\CurrentBib}

\bibitem [\protect \citeauthoryear {%
Strocchi%
}{%
Strocchi%
}{%
{\protect \APACyear {2015}}%
}]{%
Strocchi_phil}
\APACinsertmetastar {%
Strocchi_phil}%
\begin{APACrefauthors}%
Strocchi, F.%
\end{APACrefauthors}%
\unskip\
\newblock
\APACrefYearMonthDay{2015}{}{}.
\newblock
{\BBOQ}\APACrefatitle {{Symmetries, Symmetry Breaking, Gauge Symmetries}}
  {{Symmetries, Symmetry Breaking, Gauge Symmetries}}.{\BBCQ}
\newblock

\PrintBackRefs{\CurrentBib}

\bibitem [\protect \citeauthoryear {%
Teh%
}{%
Teh%
}{%
{\protect \APACyear {2016}}%
}]{%
Teh_emp}
\APACinsertmetastar {%
Teh_emp}%
\begin{APACrefauthors}%
Teh, N\BPBI J.%
\end{APACrefauthors}%
\unskip\
\newblock
\APACrefYearMonthDay{2016}{}{}.
\newblock
{\BBOQ}\APACrefatitle {Galileo’s Gauge: Understanding the Empirical
  Significance of Gauge Symmetry} {Galileo’s gauge: Understanding the
  empirical significance of gauge symmetry}.{\BBCQ}
\newblock
\APACjournalVolNumPages{Philosophy of Science}{83}{1}{93-118}.
\newblock
\begin{APACrefURL} \url{https://doi.org/10.1086/684196} \end{APACrefURL}
\newblock
\begin{APACrefDOI} \doi{10.1086/684196} \end{APACrefDOI}
\PrintBackRefs{\CurrentBib}

\bibitem [\protect \citeauthoryear {%
Thierry-Mieg%
}{%
Thierry-Mieg%
}{%
{\protect \APACyear {1980}}%
}]{%
Thierry-MiegJMP}
\APACinsertmetastar {%
Thierry-MiegJMP}%
\begin{APACrefauthors}%
Thierry-Mieg, J.%
\end{APACrefauthors}%
\unskip\
\newblock
\APACrefYearMonthDay{1980}{}{}.
\newblock
{\BBOQ}\APACrefatitle {Geometrical reinterpretation of Faddeev-Popov ghost
  particles and BRS transformations} {Geometrical reinterpretation of
  faddeev-popov ghost particles and brs transformations}.{\BBCQ}
\newblock
\APACjournalVolNumPages{Journal of Mathematical Physics}{21}{12}{2834-2838}.
\newblock
\begin{APACrefURL} \url{https://doi.org/10.1063/1.524385} \end{APACrefURL}
\newblock
\begin{APACrefDOI} \doi{10.1063/1.524385} \end{APACrefDOI}
\PrintBackRefs{\CurrentBib}

\bibitem [\protect \citeauthoryear {%
't Hooft%
}{%
't Hooft%
}{%
{\protect \APACyear {1980}}%
}]{%
thooft}
\APACinsertmetastar {%
thooft}%
\begin{APACrefauthors}%
't Hooft, G.%
\end{APACrefauthors}%
\unskip\
\newblock
\APACrefYearMonthDay{1980}{}{}.
\newblock
{\BBOQ}\APACrefatitle {{Gauge Theories and the Forces Between Elementary
  Particles}} {{Gauge Theories and the Forces Between Elementary
  Particles}}.{\BBCQ}
\newblock
\APACjournalVolNumPages{Scientific American, 242, pp. 90-166}{}{}{}.
\PrintBackRefs{\CurrentBib}

\bibitem [\protect \citeauthoryear {%
Wallace%
}{%
Wallace%
}{%
{\protect \APACyear {2014}}%
}]{%
Wallace_deflating}
\APACinsertmetastar {%
Wallace_deflating}%
\begin{APACrefauthors}%
Wallace, D.%
\end{APACrefauthors}%
\unskip\
\newblock
\APACrefYearMonthDay{2014}{}{}.
\newblock
{\BBOQ}\APACrefatitle {{Deflating the Aharonov-Bohm Effect}} {{Deflating the
  Aharonov-Bohm Effect}}.{\BBCQ}
\newblock
\APACjournalVolNumPages{arxiv: 1407.5073}{}{}{}.
\PrintBackRefs{\CurrentBib}

\bibitem [\protect \citeauthoryear {%
Wallace%
}{%
Wallace%
}{%
{\protect \APACyear {2019}}%
{\protect \APACexlab {{\protect \BCnt {1}}}}}]{%
Wallace2019a}
\APACinsertmetastar {%
Wallace2019a}%
\begin{APACrefauthors}%
Wallace, D.%
\end{APACrefauthors}%
\unskip\
\newblock
\APACrefYearMonthDay{2019{\protect \BCnt {1}}}{}{}.
\newblock
\APACrefbtitle {Isolated Systems and their Symmetries, Part I: General
  Framework and Particle-Mechanics Examples.} {Isolated systems and their
  symmetries, part i: General framework and particle-mechanics examples.}
\newblock
\begin{APACrefURL} \url{http://philsci-archive.pitt.edu/16623/}
  \end{APACrefURL}
\PrintBackRefs{\CurrentBib}

\bibitem [\protect \citeauthoryear {%
Wallace%
}{%
Wallace%
}{%
{\protect \APACyear {2019}}%
{\protect \APACexlab {{\protect \BCnt {2}}}}}]{%
Wallace2019b}
\APACinsertmetastar {%
Wallace2019b}%
\begin{APACrefauthors}%
Wallace, D.%
\end{APACrefauthors}%
\unskip\
\newblock
\APACrefYearMonthDay{2019{\protect \BCnt {2}}}{}{}.
\newblock
{\BBOQ}\APACrefatitle {{Isolated systems and their symmetries, part II: local
  and global symmetries of field theories}} {{Isolated systems and their
  symmetries, part II: local and global symmetries of field theories}}.{\BBCQ}
\newblock
\begin{APACrefURL} \url{http://philsci-archive.pitt.edu/16624/}
  \end{APACrefURL}
\PrintBackRefs{\CurrentBib}

\bibitem [\protect \citeauthoryear {%
Wallace%
}{%
Wallace%
}{%
{\protect \APACyear {2019}}%
{\protect \APACexlab {{\protect \BCnt {3}}}}}]{%
Wallace2019}
\APACinsertmetastar {%
Wallace2019}%
\begin{APACrefauthors}%
Wallace, D.%
\end{APACrefauthors}%
\unskip\
\newblock
\APACrefYearMonthDay{2019{\protect \BCnt {3}}}{}{}.
\newblock
{\BBOQ}\APACrefatitle {Observability, redundancy and modality for dynamical
  symmetry transformations} {Observability, redundancy and modality for
  dynamical symmetry transformations}.{\BBCQ}
\newblock
\APACjournalVolNumPages{Forthcoming}{}{}{}.
\newblock
\begin{APACrefURL} \url{http://philsci-archive.pitt.edu/18813/}
  \end{APACrefURL}
\newblock
\APACrefnote{Revised 3/2021 to correct a few typos and add a section on
  Noether's Theorem.}
\PrintBackRefs{\CurrentBib}

\bibitem [\protect \citeauthoryear {%
Wilkins%
}{%
Wilkins%
}{%
{\protect \APACyear {1989}}%
}]{%
YangMillsSlice}
\APACinsertmetastar {%
YangMillsSlice}%
\begin{APACrefauthors}%
Wilkins, D\BPBI R.%
\end{APACrefauthors}%
\unskip\
\newblock
\APACrefYearMonthDay{1989}{}{}.
\newblock
{\BBOQ}\APACrefatitle {Slice Theorems in Gauge Theory} {Slice theorems in gauge
  theory}.{\BBCQ}
\newblock
\APACjournalVolNumPages{Proceedings of the Royal Irish Academy. Section A:
  Mathematical and Physical Sciences}{89A}{1}{13--34}.
\newblock
\begin{APACrefURL} \url{http://www.jstor.org/stable/20489307} \end{APACrefURL}
\PrintBackRefs{\CurrentBib}

\end{thebibliography}

\end{document}